\documentclass{aa501}     
\usepackage{graphicx}
\def\cm2{\,{\rm cm^{-2}}}
\def\kms{\,{\rm {km\,s^{-1}}}}

\def\aua{{\rm A\&A} }
\def\auas{{\rm A\&AS} }
\def\apj{{\rm ApJ} }
\def\aj{{\rm AJ} }
\def\apjs{{\rm ApJS} }

\def\mnras{{\rm MNRAS} }

\begin{document}
 
\title{Neutral hydrogen in dwarf galaxies}
 
   \subtitle{II. The kinematics of HI}
 
\author{J.M. Stil
          \inst{1, 2}
           and F. P. Israel
          \inst{1}
           }
 
   \offprints{F.P. Israel}
 
  \institute{Sterrewacht Leiden, P.O. Box 9513, 2300 RA Leiden,
             The Netherlands
  \and       Physics Department, Queen's University, Kingston ON K7L 4P1, 
	     Canada}
 
\date{Received ????; accepted ????}
 
\abstract{This paper is the second in a series presenting a sample of 
29 late-type dwarf galaxies observed with the Westerbork Synthesis Radio 
Telescope in the 21-cm line of neutral hydrogen (HI). Here we present 
rotation curves, maps of the velocity field and maps of the velocity 
dispersion across the sample galaxies.
}

\maketitle
 
\section{Introduction.}        

In this paper we we present the kinematical information extracted from
HI maps of the dwarf galaxy sample described by Stil $\&$ Israel (2002; 
hereafter Paper I). As neutral atomic hydrogen is one of the most 
extended observable components of a galaxy, its line emission provides 
an excellent tool to probe galaxy mass distributions. The dopplershift 
of a line profile is a direct measure of the projected rotation velocity 
at the position sampled. Its linewidth is a measure of the macroscopic 
chaotic motion of the gas and reflects its physical condition. Both 
rotation and velocity dispersion can be measured out to the edge of the 
HI distribution, which usually extends far beyond the stellar 
distribution. In the outer regions, rotational velocities are more or 
less constant, implying the presence of large amounts of matter even at 
the largest distances to the center. These rotation velocities are about 
three times higher than expected if only stars and gas would contribute 
to the mass; thus the amount of `unseen' or `dark' mass is about an order 
of magnitude higher than that associated with luminous matter.

%
\begin{table*}
\caption{HI disk parameters from tilted-ring fits; inclination fitted}
\begin{center}
\tabcolsep=1.0mm
\begin{tabular}{| l | c | r  r |  c  c | r | r | c |} 
\hline  
Name    & resolution &\multicolumn{2}{| c |}{kinematic center}  & incl. &fixed/free & PA\ \ \ \ \  & $v_{\rm sys}$\ \ \ \ \  & scale \\
\hline
        & arcsec     &$\alpha_{1950}$\ \ \ \ \ \ \ \ \ \ \ \  &$\delta_{1950}$\ \ \ \ \ \ \ \ \ \ \  &    degr. &         & degr.&$\rm km\ s^{-1}$&kpc/$'$\\
\hline
\ \ \ \ \ [1]  &[2] &[3] \ \ \ \ \ \ \ \ \ \ \ \ \ &[4]\ \ \ \ \ \ \ \ \ \ \ \ \ &   [5]  &  [6]    &[7]\ \ \ \ \ &[8]\ \ \ \ \ \ &  [9]  \\    
\hline
DDO 46  &13.5&$\rm 7^h38^m 00^s.9\ \pm 0^s.4$          &$\rm 40^\circ13'37''\ \pm\ \ 5''$  &$\rm 45$    &fixed&$\rm 270\pm5$&$362 \pm 2$&1.4 \\
DDO 47  &13.5&$\rm 7^h39^m 03^s.1$ \ \ \ \ \ \ \ \ \ \ &$\rm 16^\circ55'13''$ \ \ \ \ \ \ \ \ \ &$\rm 30$    &fixed&$\rm 318\pm8$&$272 \pm 2$&0.58\\
        &30  &$\rm 7^h39^m 03^s.1$ \ \ \ \ \ \ \ \ \ \ &$\rm 16^\circ55'13''$ \ \ \ \ \ \ \ \ \ &$\rm 30$    &fixed&$\rm 316\pm9$&$272 \pm 2$&    \\
DDO 48  &13.5&$\rm 7^h54^m 46^s.4\ \pm 0^s.2$          &$\rm 58^\circ10'43''\ \pm\ \ 3''$     &$\rm 75\pm1$&free &$\rm 356\pm1$&$1087\pm 1$&4.6 \\
        &30  &$\rm 7^h54^m 46^s.4\ \pm 0^s.3$          &$\rm 58^\circ10'43''\ \pm10''$        &$\rm 80\pm5$&free &$\rm 356\pm2$&$1088\pm 2$&    \\
NGC 2537&30  &$\rm 8^h09^m 42^s.6\ \pm 0^s.4$          &$\rm 46^\circ08'40''\ \pm\ \ 4''$     &$\rm 42\pm3$&free &$\rm 174\pm6$&$444 \pm 1$&1.9 \\
UGC 4278&13.5&$\rm 8^h10^m 27^s.4\ \pm 0^s.5$          &$\rm 45^\circ53'52''\ \pm\ \ 4''$     &$\rm 80\pm3$&free &$\rm 351\pm1$&$564 \pm 1$&1.9 \\
NGC 2976&13.5&$\rm 9^h43^m 08^s.0\ \pm 1^s.0$          &$\rm 68^\circ08'57''\ \pm\ \ 5''$     &$\rm 65\pm3$&free &$\rm 326\pm2$&$4   \pm 2$&1.0 \\
        &30  &$\rm 9^h43^m 07^s.7\ \pm 0^s.4$          &$\rm 68^\circ08'52''\ \pm\ \ 4''$     &$\rm 62\pm2$&free &$\rm 325\pm4$&$4   \pm 1$&    \\
DDO 83  &13.5&$\rm 10^h33^m54^s.4\ \pm 0^s.1$          &$\rm 31^\circ48'24''\ \pm\ \ 2''$     &$\rm 66\pm2$&free &$\rm 59 \pm3$&$582 \pm 2$&2.6 \\
DDO 87  &30  &$\rm 10^h46^m17^s.0\ \pm 0^s.5$          &$\rm 65^\circ47'35''\ \pm\ \ 5''$     &$\rm 63\pm4$&free &$\rm 239\pm3$&$338 \pm 2$&1.0 \\
DDO 123 &13.5&$\rm 12^h23^m47^s.0\ \pm 0^s.3$          &$\rm 58^\circ35'51''\ \pm\ \ 7''$     &$\rm 25$    &fixed&$\rm 201\pm4$&$722 \pm 1$&3.3 \\
DDO 133 &30  &$\rm 12^h30^m26^s.8\ \pm 1^s.2$          &$\rm 31^\circ48'46''\ \pm15''$        &$\rm 20$    &fixed&$\rm 353\pm4$&$330 \pm 3$&1.5 \\
DDO 168 &30  &$\rm 13^h12^m15^s.0\ \pm 1^s.5$          &$\rm 46^\circ11'24''\ \pm\ \ 6''$     &$\rm 63\pm3$&free &$\rm 276\pm1$&$190 \pm 3$&1.0 \\
DDO 185 &30  &$\rm 13^h52^m53^s.3\ \pm 0^s.5$          &$\rm 54^\circ08'30''\ \pm\ \ 4''$     &$\rm 66\pm5$&free &$\rm 18 \pm2$&$140 \pm 2$&2.0 \\
DDO 217 &30  &$\rm 23^h27^m32^s.1\ \pm 1^s.1$          &$\rm 40^\circ42'56''\ \pm 15''$       &$\rm 46\pm5$&free &$\rm 35 \pm3$&$428 \pm 2$&2.7 \\
\hline
\end{tabular}
\end{center}
Column designations: [1] Object name; [2] resolution of dataset used;
[3] and [4] right ascension and declination (epoch 1950) of kinematic center 
and their r.m.s. scatter between radii;
[5] inclination in degrees and its error if it was a free parameter in the fit;
[6] flag indicating whether the inclination was a free parameter;
[7] position angle in degrees and its r.m.s. scatter between radii;
[8] heliocentric systemic velocity in $\kms$ and its r.m.s. scatter between 
radii;
[9] radial scale of the rotation curve in kpc/arcmin.\\
\\
Notes: DDO~168 position angle is mean for radii $150''$ and $180''$; 
DDO~217 position angle is mean for radii larger than $150$ arcseconds.
\label{cnvrc1}
\end{table*}

%
\begin{table*}
\caption{Rotation velocities ($\kms$) from tilted ring fits}
\begin{center}
\tabcolsep=1.0mm
\begin{tabular}{| c  | c | c |  c  | c | c |  c  | c | c | c |} 
\hline  
radius & DDO\,46    & DDO\,47    & DDO\,47    & DDO\,48     & DDO\,48   & NGC\,2537  & UGC\,4278  & NGC\,2976        & NGC\,2976    \\
$''$   & HR         & HR         & LR         & HR          & LR        & LR         & HR         & HR               & LR           \\ 
\hline
   15  &$15.8\pm5.2$&$20.6\pm0.1$&            &$30.0\pm7.6$ &            &            &\ \ $8\pm5$ &$20.9\pm5$\ \ \ \ &                    \\ 
   30  &$36.0\pm4.4$&$24.5\pm1.5$&$24.4\pm0.4$&$47.5\pm0.3$ &$40.9\pm3.9$&$46.2\pm2.0$&$25.6\pm1.5$&$28.8\pm3$\ \ \ \ &$29.6\pm1.3$ \\
   45  &$41.2\pm3.2$&$27.1\pm3.0$&            &$59.4\pm0.3$ &            &            &$40.1\pm0.5$&$40.4\pm1.8$      &                    \\ 
   60  &$44.3\pm3.7$&$25.7\pm1.6$&$27.1\pm0.5$&$66.8\pm0.4$ &$63.3\pm9.8$&$53.4\pm4.3$&$50.1\pm1.4$&$49.6\pm2.8$      &$51.6\pm0.3$ \\ 
   75  &$46.0\pm3.7$&$30.4\pm1.0$&            &$72.0\pm0.2$ &            &            &$59.3\pm0.5$&$61.5\pm0.4$      &                    \\ 
   90  &$44.8\pm6.6$&$36.1\pm0.5$&$36.4\pm0.5$&$74.9\pm0.1$ &$74.3\pm6.7$&$63.2\pm3.3$&$68.5\pm1.1$&$69.1\pm1.2$      &$69.2\pm0.1$ \\ 
  105  &            &$41.7\pm0.5$&            &$76.6\pm0.5$ &            &            &$76.4\pm0.3$&$71.1\pm0.4$      &                    \\ 
  120  &            &$48.2\pm1.1$&$47.9\pm0.5$&             &            &$71.9\pm5.2$&$82.5\pm2.5$&$71.5\pm0.2$      &$71.7\pm0.3$ \\ 
  135  &            &$55.0\pm0.5$&            &             &            &            &$87.4\pm2.1$&                  &                    \\
  150  &            &$62.0\pm0.8$&$60.6\pm0.4$&             &            &            &$87.6\pm1.8$&                  &                    \\
  165  &            &$66.4\pm0.7$&            &             &            &            &            &                  &                    \\
  180  &            &            &$68.0\pm1.3$&             &            &            &            &                  &                    \\
\hline
radius & DDO\,83    & DDO\,87    & DDO\,123   & DDO\,133    & DDO\,168   & DDO\,185   & DDO\,217   & & \\ 
$''$   & HR         & LR         & HR         & LR          & LR         & LR         & LR         & & \\
\hline
   15  &$23.7\pm7.5$&            &$26.1\pm0.9$&             &            &            &            & & \\
   30  &$35.6\pm0.4$&17.6$\pm$1.3&$35.5\pm4.8$&$34.6\pm5.7$ &$11.2\pm2.0$&$17.8\pm4.7$&$26.1\pm7.5$& & \\ 
   45  &$46.1\pm0.2$&            &$50.8\pm0.4$&             &            &            &            & & \\
   60  &$51.7\pm0.3$&30.0$\pm$0.6&$41.8\pm0.2$&$50.3\pm14$\ &$26.6\pm2.0$&$28.1\pm0.4$&$58.7\pm6.0$& & \\
   75  &$51.0\pm3.0$&            &$49.3\pm0.8$&             &            &            &            & & \\
   90  &            &32.3$\pm$0.4&$55.5\pm0.2$&$65.8\pm11$\ &$29.7\pm1.1$&$40.1\pm0.5$&$66.2\pm3.8$& & \\
  105  &            &            &$59.9\pm0.7$&             &            &            &            & & \\
  120  &            &34.4$\pm$1.8&$63.1\pm3.0$&$75.7\pm2.1$ &$39.4\pm1.4$&$47.3\pm0.3$&$68.1\pm3.3$& & \\
  150  &            &            &            &$75.1\pm4.2$ &$42.4\pm1.0$&$51.9\pm2.3$&$72.5\pm2.8$& & \\
  180  &            &            &            &             &$44.3\pm0.5$&    	      &$74.9\pm3.3$& & \\
  210  &            &            &            &             &   	 &	      &$77.0\pm0.8$& & \\
  240  &            &            &            & 	    &  	         &            &$78.1\pm2.1$& & \\
\hline
\end{tabular}
\end{center}
\centerline{Note: HR is based on high-resolution (13.5$''$) fits; 
LR is based on low-resolution (30$''$) fits.}
\label{cnvrc2}
\end{table*}

It is interest to determine the detailed kinematical conditions governing 
late-type dwarf galaxies. These galaxies are gas-rich, dynamically simple 
and relatively easy to observe.  Their rotation curves trace the distribution 
of dark matter more directly than those of more massive galaxies, where the 
mass-to-light ratio of the stellar disk and bulge is a critical parameter.
Interestingly, it has been suggested that in some dwarf galaxies, the 
stellar contribution to the total mass is quite small (Broeils 1992),
and that this contribution decreases with decreasing maximum rotation curve 
velocity (Broeils 1992, Swaters 1999), 

\section{Velocity field and rotation curve}                       

We determined intensity-weighted mean velocities in order to make maximum
use of the information contained in spectra with limited signal-to-noise 
ratios. Noise was further suppressed by only using data from the areas 
delineated by the cleaning masks (cf. Paper I). The intensity-weighted 
velocity fields of our galaxy sample are shown in Fig.~4.

We have determined rotation curves by iteratively fitting to the observed 
velocity field the parameters found with the tilted-ring method (Warner et 
al. 1973), incorporated in the GIPSY packages as the task ROTCUR (Begeman 
1987). Note that the inclination and the location of the kinematic center 
can be determined only if the rotation curve flattens at the outer radii (i.e. 
shows the onset of differential rotation). In order to obtain robust 
solutions, we performed a large number of fits where each parameter was 
free in 25 to 30 fits, allowing calculation of the r.m.s. scatter of each 
parameter at all radii. Solution of $v(r)\sin\,i$ can also directly be 
compared to observed major axis position-velocity maps (Sect.~3). 
Although the product $v(r)\sin\,i$ is well-constrained, the separate 
solutions for $v(r)$ and $i$ in general are not unique. In particular, at 
inclination angles $\leq 50^\circ$, $v(r)$ and $i$ cannot be fitted 
independently (Begeman 1987). 

Depending on galaxy HI extent and surface brightness, we performed fits
on full-resolution ($13.5''$) or low-resolution (30$''$) data. As a
consistency check, we used both low and full resolution data for {\small
DDO\,}47, {\small DDO\,}48 and {\small NGC\,}2976. The results are 
presented in Table~\ref{cnvrc1} and \ref{cnvrc2}.
In total, 13 out of 29 galaxies in the sample could so be fitted. 
For the remainder, a complete analysis was not feasible. There, we 
required the rotation center to coincide with the HI center of mass
which was found to be correct at least for all the objects listed in 
Tables~\ref{cnvrc1}. In addition, we assigned to these objects a low 
($30^\circ$), average ($60^\circ$) or high ($80^\circ$) inclination 
based on the appearance of the HI isophotes. The results of fits
to the high-resolution data, but restricted by these assumptions, are 
given in Table~\ref{tworing1} and Table~\ref{tworing2}. The center
coordinates in Table~\ref{tworing1} are those of the HI-intensity-weighted
mean position in $\alpha_{1950}$ and $\delta_{1950}$. They typically 
change only by a few arcseconds if different intensity thresholds are 
applied. The errors quoted for position angle and systemic velocity 
represent their scatter between different radii. Note that the 
velocities listed in Table~\ref{tworing2} have {\it not} been corrected 
for inclination. These $v(r)\sin i$ values, as indeed also the position 
angle, depend only weakly on the assumed inclination.

%
\begin{table*}
\caption{HI disk parameters from restricted tilted-ring/HI isophote fits; inclination assumed}
\begin{center}
\tabcolsep=1.0mm
\begin{tabular}{| l  | r  r |  c | r | r | c | c | r | c |} 
\hline  
Name    &\multicolumn{2}{| c |}{HI center of mass}  & $i$ & $\rm PA_{kin}$ & $v_{\rm sys}$\ \ \ \ \  & $D_{\rm HI}$ & $q_{\rm HI}$ & $\rm PA_{HI}$ & Scale \\
\hline
        &$\alpha_{1950}$\ \ \ \ \ \ \ &$\delta_{1950}$\ \ \ \ \ \ & $^\circ$ & $^\circ$\ \ \ \ \ &$\kms$& arcsec &  & $^\circ$\ \ \ \  &kpc/$'$\\
\hline
\ \ \ \ \ [1]   &[2] \ \ \ \ \ \ \ \ &[3]\ \ \ \ \ \ \ \ &   [4]  &  [5]\ \ \ \ &[6]\ \ \ \ \ &[7]\ \ & [8] & [9]\ \ \ & [10] \\    
\hline
D\,22 &$\rm 2^h29^m47^s.2$&$\rm 38^\circ27'35''$&80&$\rm 178\pm 1 $\ \ \ &$\rm 564 \pm 2$ &157 $\pm$ 9\ \ \  &0.47 $\pm$ 0.04&181 $\pm$ 2\ \ \ &2.9 \\
D\,43 &$\rm 7^h24^m50^s.2$&$\rm 40^\circ52'19''$&30&$\rm 296\pm 4 $\ \ \ &$\rm 355 \pm 1$ &145 $\pm$ 3\ \ \  &0.93 $\pm$ 0.03 &335 $\pm$ 11&1.4 \\
D\,46$^{*}$ &$\rm 7^h38^m00^s.9$&$\rm 40^\circ13'30''$&30&$\rm 273\pm 4 $\ \ \ &$\rm 363 \pm 2$ &189 $\pm$ 5\ \ \  &0.97 $\pm$ 0.05 &226 $\pm$ 45&1.4 \\
D\,47$^{*}$ &$\rm 7^h39^m03^s.1$&$\rm 16^\circ55'13''$&30&$\rm 319\pm 8$\ \ \ &$\rm 272 \pm 1$ &403 $\pm$ 4\ \ \  &0.85 $\pm$ 0.01 &266 $\pm$ 3\ \ \ &0.58 \\
D\,48$^{*}$ &$\rm 7^h54^m46^s.8$&$\rm 58^\circ10'39''$&80&$\rm 357\pm 3 $\ \ \ &$\rm 1086 \pm 1$ &245 $\pm$ 12 &0.33 $\pm$ 0.02 &354 $\pm$ 2\ \ \ &4.6 \\
N\,2537$^{*}$ &$\rm 8^h09^m42^s.3$&$\rm 46^\circ08'32''$&30&$\rm 171\pm 2 $\ \ \  &$\rm 446 \pm 1$ &219 $\pm$ 5\ \ \  &0.80 $\pm$ 0.04 &261 $\pm$ 6\ \ \ &1.9 \\
D\,52 &$\rm 8^h25^m06^s.4$&$\rm 42^\circ01'17''$&60&$\rm 5 \pm 3$\ \ \  &$\rm 394 \pm 1$ &204 $\pm$ 9\ \ \  &0.58 $\pm$ 0.04 &359 $\pm$ 3\ \ \ &1.5 \\
D\,63 &$\rm 9^h36^m01^s.9$&$\rm 71^\circ25'07''$&30&$\rm 38\pm 15 $ &$\rm 140 \pm 2$ &307 $\pm$ 6\ \ \  &0.76 $\pm$ 0.02 &5 $\pm$ 2\ \ \ &1.0 \\
N\,2976$^{*}$ &$\rm 9^h43^m11^s.7$&$\rm 68^\circ09'06''$&60&$\rm 321 \pm 2$\ \ \  &$\rm   3 \pm 2$ &316 $\pm$ 5\ \ \  &0.62 $\pm$ 0.01 &326 $\pm$ 1\ \ \ &1.0 \\
D\,64 &$\rm 9^h47^m26^s.3$&$\rm 31^\circ43'19''$&60&$\rm 97\pm 7$\ \ \  &$\rm 517 \pm 1$ &210 $\pm$ 11 &0.56 $\pm$ 0.05 &84 $\pm$ 4\ \ \ &1.8 \\
D\,68 &$\rm 9^h53^m52^s.8$&$\rm 29^\circ03'42''$&60&$\rm 29\pm 10 $ &$\rm  504 \pm 3$ &300 $\pm$ 9\ \ \  &0.56 $\pm$ 0.03 &19 $\pm$ 2\ \ \ &1.8\\
D\,73 &$\rm 10^h06^m39^s.5$&$\rm 30^\circ23'50''$&60&$\rm 66\pm 3$\ \ \  &$\rm  1378 \pm 2$ &135 $\pm$ 3\ \ \  &0.94 $\pm$ 0.03 &38 $\pm$ 15&5.2\\
D\,83$^{*}$ &$\rm 10^h33^m54^s.7$&$\rm 31^\circ48'24''$&60&$\rm 55\pm 3$\ \ \  &$\rm  584 \pm 1$ &190 $\pm$ 4\ \ \  &0.55 $\pm$ 0.02 &56 $\pm$ 2\ \ \ &2.6 \\
D\,87$^{*}$ &$\rm 10^h46^m16^s.3$&$\rm 65^\circ47'34''$&60&$\rm 239\pm 5$\ \ \  &$\rm 339 \pm 1$ &246 $\pm$ 5\ \ \  &0.86 $\pm$ 0.03 &224 $\pm$ 6\ \ \ &1.0  \\
M\,178 &$\rm 11^h30^m45^s.1$&$\rm 49^\circ31'06''$&-- &-- \ \ \ \ \ &-- \ \ \ \ &91 $\pm$ 4\ \ &0.70 $\pm$ 0.06 &146 $\pm$ 7\ \ \ &1.5  \\
N\,3738 &$\rm 11^h33^m04^s.7$&$\rm 54^\circ48'10''$&60&$\rm   270$\ \ \ \ \ \ &$\rm  225 \pm 4$ &182 $\pm$ 6\ \ \  &0.80 $\pm$ 0.05 &274 $\pm$ 10 &1.5  \\
D\,101 &$\rm 11^h53^m07^s.1$&$\rm 31^\circ47'41''$&--&-- \ \ \ \ \ &-- \ \ \ \ &--\ \ \ &-- &-- \ \ \ \ \ \ &2.1  \\
D\,123$^{*}$ &$\rm 12^h23^m45^s.9$&$\rm 58^\circ35'52''$&18&$\rm   198 \pm 5$\ \ \ &$\rm 723 \pm 1$ &240 $\pm$ 2\ \ \  &0.97 $\pm$ 0.01 &197 $\pm$ 13 &3.3\\
M\,209 &$\rm 12^h23^m50^s.2$&$\rm 48^\circ46'25''$&30&$\rm 225 $\ \ \ \ \ \ &$\rm 285 \pm 8$ &125 $\pm$ 5\ \ \  &0.86 $\pm$0.08 &206 $\pm$ 20 &1.4 \\
D\,125 &$\rm 12^h25^m14^s.0$&$\rm 43^\circ46'17''$&60&$\rm   135\pm 2$\ \ \ &$\rm  195 \pm 1$ &281 $\pm$ 3\ \ \  &0.66 $\pm$ 0.02 &118 $\pm$ 2\ \ \ &1.3 \\
D\,133$^{*}$ &$\rm 12^h30^m27^s.6$&$\rm 31^\circ48'54''$&30&$\rm    350 \pm 5 $\ \ \  &$\rm 331 \pm 1$&296 $\pm$ 5\ \ \  &0.86 $\pm$ 0.02 &18 $\pm$ 5\ \ \ &1.5 \\
D\,165 &$\rm 13^h04^m40^s.2$&$\rm 67^\circ58'20'' $&--&$135$\ \ \  &$22$\ \ \ \ \ \ &253 $\pm$ 5\ \ \  &0.79 $\pm$ 0.03 &93  $\pm$ 4\ \ \ &1.3 \\
D\,166 &$\rm 13^h11^m00^s.0$&$\rm 36^\circ28'36''$&30&$\rm 41\pm 9$\ \ \ &$\rm 942 \pm 3$&200 $\pm$ 3\ \ \  &0.75 $\pm$ 0.02 &46 $\pm$ 3\ \ \ &4.7 \\
D\,168$^{*}$ &$\rm 13^h12^m14^s.7$&$\rm 46^\circ11'07''$&60&$\rm 276 \pm 1 $\ \ \ &$\rm 191 \pm 2$&366 $\pm$ 5\ \ \ &0.82 $\pm$ 0.02 &304 $\pm$ 3\ \ \ &1.0 \\
D\,185$^{*}$ &$\rm 13^h52^m53^s.5$&$\rm 54^\circ08'24''$&80&$\rm 39\pm 5$\ \ \ &$\rm 137 \pm 3$&302 $\pm$ 13 &0.37 $\pm$ 0.02 &17 $\pm$ 1\ \ \ &2.0 \\
D\,190 &$\rm 14^h22^m48^s.2$&$\rm 44^\circ45'09''$&60&$\rm 149\pm 6$\ \ \ &$\rm  149 \pm 1$ &201 $\pm$ 4\ \ \  &0.86 $\pm$ 0.03 &150 $\pm$ 7\ \ \ &1.7  \\
D\,216 &$\rm 23^h26^m02^s.9$&$\rm 14^\circ28'02'' $&60&$135$\ \ \ \ \ \ &-- \ \ \ \ &286 $\pm$ 17&0.32 $\pm$ 0.02 &130 $\pm$ 1\ \ \ &0.29  \\
D\,217$^{*}$ &$\rm 23^h27^m32^s.8$&$\rm 40^\circ43'03'' $&30&$\rm 43\pm 8$\ \ \ &$\rm 432 \pm 2$&472 $\pm$ 5\ \ \  &0.75 $\pm$ 0.01 &27 $\pm$ 2\ \ \ &2.7  \\
\hline
\end{tabular}
\end{center}
Column designations
[1] Object name;
[2] and [3] right asccension and declination (epoch 1950) of kinematic center and its r.m.s. scatter between radii;
[4] inclination in degrees (see text for explanation);
[5] position angle of kinematic major axis in degrees and its r.m.s. scatter between radii;
[6] systemic velocity in $\kms$ and its r.m.s. scatter between radii;
[7] major axis of ellipse fitted to the $N_{\rm HI}=3 \cdot 10^{20}\ \cm2$ contour in the HI column density map;
[8] axial ratio of the ellipse in [7];
[9] position angle of the ellipse major axis  in [7]; 
[10] radial scale of the rotation curve in kpc/arcmin. \\
\\
Notes: If no error is given, the value is a best estimate.
For galaxies marked with an asterisk, more accurate values are given in 
Table~\ref{cnvrc1}; they are included here only for comparison purposes.
\label{tworing1}
\end{table*}

%
\begin{table*}
\caption{Velocity $v(r) \sin(i)$ from restricted rotation curve fits}. 
\begin{center}
\begin{tabular}{| c | c | c |  c |  c  | c | c | c | c | c |} 
\hline 
radius &DDO\,22     & DDO\,43    &DDO\,46$^{*}$&DDO\,47$^{*}$&DDO\,48$^{*}$&NGC\,2537$^{*}$& DDO\,52    & DDO\,63    &NGC\,2976$^{*}$\\ 
$''$   &            &            &            &            &            &            &            &            &             \\
\hline 
 30    &$ 8.1\pm2.4$&$11.7\pm2.0$&$23.7\pm3.3$&$12.2\pm2.0$&$42.4\pm6.5$&$33.6\pm3.2$&$23.7\pm3.6$&$ 6.3\pm2.3$&$25.6\pm5.6$ \\ 
 60    &$18.7\pm2.8$&$15.7\pm2.8$&$29.9\pm2.1$&$13.4\pm2.3$&$62.7\pm5.0$&$38.9\pm2.5$&$32.7\pm2.6$&$ 8.0\pm1.9$&$44.7\pm3.8$ \\ 
 90    &   	    &$17.5\pm3.9$&$30.2\pm3.6$&$18.3\pm1.7$&$73.0\pm5.0$&$43.2\pm3.5$&$38.2\pm2.5$&$ 7.1\pm2.6$&$57.2\pm5.3$ \\ 
120    &   	    &   	 &            &$24.1\pm2.2$&		&   	     &$43.3\pm5.0$&$ 8.6\pm3.2$&$61.0\pm7.6$ \\ 
150    &   	    &            &   	      &$30.6\pm2.1$&   		&   	     &            &$ 8.0\pm5.3$&$54.0\pm\ 11$ \\ 
\hline 
radius & DDO\,64    & DDO\,68    & DDO\,73    &DDO\,83$^{*}$&DDO\,87$^{*}$& NGC\,3738  &DDO\,123$^{*}$& Mk\,209    & DDO\,125    \\ 
$''$   &            &            &            &            &            &            &            &            &             \\
\hline 
 30    &$22.0\pm3.1$&$20.5\pm4.3$&$20.0\pm1.6$&$32.6\pm4.8$&$16.9\pm3.3$&$39.0\pm6.2$&$14.0\pm2.5$&$14.3\pm5.0$&$ 5.4\pm1.8$ \\ 
 60    &$33.4\pm2.9$&$28.4\pm5.7$&$26.6\pm2.4$&$43.7\pm2.5$&$26.7\pm1.8$&            &$18.7\pm1.7$&$17.1\pm5.0$&$ 8.7\pm1.1$ \\ 
 90    &$41.8\pm7.4$&$42.9\pm5.6$&   	      &$46.7\pm3.3$&$28.8\pm2.9$& 	     &$22.6\pm2.6$&            &$ 9.8\pm1.6$ \\ 
120    & 	    &$45.9\pm5.6$&   	      & 	   &$30.2\pm4.0$& 	     &$27.5\pm2.7$&   	       &$11.2\pm2.7$ \\ 
150    &	    &            &   	      &  	   &$31.0\pm5.8$&            & 	          & 	       & 	     \\ 
\hline 
radius &DDO\,133$^{*}$& DDO\,165   & DDO\,166   &DDO\,168$^{*}$&DDO\,185$^{*}$& DDO\,190   & DDO\,216   &DDO\,217$^{*}$&             \\ 
$''$   &            &            &            &            &            &            &            &            &             \\
\hline 
 30    &$11.5\pm3.0$&  9         &$22.6\pm4.3$&$10.7\pm2.5$&$11.7\pm2.2$&$12.3\pm1.8$&$ 1\pm1$    &$24.1\pm4.7$&             \\
 60    &$17.7\pm1.9$& 17  	 &$30.2\pm3.1$&$20.3\pm2.4$&$22.9\pm3.0$&$19.4\pm2.1$&$ 4\pm2$    &$38.1\pm3.1$&             \\ 
 90    &$22.9\pm2.8$& 26  	 &$35.4\pm5.0$&$24.8\pm2.2$&$36.2\pm3.2$&$24.7\pm2.5$&$ 7\pm5$    &$41.3\pm2.8$&             \\ 
120    &$26.6\pm2.2$&   	 & 	      &$35.6\pm5.5$& 	     	&            &$ 7\pm5$    &$43.4\pm2.9$&             \\ 
150    &   	    &	         &   	      &$37.3\pm6.7$&	 	&	     &   	  &$47.0\pm2.0$&             \\ 
\hline 
\end{tabular} 
\end{center}
Notes: See text for fit procedure details. Velocities have 
{\it not} been corrected for inclination, as opposed
to results presented in Table~\ref{cnvrc2}. For galaxies marked with 
an asterisk, more accurate values are given in Table~\ref{cnvrc2};
values given here are only for comparison purposes.
The rotation velocity of {\small DDO\,}165 was determined from the
position-velocity map in position angle $135^\circ$.
\label{tworing2}
\end{table*}

\section{Major axis position-velocity diagrams}                       

The projected rotation curve, $v(r) \sin\,i$, can be visualized as a 
position-velocity slice. We include such maps in Fig.~4, which show 
intensity as a function of position along the kinematic major axis 
and heliocentric velocity. Position angles and central positions were 
taken from Tables~\ref{cnvrc1} or~\ref{tworing1}, in order of preference. 
We have also included in Fig.~4 the rotation curve points listed in 
Tables~\ref{cnvrc2} and ~\ref{tworing2}.

%
\begin{figure}
\begin{minipage}[t]{8.5cm}
\mbox{}\\
\resizebox{\hsize}{!}{\includegraphics[angle=-90]{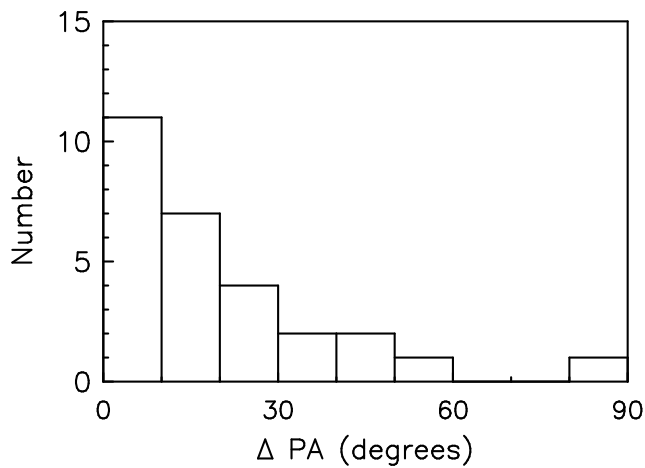}}
\end{minipage}
\caption{\small Histogram of the difference in position angle between the
major axis of the $N_{\rm HI}= 3 \cdot 10^{20} \cm2$ isophote
and the velocity gradient for 27 objects (Mkn\,178 and {\small
DDO\,}101 were excluded because of missing data). The angle 
$\Delta\rm PA$ is the smallest angle between the velocity gradient and the
major axis. The median difference is $\Delta\rm
PA=15^\circ$. The object with $\Delta\rm PA=90^\circ$ is {\small NGC\,}2537.}
\label{PA-hist}
\end{figure}

Virtually all of the observed galaxies show a velocity gradient. However,
in poorly-ordered velocity fields, it is not clear whether or not this 
represents. In  Fig.~\ref{PA-hist} we
show, however, that observed velocity gradients tend to align with the 
major axis of the HI isophotes, although a few significant misalignments 
exist suggesting a rotating disk (e.g. {\small DDO\,}47, {\small DDO\,}63, 
{\small DDO\,}165 and also Sextans A as shown by Skillman et al. 1988). 
We suspect that much of the width of the distribution in Fig.~\ref{PA-hist} 
is caused by errors in the determination of major axis position angle from 
the HI distribution, although the magnitude of $\Delta$ PA is not correlated 
with the total brightness of the galaxy in HI (or in blue light).

The position-velocity maps of three very-low luminosity dwarf galaxies
($M_{\rm B} > -14$) betray high ratio of rotational to random velocities: 
those of {\small DDO\,}47, {\small DDO\,}52 and {\small DDO\,}87. The 
existence of rotationally supported dwarfs of such low luminosity is 
remarkable if simple arguments based on the luminosity-linewidth relation 
for large spirals are applied to dwarf galaxies (Lo et al. 1993, Stil 
$\&$ Israel, in preparation). In addition, the rotation curves of 
{\small DDO\,}52 and {\small DDO\,}87 show clear signs of flattening. 
They illustrate the increase in the number of extremely low luminosity 
galaxies known to be largely supported by rotation (cf. Carignan \& 
Beaulieu 1989, C\^ot\'e 1997). Indeed, with the exception of 
{\small DDO\,}47 Fig.~4 as well as Tables~\ref{cnvrc2}~and~\ref{tworing2} 
include few convincing examples of pure solid-body rotation. 

\section{Velocity dispersion}                       
\label{disp-sec}

The linewidth in a single synthesized beam carries information about the 
velocity dispersion of the ensemble of HI clouds contained within the beam. 
This information is, unfortunately, confused with the signatures of
other effects such as warps or flares of the HI layer, the extent to 
which HI structures are resolved, and the rotation velocity gradient 
over the beam. Warps and flares bring HI from different galactocentric 
radii into the line of sight, but they only cause confusion in highly 
inclined galaxies. 

%
\begin{figure}
\resizebox{9cm}{!}{\includegraphics[angle=0]{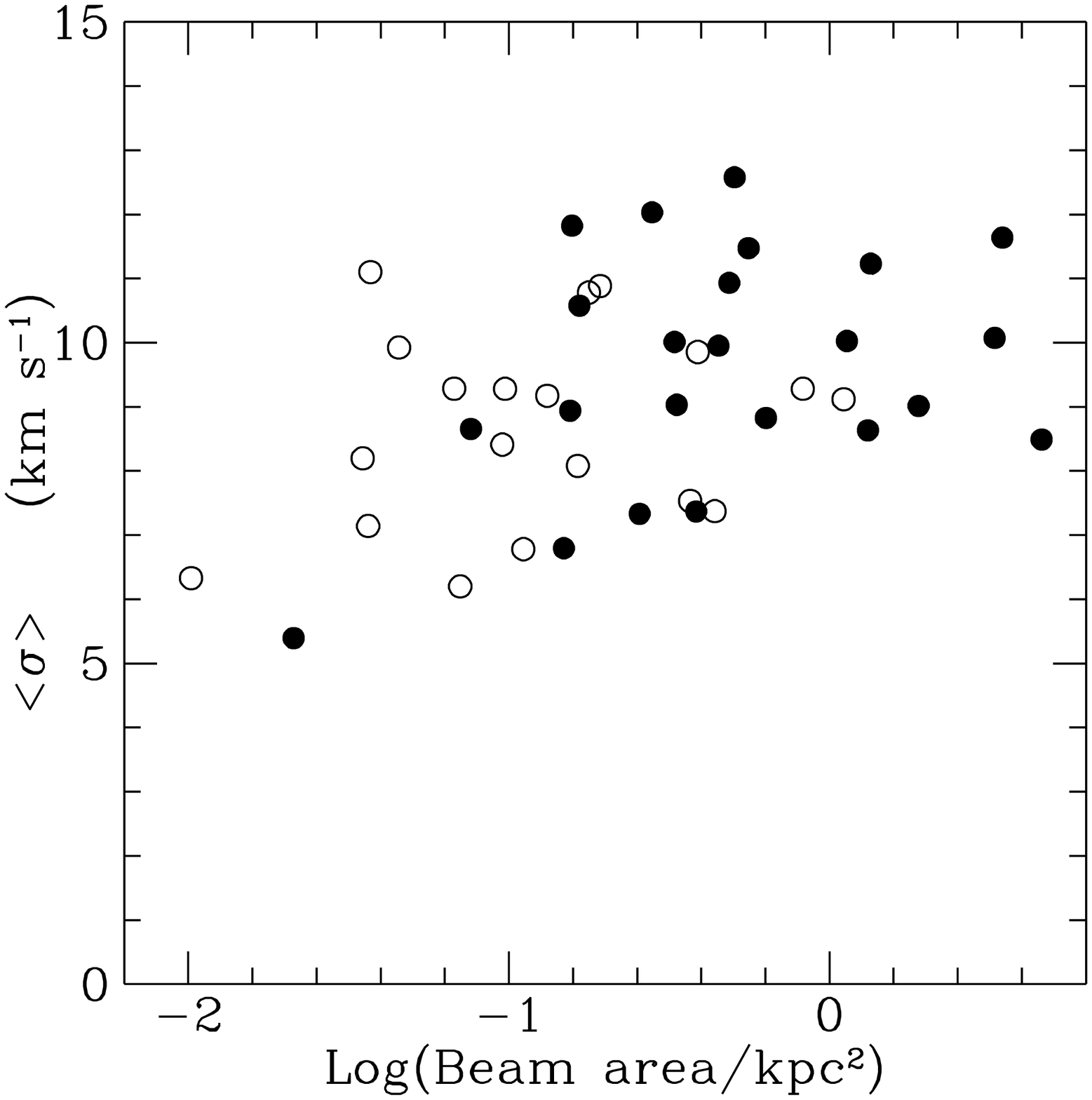}}
\caption{\small The mean velocity dispersion in the low resolution maps is
plotted as a function of the area of the synthesized beam.  The
velocity dispersions have been corrected for instrumental resolution
and the velocity gradient over the beam according to
Equation~\ref{disp-cor}.  The beam area is defined as ${1 \over 4} \pi
b_{\alpha} b_{\delta}$, where $b_{\alpha}$ and $b_{\delta}$ are the
FWHM beamsize in right ascention and declination. Open symbols
represent the $13''$ resolution data, filled symbols are the $27''$
resolution data. The pair of points with log(Beam area)$< -1.6$ represents 
the Local Group dwarf {\small DDO\,}216. 
\label{disp-beamarea}} 
\end{figure}

%
\begin{figure}
\centerline{\resizebox{9cm}{!}{\includegraphics[angle=0]{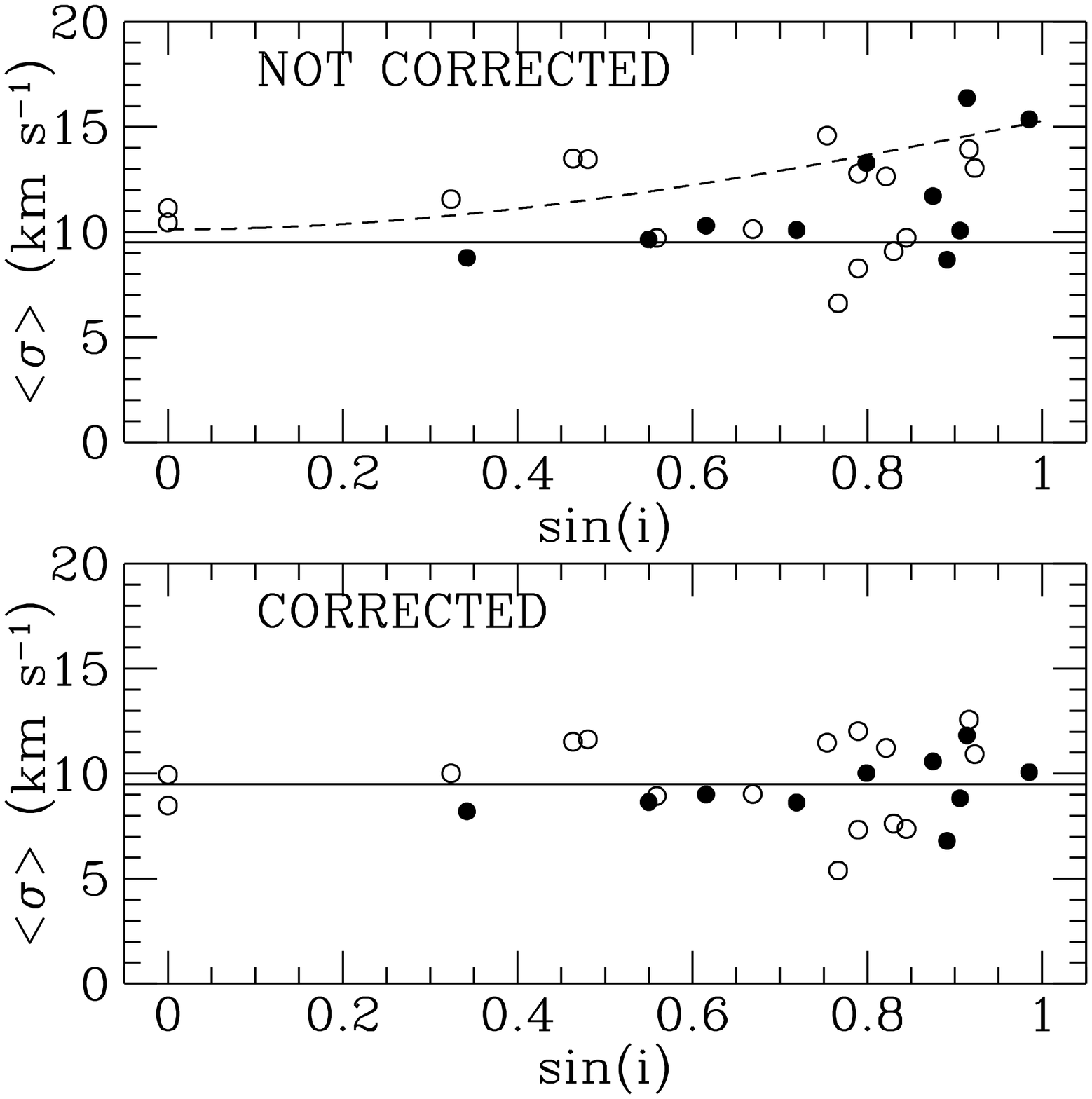}}}
\caption{\small The mean velocity dispersion as a function of
$\sin(i)$ for the low resolution maps.  Closed symbols are objects for
which the inclinations were determined with a tilted ring
fit. The inclination of the remaining objects (open symbols) was
calculated from the optical axial ratios listed in Melisse and Israel
(1994) and an intrinsic axial ratio of 0.15. The upper panel shows
the uncorrected mean velocity dispersions. The lower panel shows the
corrected values. The dashed line in the upper panel is the relation
expected for a galaxy with a solid body rotation curve with a slope
$1 \kms$ arcsec$^{-1}$ (if seen edge-on) and a velocity
dispersion of $9.5 \kms$.
\label{disp-sini}}
\end{figure}

Excluding lines of sight with low signal-to-noise ratios, we have used 
Gaussian fitting to obtain the mean velocity dispersions tabulated in
Table~\ref{veldisp-tab}. To first order, we corrected the maps for finite 
velocity resolution and linear velocity gradients over the beam according 
to the relation
$$
\sigma^2 = \sigma^2_{obs} - \sigma^2_{inst} - {1 \over 2} b^2 \Bigl( 
\nabla v \Bigr)^2 \label{disp-cor}
$$
where $\sigma_{obs}$ is the dispersion of a Gaussian 
$e^{-{1\over2}v^2/\sigma^2_{obs}}$ fitted to the
line profile at each position, $\sigma_{inst}= 2 \cdot 0.8493 \cdot
\Delta v$ is the dispersion of a Gaussian corresponding to the
velocity resolution of the Hanning-smoothed data, and $\nabla v$ is
the local velocity gradient over the beam, assumed to be
of the form $e^{-{x^2 / b^2}}$. We calculated the velocity gradient
at each position from model velocity fields constructed with the 
rotation curves presented in the previous section. No correction for 
galaxy inclination was applied. The procedure is described in more 
detail in Appendix~A.  

%
\begin{table*}
\caption{Mean velocity dispersions from single gaussian fits} 
\begin{center}
\begin{tabular}{|  l  c  c  | l c c | l c c |} 
\hline 
Name & $<\sigma>_{13''}$ & $<\sigma>_{27''}$ & Name & $<\sigma>_{13''}$ & $<\sigma>_{27''}$ & Name & $<\sigma>_{13''}$ & $<\sigma>_{27''}$  \\  
\  [1]   &[2]  & [3] & [1] & [2] & [3] & [1] & [2] & [3] \\ 
\hline 
      &$\kms$&$\kms$&  & $\kms$ & $\kms$&  & $\kms$ & $\kms$ \\ 
\hline
DDO\,22   &    12.0    &    11.2    & DDO\,64    &10.8$\pm$3.3&10.9$\pm$2.5& DDO\,125 &    6.2    & 7.3$\pm$1.5\\ 
DDO\,43   & 8.4$\pm$2.2& 9.0$\pm$1.7& DDO\,68    &10.9$\pm$3.8&12.6$\pm$4.4& DDO\,133 &    ---    & 8.2$\pm$2.0 \\
DDO\,46   & 9.3$\pm$2.3&10.0$\pm$2.2& DDO\,73    & 5.8        & 8.5$\pm$2.8& DDO\,165 &9.3$\pm$2.9&12.0$\pm$3.8\\
DDO\,47   & 8.2$\pm$1.9& 8.7$\pm$3.2& DDO\,83    & 9.9$\pm$2.4&10.0$\pm$2.5& DDO\,166 &    9.1    &11.6$\pm$3.4\\
DDO\,48   & 9.3$\pm$3.1&    10.1    & DDO\,87    & --         & 6.0$\pm$2.0& DDO\,168 &9.9$\pm$2.9&10.6$\pm$3.3\\ 
NGC\,2537 & 7.2        &11.5$\pm$5.2& Mkn~178    & --         &     7.6    & DDO\,185 &8.1$\pm$1.6& 8.8$\pm$1.7\\ 
DDO\,52   & 6.8$\pm$2.8& 7.4$\pm$2.3& NGC\,3738  & (12.2)     &   (18.2)   & DDO\,190 &9.2$\pm$2.4&10.0$\pm$2.4 \\ 
DDO\,63   & 7.1$\pm$2.1& 8.9$\pm$2.3& DDO\,123   & 7.4$\pm$2.5& 9.0$\pm$1.9& DDO\,216 &6.3$\pm$1.7& 5.4$\pm$2.6 \\
NGC\,2976 &11.1$\pm$3.5&11.8$\pm$3.2& Mkn~209    & --         &    11.5    & DDO\,217 &7.5$\pm$2.7& 8.6$\pm$2.7 \\ 
\hline 
\end{tabular} 
\end{center}
Notes: Results are corrected for finite velocity resolution and velocity
gradients over the beam as discussed in the text. Velocity dispersion 
scatter values are given only if the area considered was at at least 
ten times the synthesized beam area. Column [2] gives mean velocity dispersion 
and r.m.s. scatter of the velocity dispersion over the galaxy from the
full-resolution data, and column [3] the corresponding mean from the 
low-resolution data.
\label{veldisp-tab}
\end{table*}

The high velocity dispersion in {\small NGC\,}3738 is probably an artifact 
resulting from too small a beam/velocity gradient correction caused by
the marginally resolved steep velocity gradient of the galaxy. Excluding 
{\small NGC\,}3738, we find a mean velocity dispersion of $8.6\pm0.34 
\kms$ at $13''$ resolution and $9.5\pm0.38 \kms$ at $27''$ resolution. 
Although the difference between the two results is significant, its 
magnitude is only 10$\%$. Moreover, these values are consistent with those 
in the literature (Shostak $\&$ Van der Kruit 1984, Skillman et al. 1988). 
As the physical area contributing to the measured velocity dispersion 
increases with galaxy distance squared, we show in Fig.~\ref{disp-beamarea} 
mean velocity dispersions as a function of beam surface area for both 
the $13''$ and $27''$ resolution maps. The systematic increase in 
mean velocity dispersion with physical beam area increasing by two orders 
of magnitude is no more than about $2 \kms$. As our results are thus
effectively insensitive to linear resolution on scales of $\sim$ 0.1 kpc 
and larger, we may compare the velocity dispersions of galaxies at 
various distances without fear of introducing large systematic effects. 

Finally, we show mean velocity dispersions as a function of inclination 
in Fig.~\ref{disp-sini}. We used tilted-ring inclinations and, lacking
these, inclinations estimated from optical axial ratios assuming an
intrinsic axial ratio of 0.15. The result is not sensitive to the 
exact value of this intrinsic axial ratio. For instance, use of the higher 
values suggested by the work of Staveley-Smith et al. (1992), 
increases $\sin\,i$ values by at most 0.08. The upper and lower panels 
in Fig.~\ref{disp-sini} show velocity dispersions before and after the 
correction for inclination. The upper panel shows velocity dispersions
increasing at the highest inclinations ($\sin\,i>0.9$; $i > 65^\circ$). 
This increase has disappeared completely in the corrected set in the
lower panel. The widths of local line profiles depend on inclination only
through the observed velocity gradient over the beam, which is
steeper on average for high inclination angles.
Thus, {\it all} observed galaxies, irrespective their absolute luminosity 
(-12.8 mag $>$ M$_{\rm B} >$ -17.6 mag) are have mean velocity dispersions 
of about 10$\kms$, very similar to that of spiral galaxy disks. We
will return to this result in a forthcoming paper.  

The velocity dispersion maps shown in Fig.~4 are corrected for 
the local velocity gradient over the beam. Note that the steep inner 
rotation curves of {\small NGC\,}2537 and {\small NGC\,}3738 are not 
completely resolved, resulting in artificially large linewidths. 

\section{Individual objects}

\indent
{\small DDO\,}22: The HI axial ratio suggests that this galaxy is 
seen at a high inclination. A region of high HI column density is 
present in the southern side of the galaxy. The kinematic and HI 
major axes appear to be misaligned.

{\small DDO\,}43: The HI axial ratio indicates a face-on orientation. 
Parallel velocity contours in a regular velocity field mark a rapidly 
rising rotation curve. The kinematic and optical major axes are 
roughly perpendicular. 

{\small DDO\,}46: The HI distribution shows a U-shaped high-column density 
ridge. The velocity field is reasonably symmetric with respect to the
kinematic minor axis, but asymmetric with respect to the major axis. 
The rotation curve is almost flat at the edge of the HI distribution.

{\small DDO\,}47: Spiral structure in the outer HI disk was reported 
by Puche $\&$ Westpfahl (1994). No optical emission appears associated 
with the spiral arms. Regularly shaped, elliptical outer HI isophotes 
suggest a disk seen at low inclination. A number of high column density 
regions are distributed evenly over the disk. There is a deep hole in the HI
distribution at $\alpha=\rm 7^h39^m8^s.1,\,\delta=\rm 16^\circ 54' 31''$. 
The velocity field presents one of the few clear examples of solid-body 
rotation, as does the major axis XV diagram.

{\small DDO\,}48: This is probably a nearly edge-on disk with a regular
velocity field. The rotation curve rises rapidly
near the center and flattens gradually outwards. 

\begin{figure*}
\begin{minipage}[b]{5.7 cm}
\resizebox{5.7cm}{!}{\includegraphics[angle=-90]{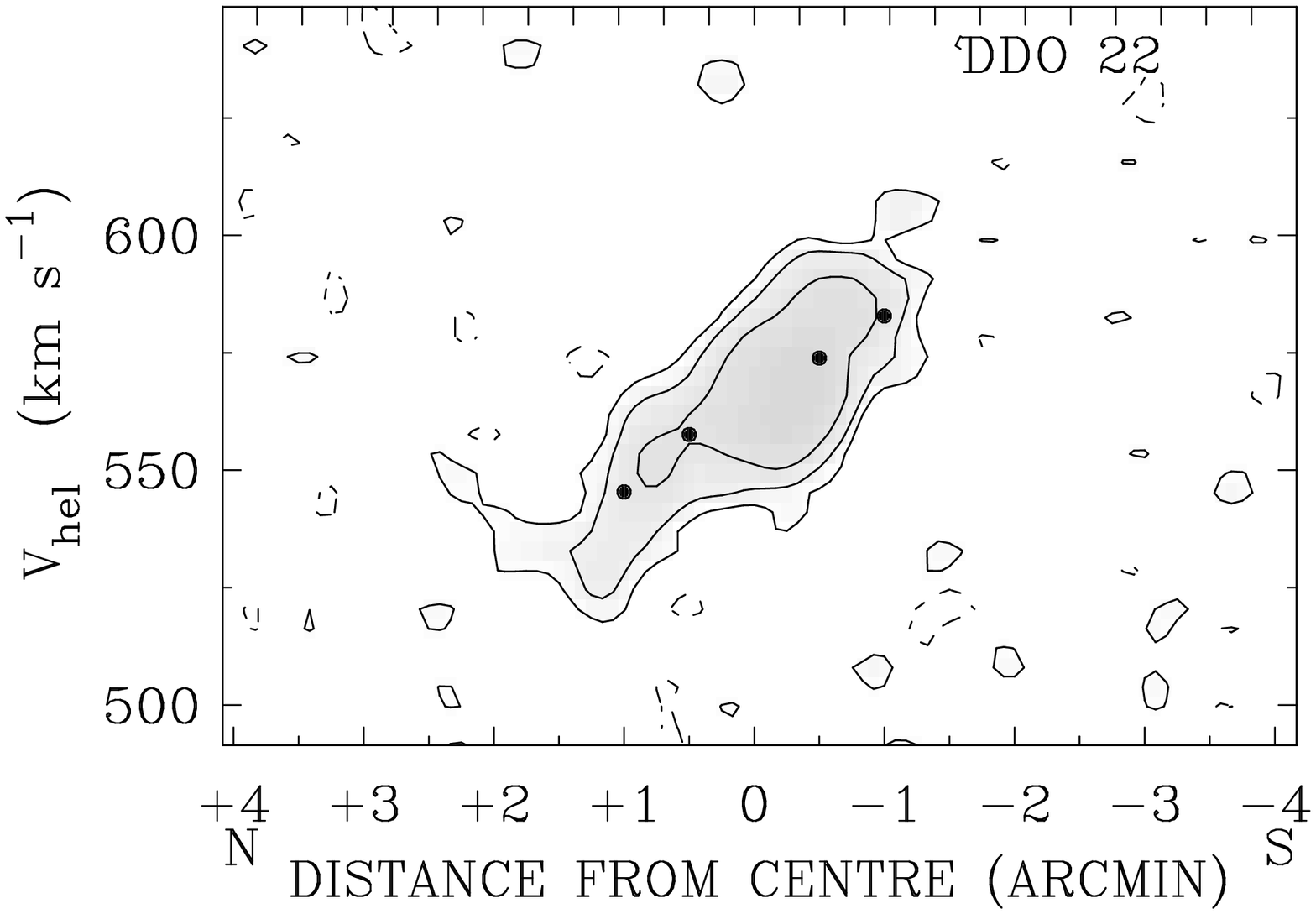}}
\end{minipage}
\hfill
\begin{minipage}[b]{5.7 cm}
\resizebox{5.7cm}{!}{\includegraphics[angle=-90]{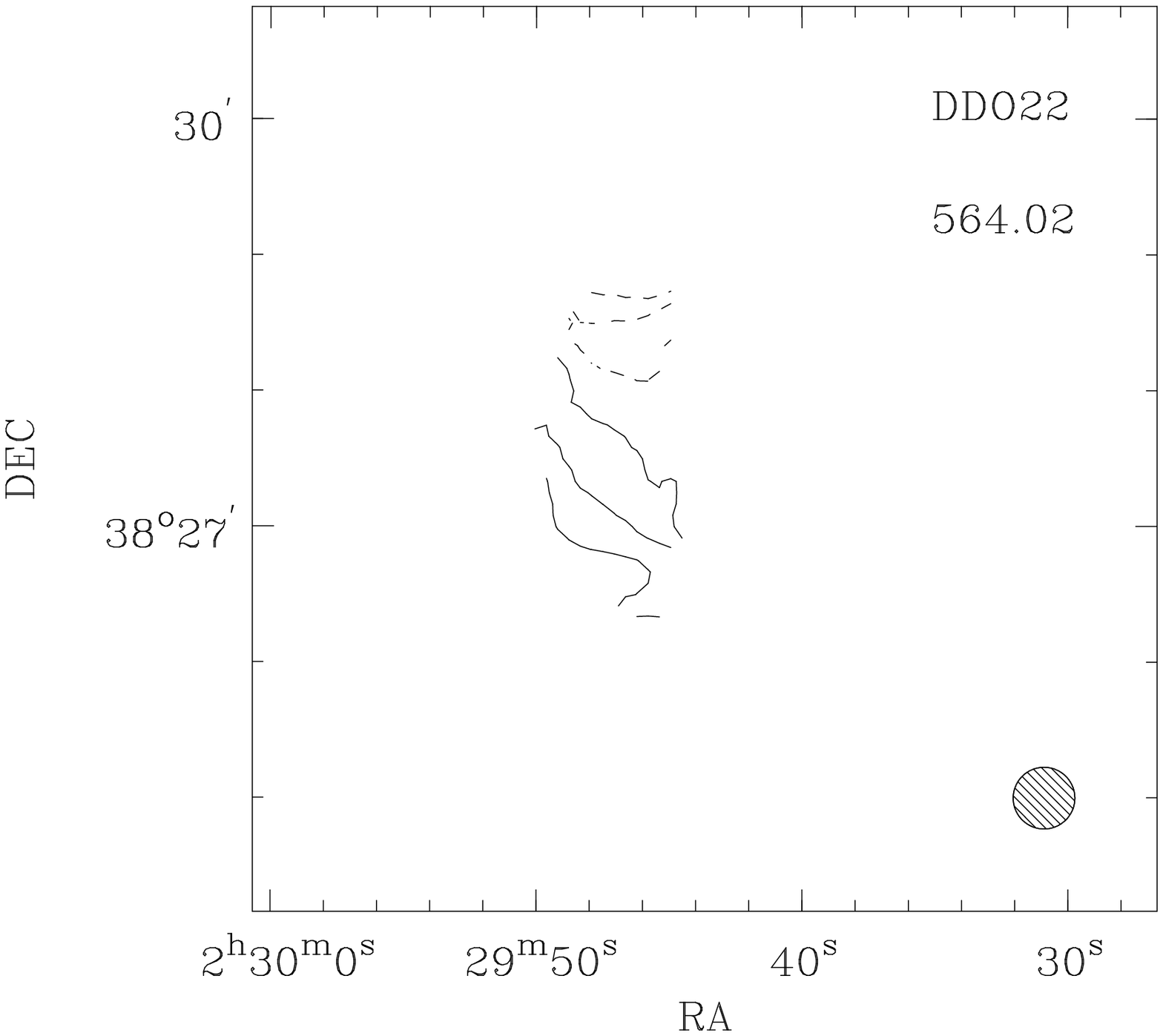}}
\end{minipage}
\hfill
\begin{minipage}[b]{5.7 cm}
\resizebox{5.85cm}{!}{\includegraphics[angle=-90]{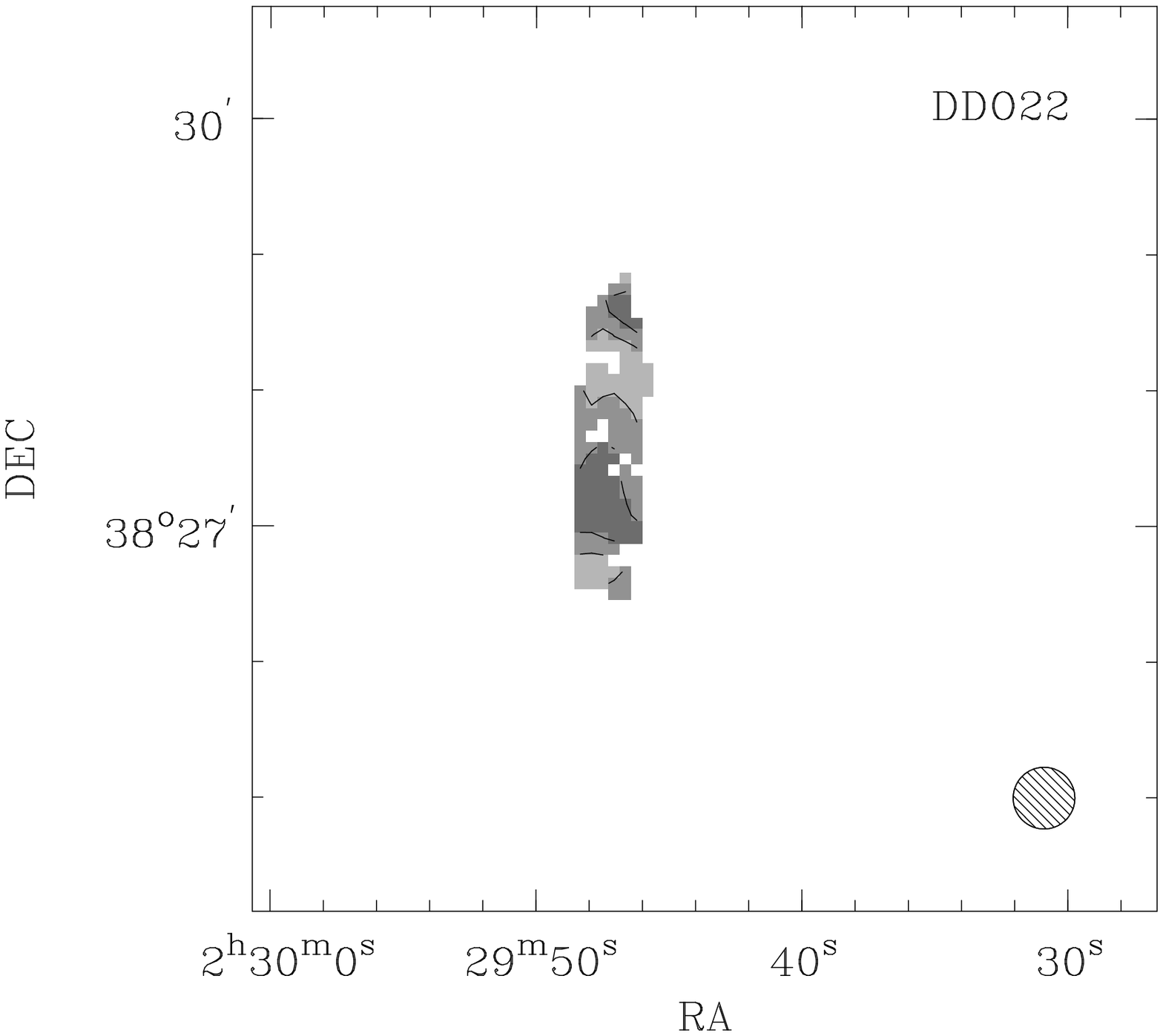}}
\end{minipage}
\begin{minipage}[b]{5.7 cm}
\resizebox{5.7cm}{!}{\includegraphics[angle=-90]{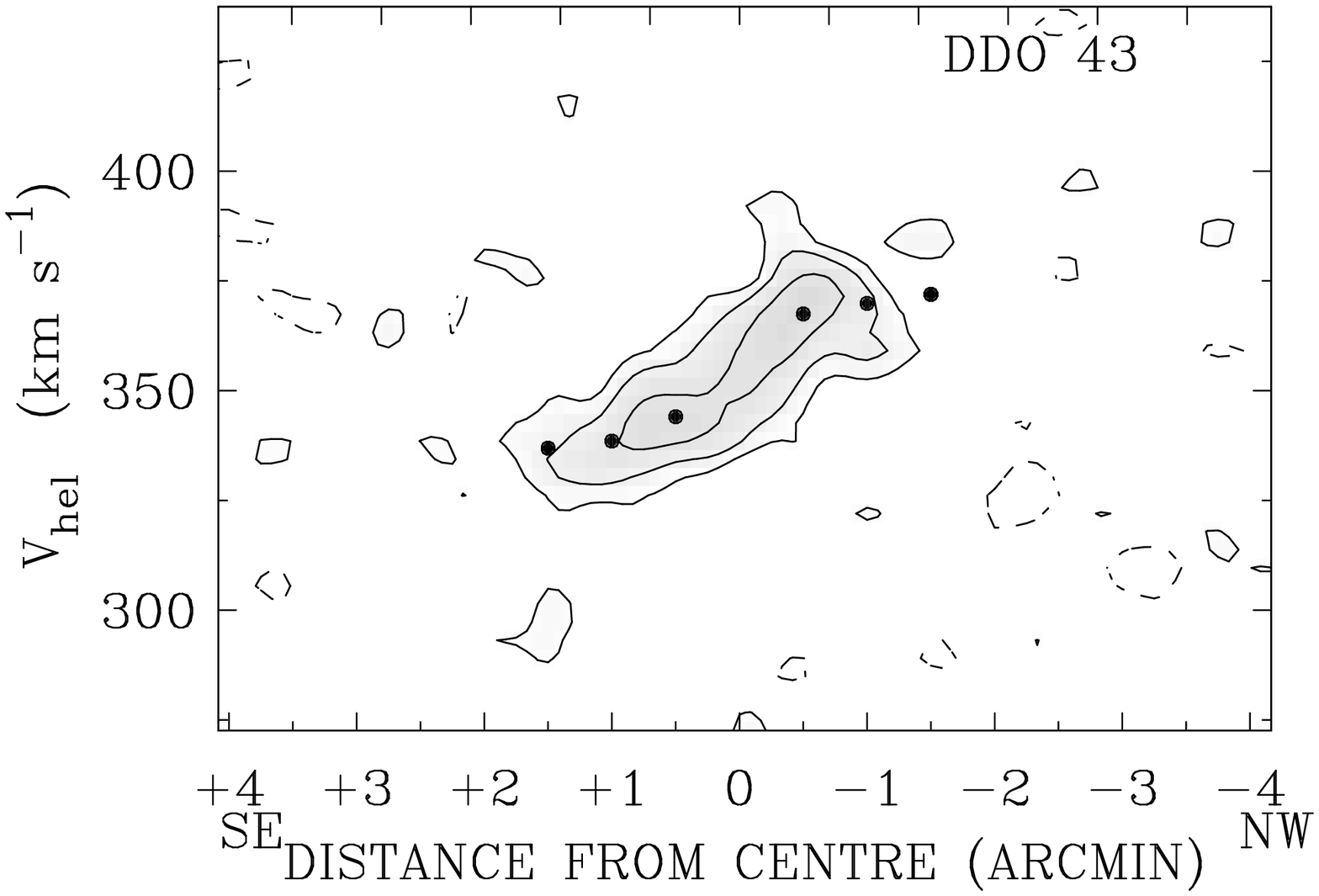}}
\end{minipage}
\hfill
\begin{minipage}[b]{5.7 cm}
\resizebox{5.7cm}{!}{\includegraphics[angle=-90]{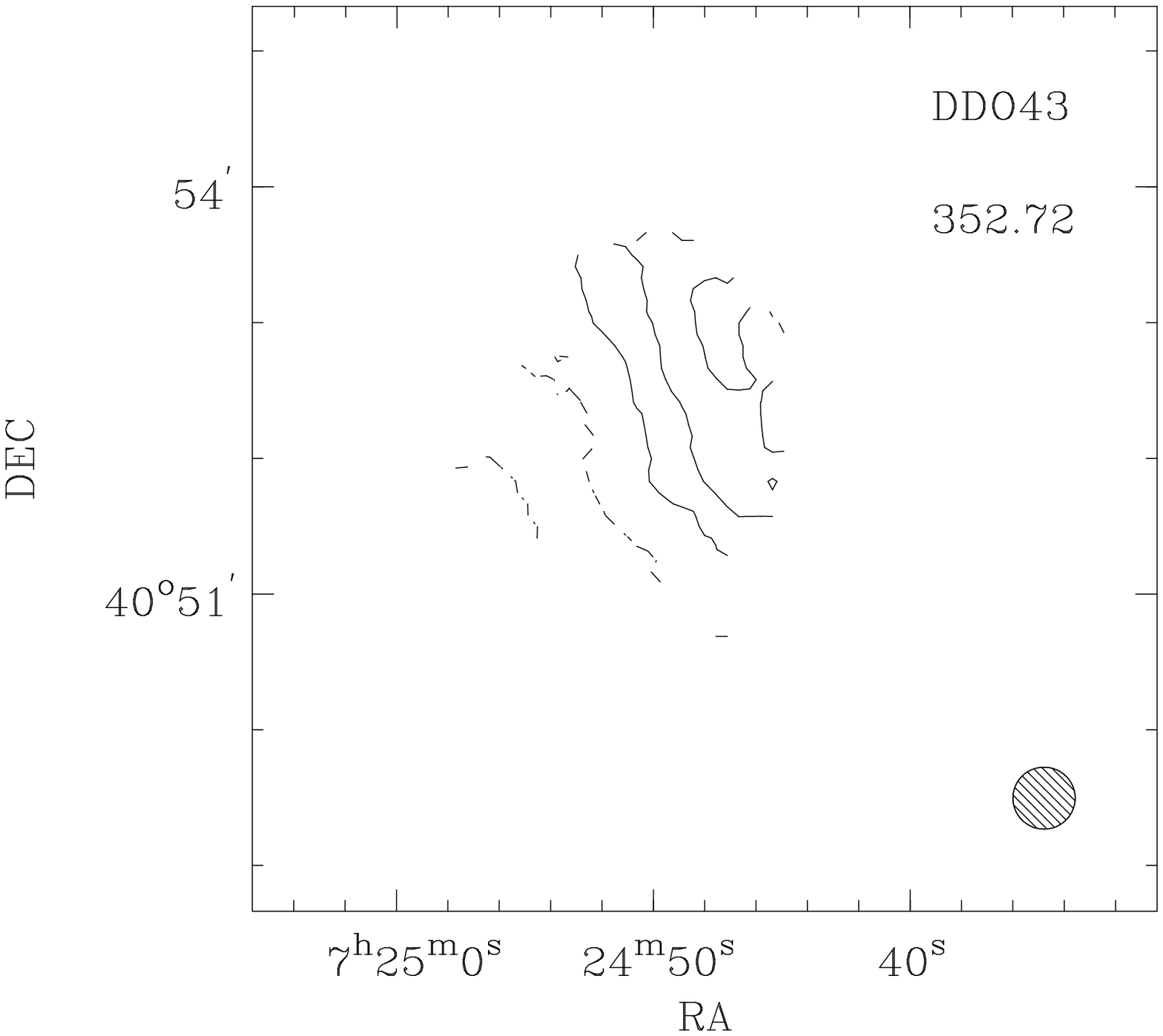}}
\end{minipage}
\hfill
\begin{minipage}[b]{5.7 cm}
\resizebox{5.85cm}{!}{\includegraphics[angle=-90]{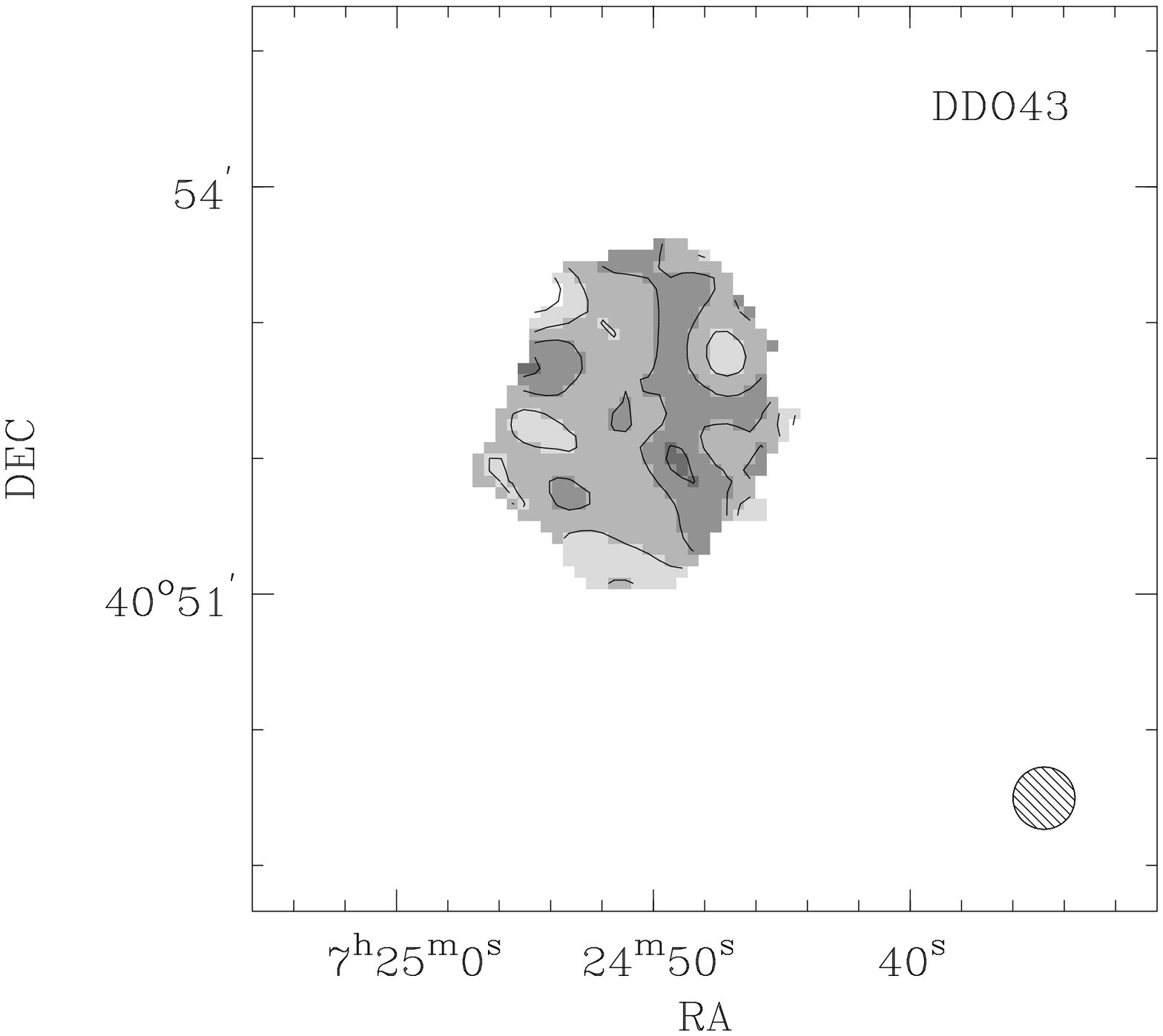}}
\end{minipage}
\begin{minipage}[b]{5.7 cm}
\resizebox{5.7cm}{!}{\includegraphics[angle=-90]{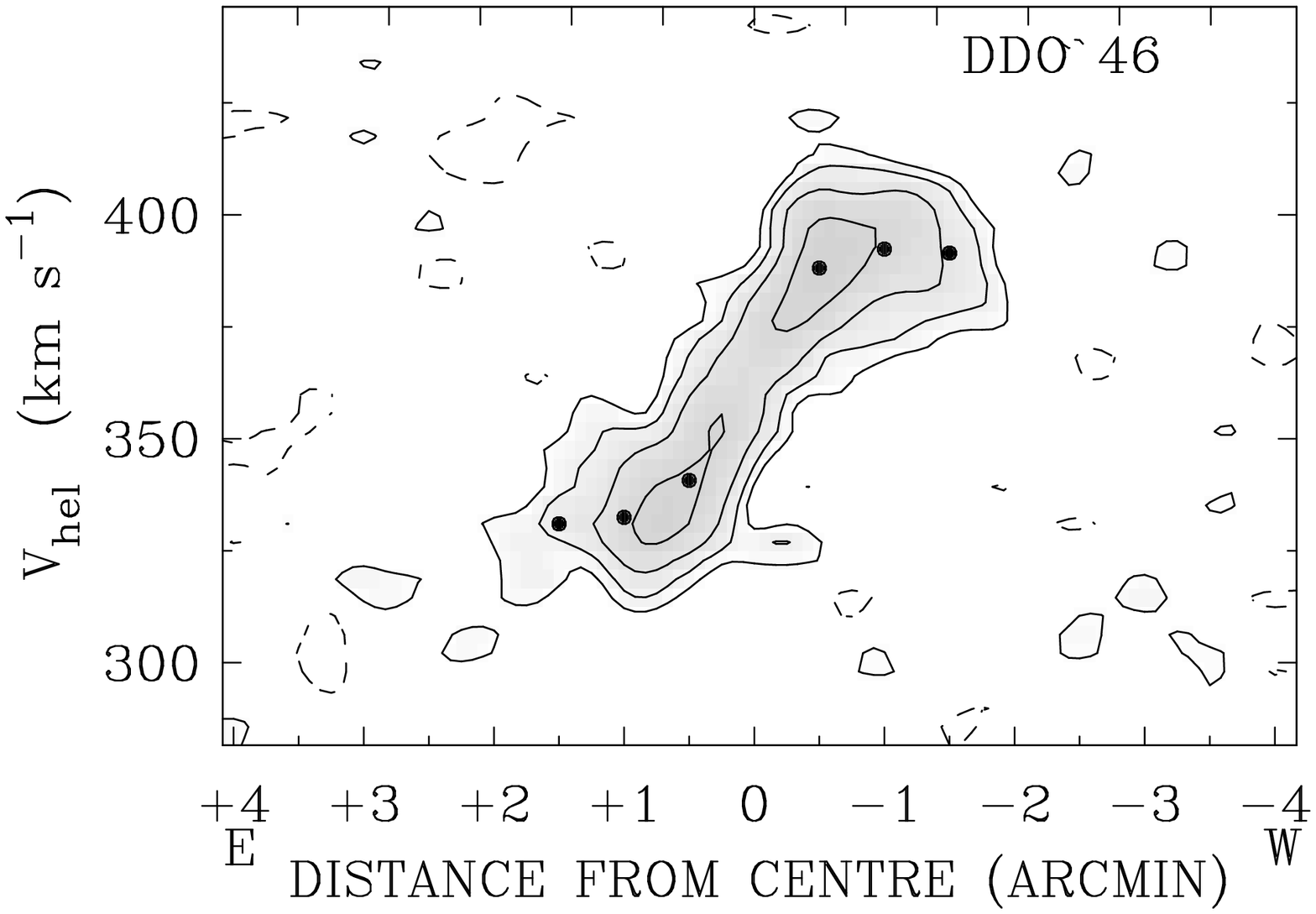}}
\end{minipage}
\hfill
\begin{minipage}[b]{5.7 cm}
\resizebox{5.7cm}{!}{\includegraphics[angle=-90]{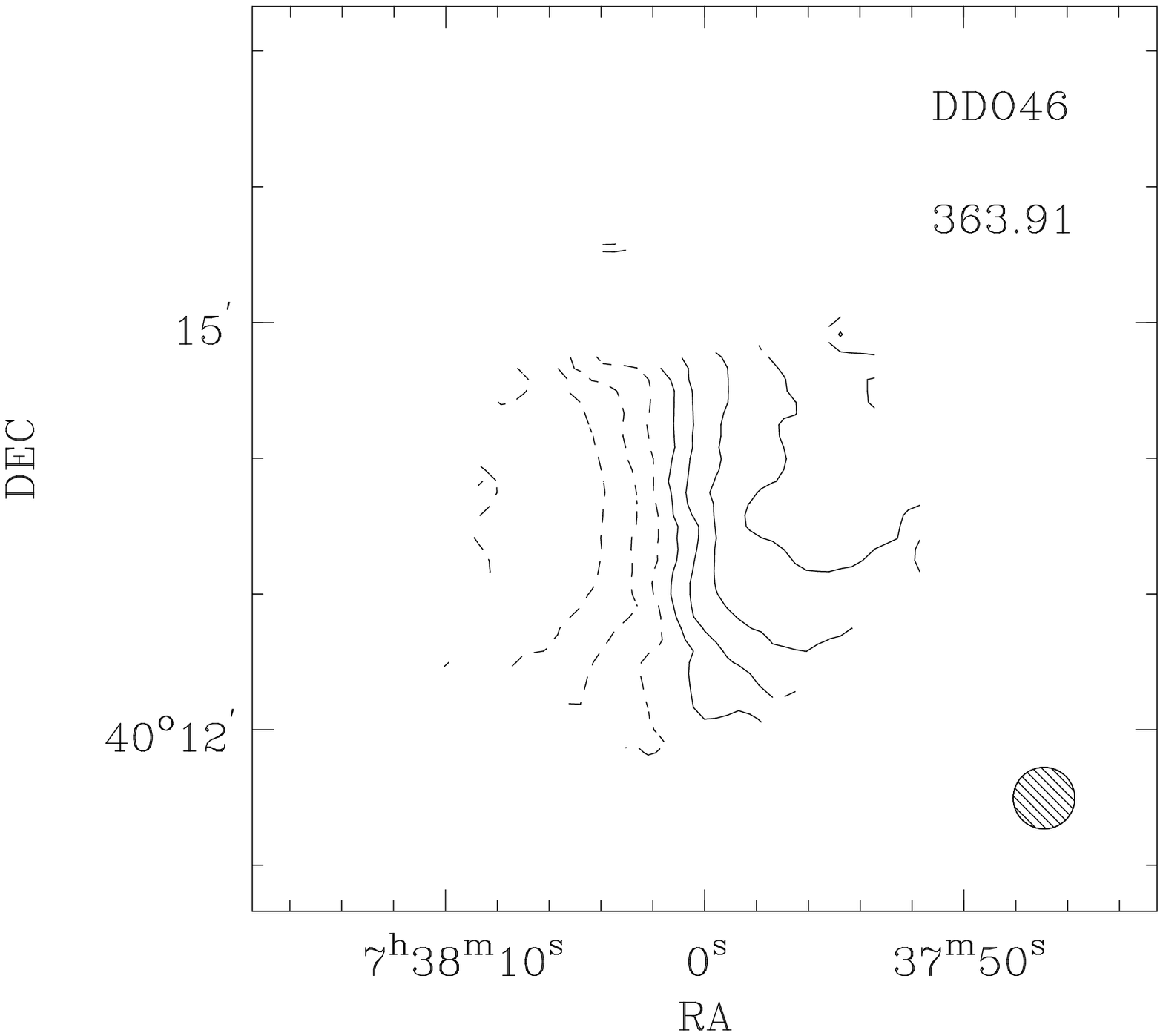}}
\end{minipage}
\hfill
\begin{minipage}[b]{5.7 cm}
\resizebox{5.85cm}{!}{\includegraphics[angle=-90]{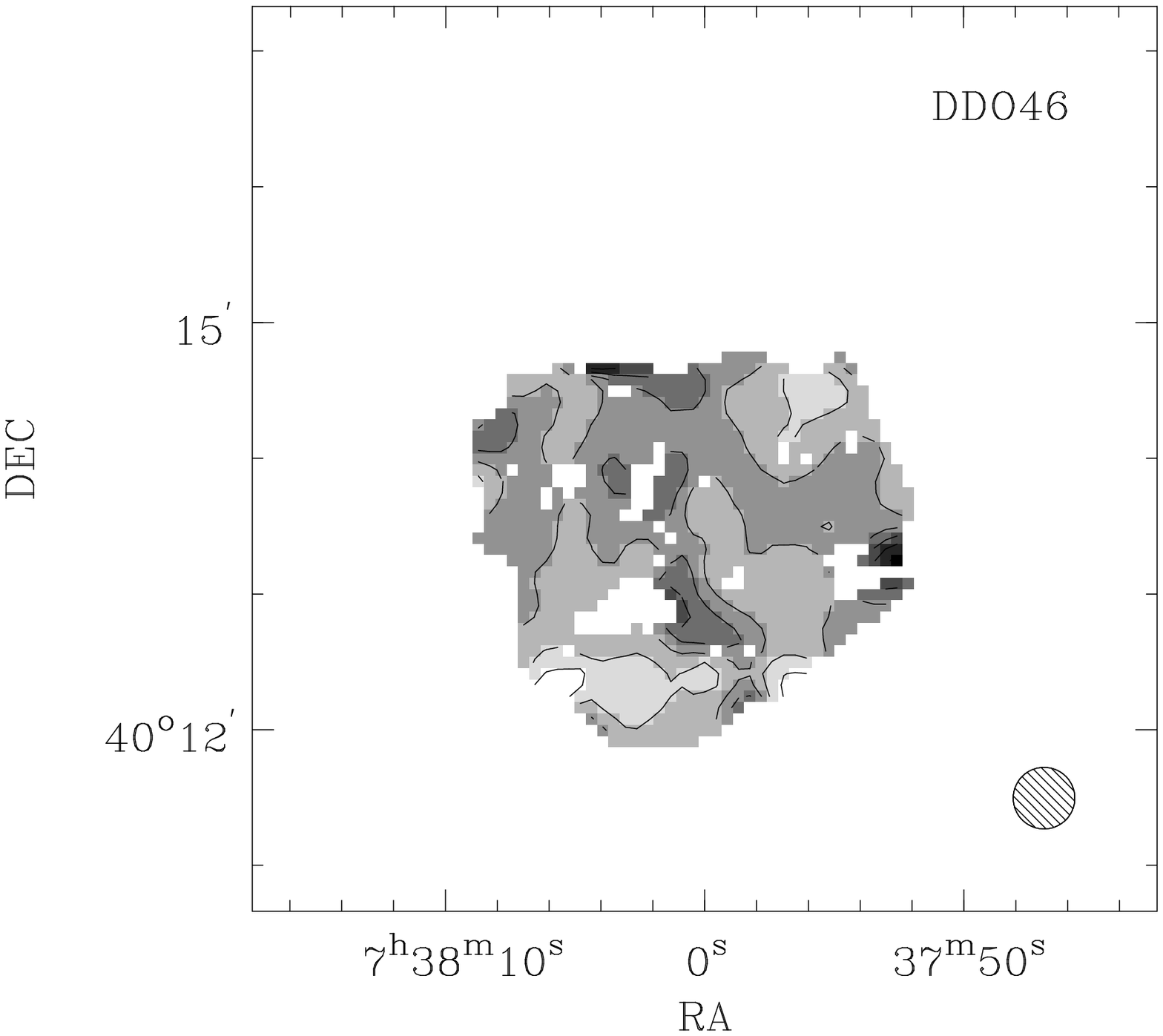}}
\end{minipage}
\caption{
\small For each galaxy, all maps at 27$''$ resolution: 
(left) major axis position-velocity maps with contours at 
$-2\sigma\, 2^{N} \sigma$, with $N=0,\ 1,\ 2,\ \ldots $ and $\sigma$ the 
r.m.s. noise in empty channel maps; grayscales are logarithmic; tickmarks 
on the upper horizontal axis mark intervals of 1 kpc.
(center) velocity fields with contour intervals of $8 \kms$; 
dashed contours mark approaching
velocities;  the first solid contour is at the system velocity which
is printed under each object's name; WSRT beamsize is indicated by the 
hatched ellipse.
(right): velocity dispersion maps with contour levels at 5, 7.5, 10, 12.5
$\ldots \kms$; greyscales are linear from $5 \kms$ to $20 \kms$.
}
\end{figure*}

\begin{figure*}
\addtocounter{figure}{-1}
\begin{minipage}[b]{5.7 cm}
\resizebox{5.7cm}{!}{\includegraphics[angle=-90]{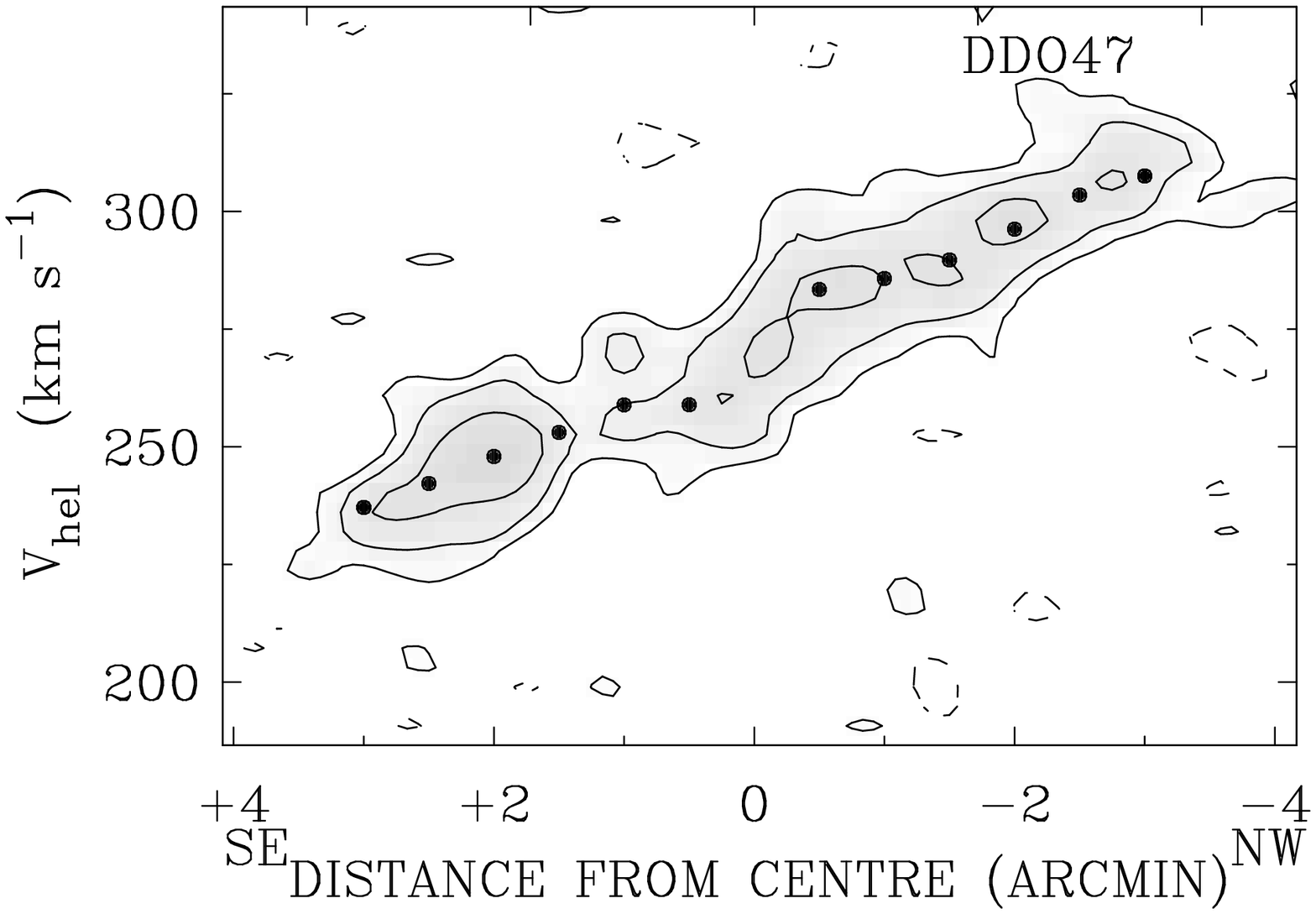}}
\end{minipage}
\hfill
\begin{minipage}[b]{5.7 cm}
\resizebox{5.7cm}{!}{\includegraphics[angle=-90]{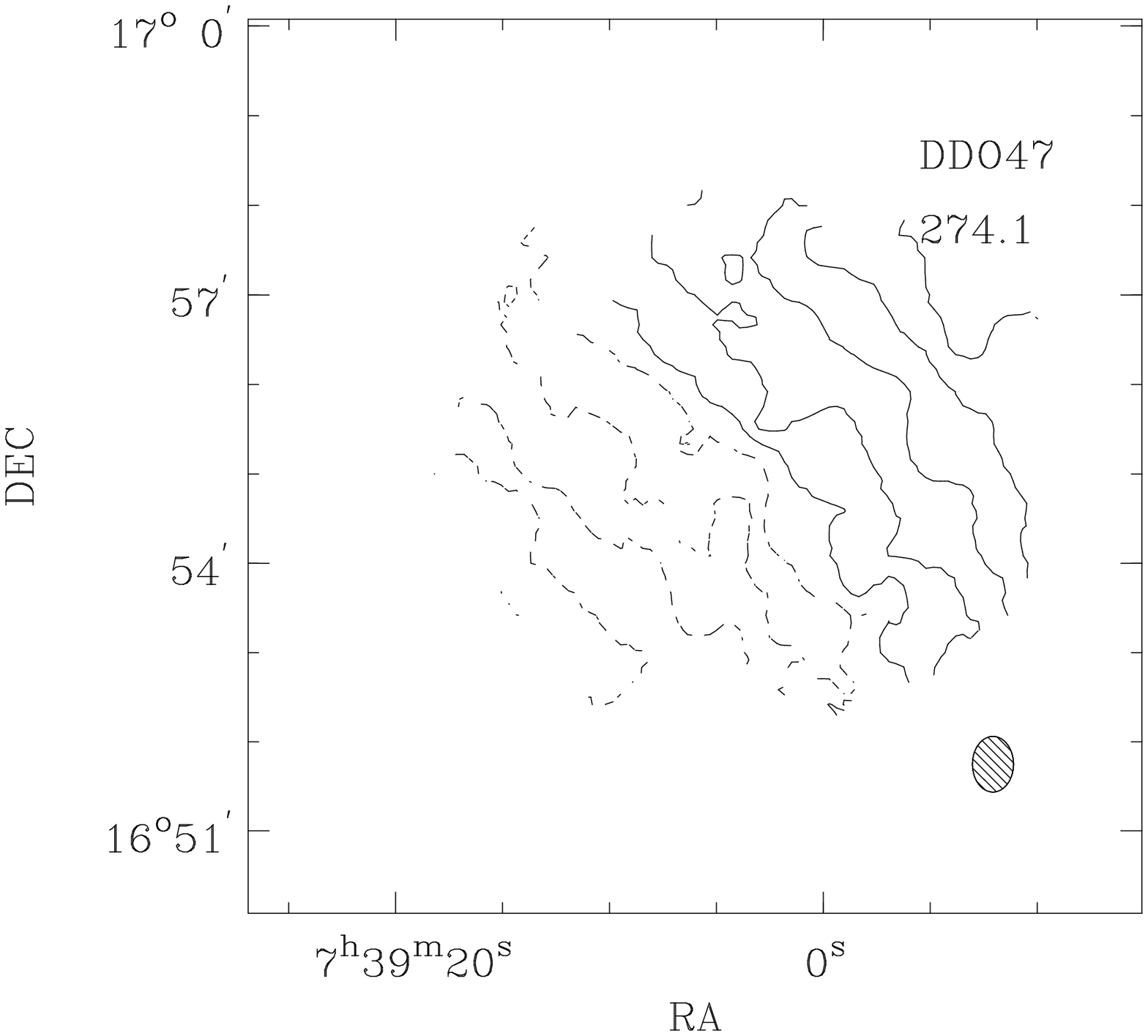}}
\end{minipage}
\hfill
\begin{minipage}[b]{5.7 cm}
\resizebox{5.85cm}{!}{\includegraphics[angle=-90]{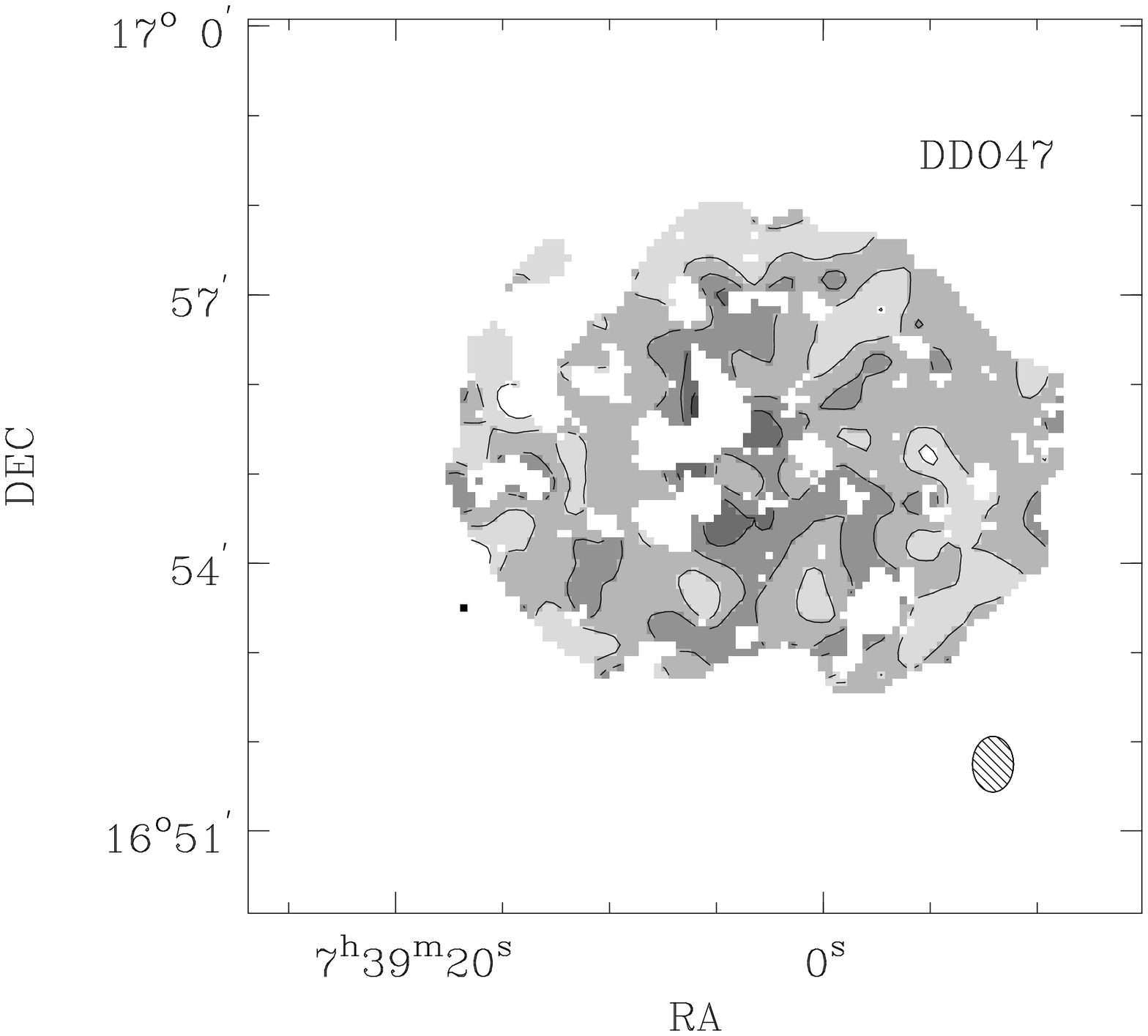}}
\end{minipage}
\begin{minipage}[b]{5.7 cm}
\resizebox{5.7cm}{!}{\includegraphics[angle=-90]{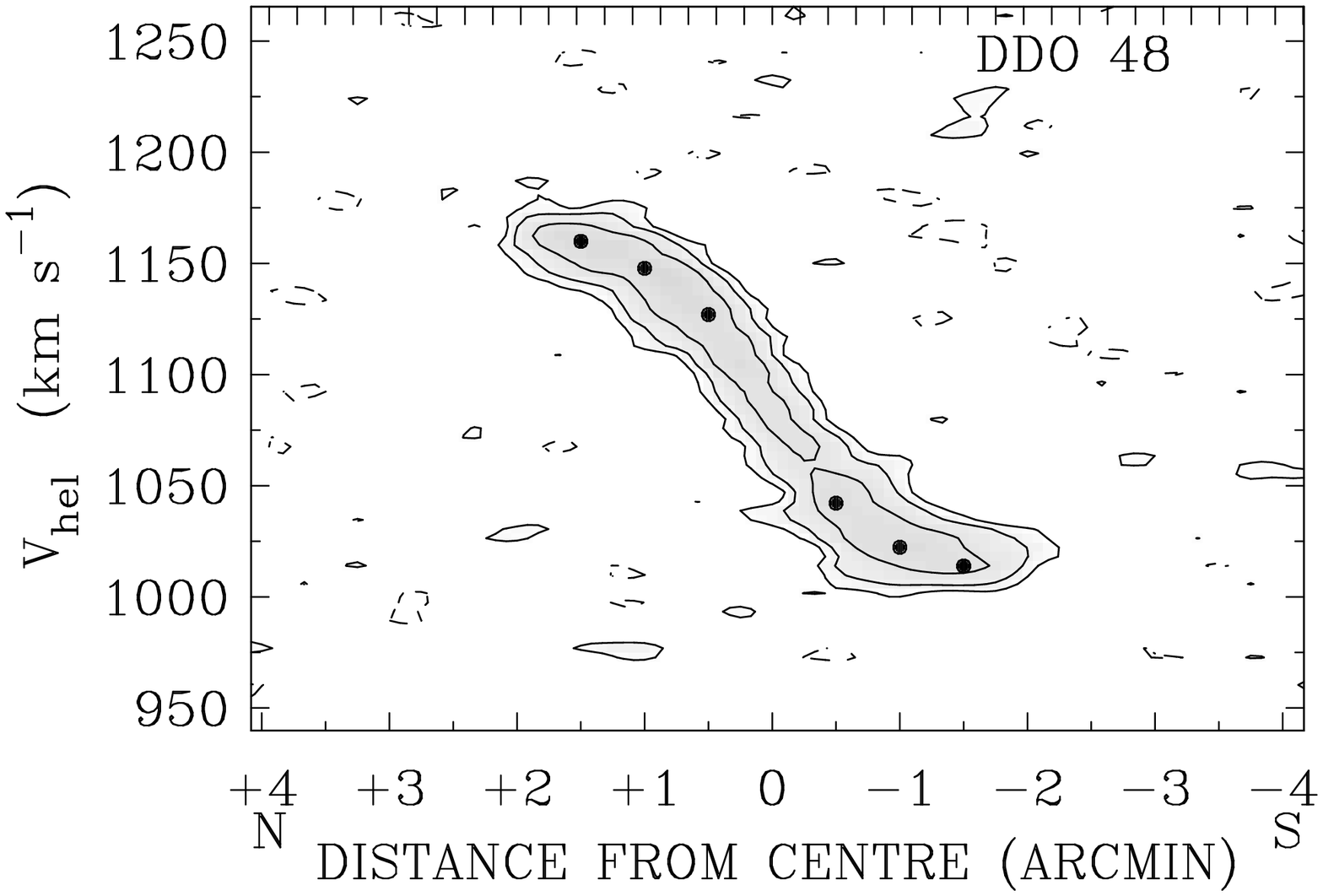}}
\end{minipage}
\hfill
\begin{minipage}[b]{5.7 cm}
\resizebox{5.7cm}{!}{\includegraphics[angle=-90]{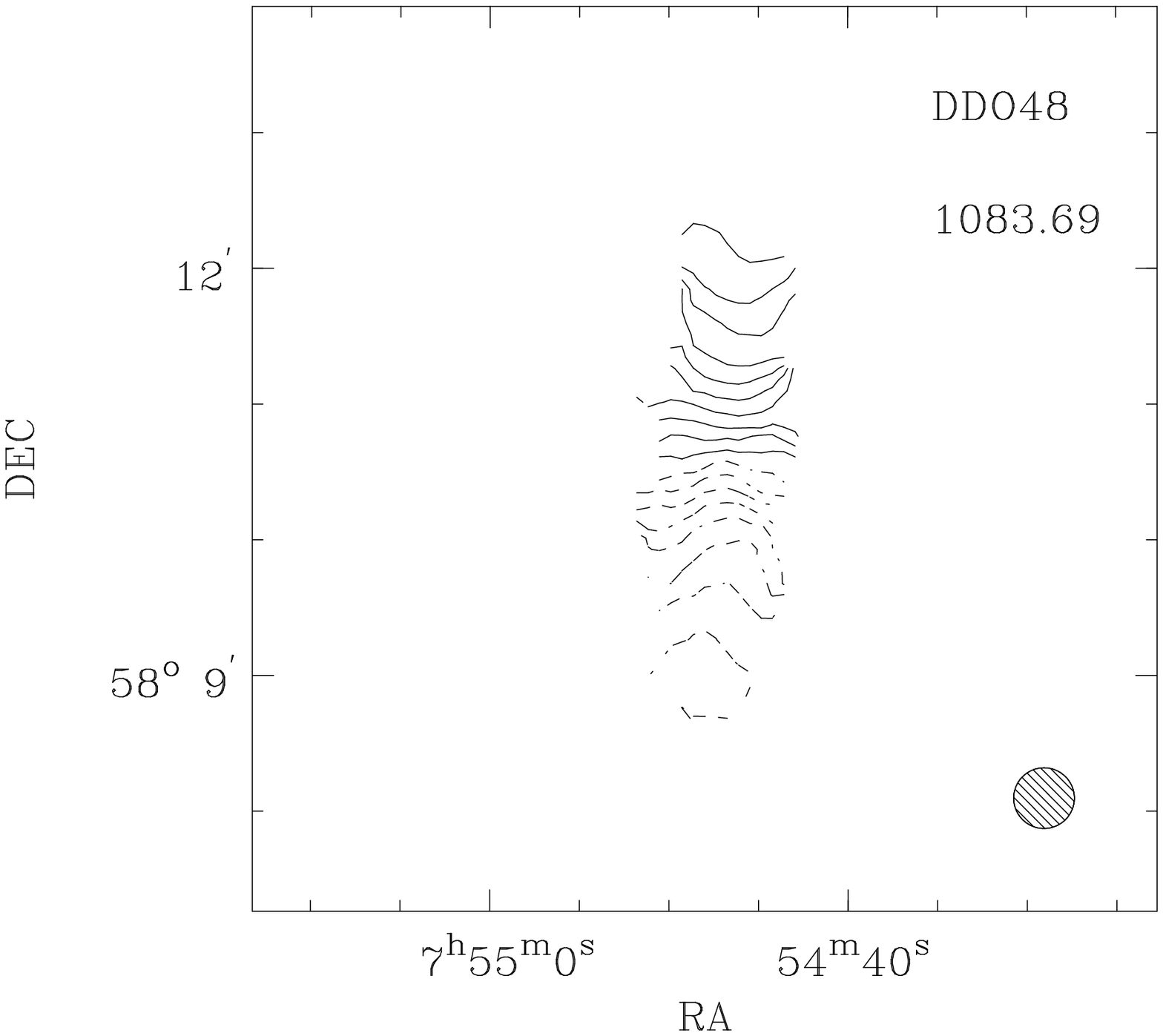}}
\end{minipage}
\hfill
\begin{minipage}[b]{5.7 cm}
\resizebox{5.85cm}{!}{\includegraphics[angle=-90]{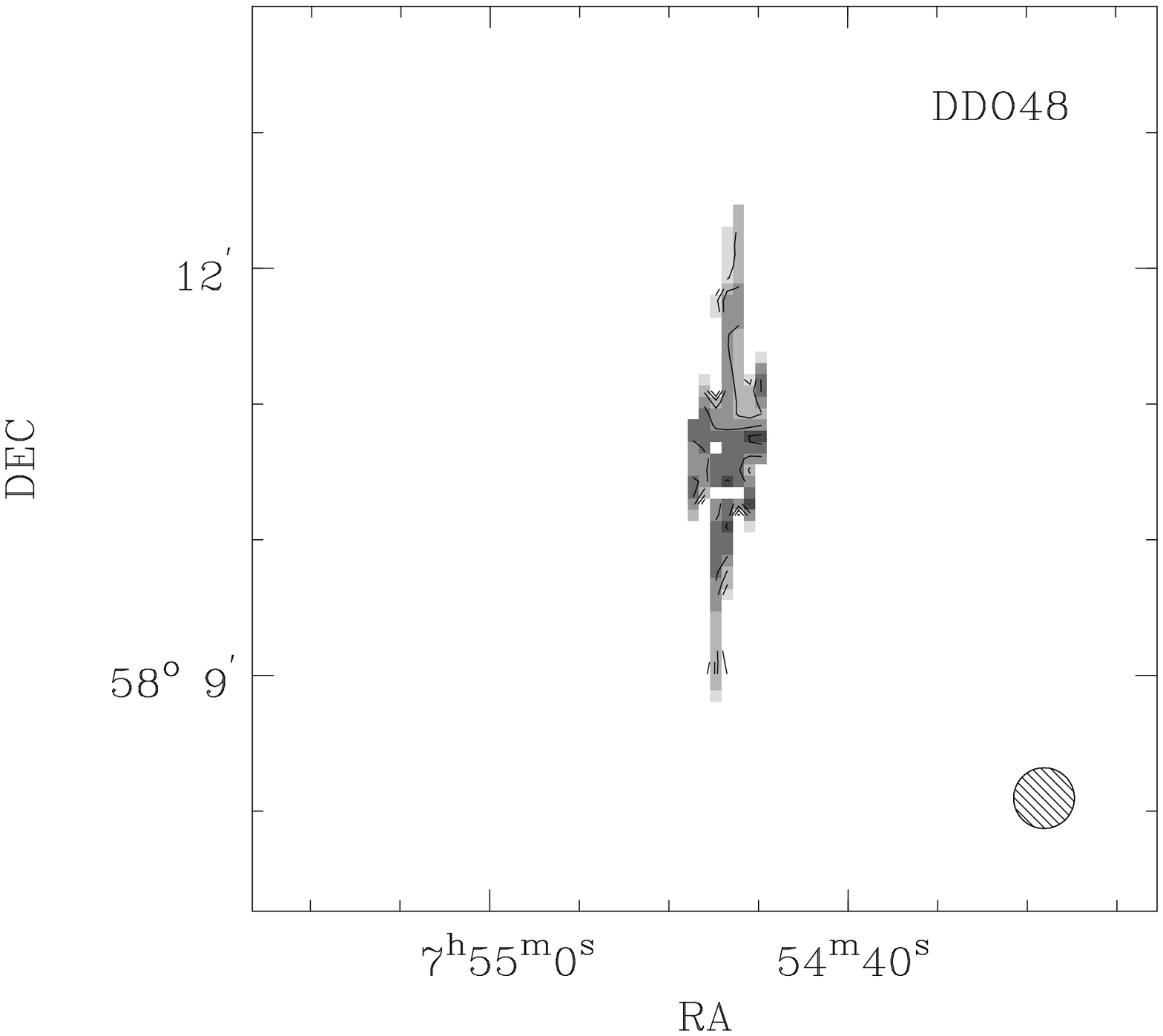}}
\end{minipage}
\begin{minipage}[b]{5.7 cm}
\resizebox{5.7cm}{!}{\includegraphics[angle=-90]{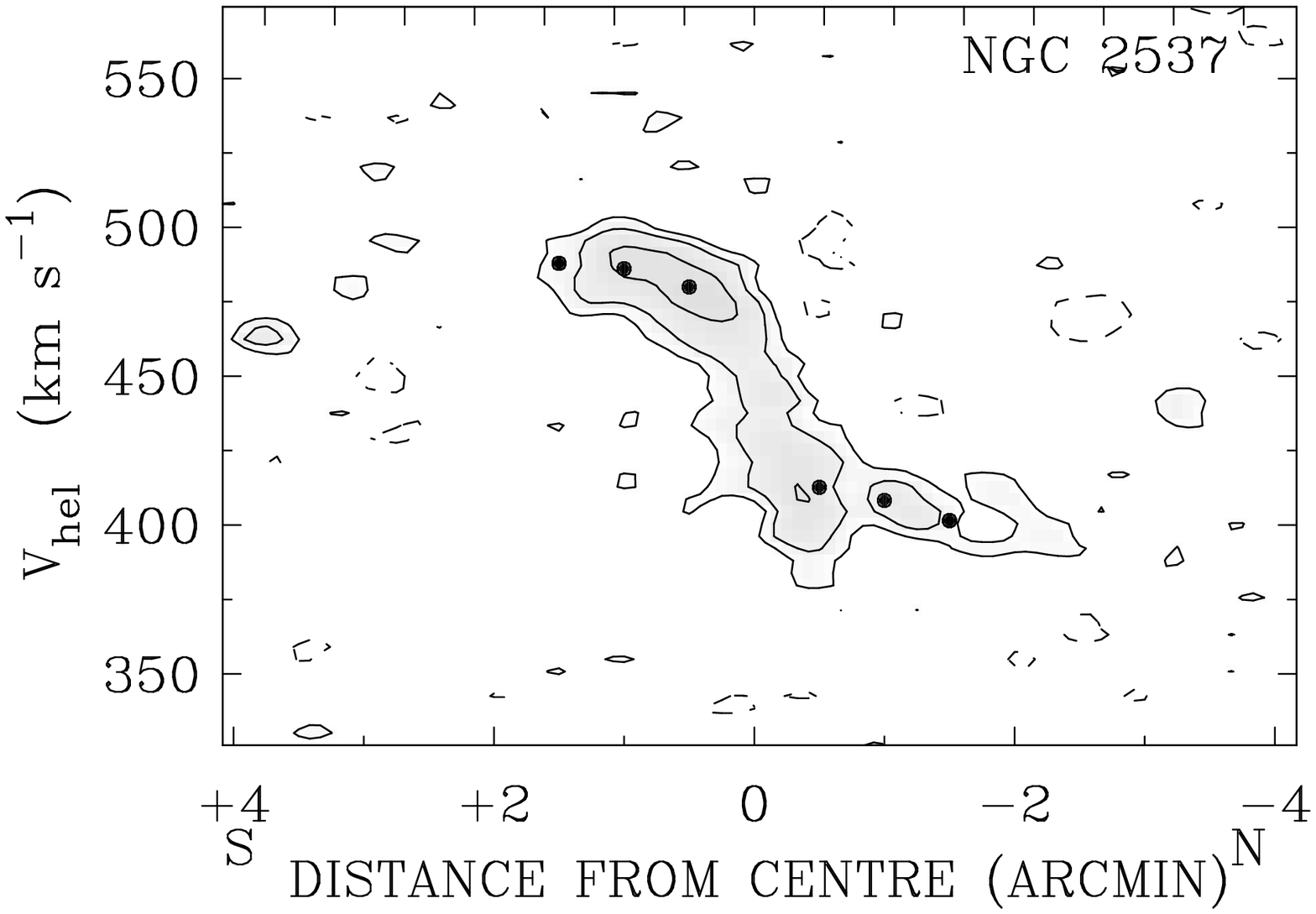}}
\end{minipage}
\hfill
\begin{minipage}[b]{5.7 cm}
\resizebox{5.7cm}{!}{\includegraphics[angle=-90]{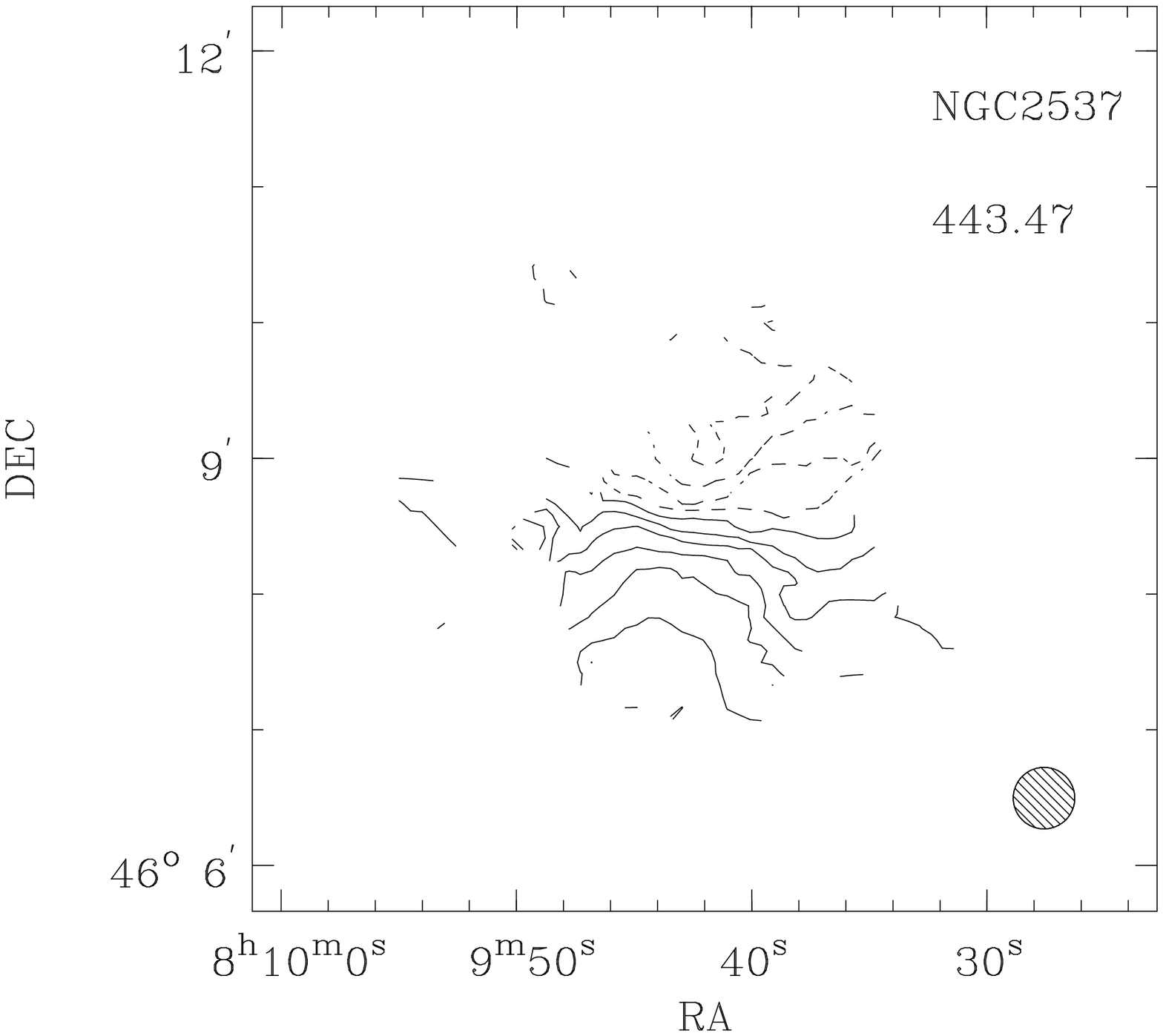}}
\end{minipage}
\hfill
\begin{minipage}[b]{5.7 cm}
\resizebox{5.85cm}{!}{\includegraphics[angle=-90]{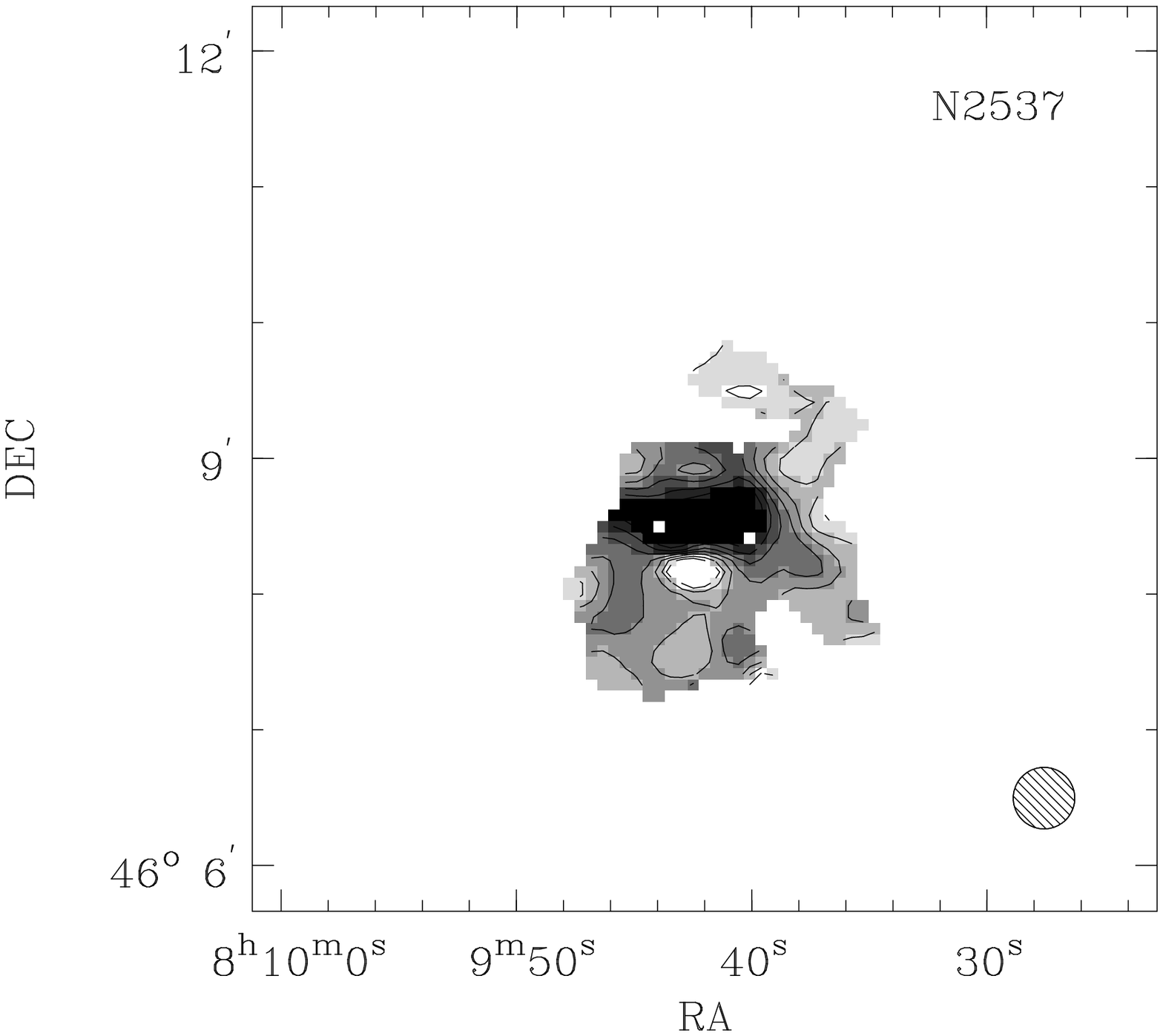}}
\end{minipage}
\begin{minipage}[b]{5.7 cm}
\resizebox{5.7cm}{!}{\includegraphics[angle=-90]{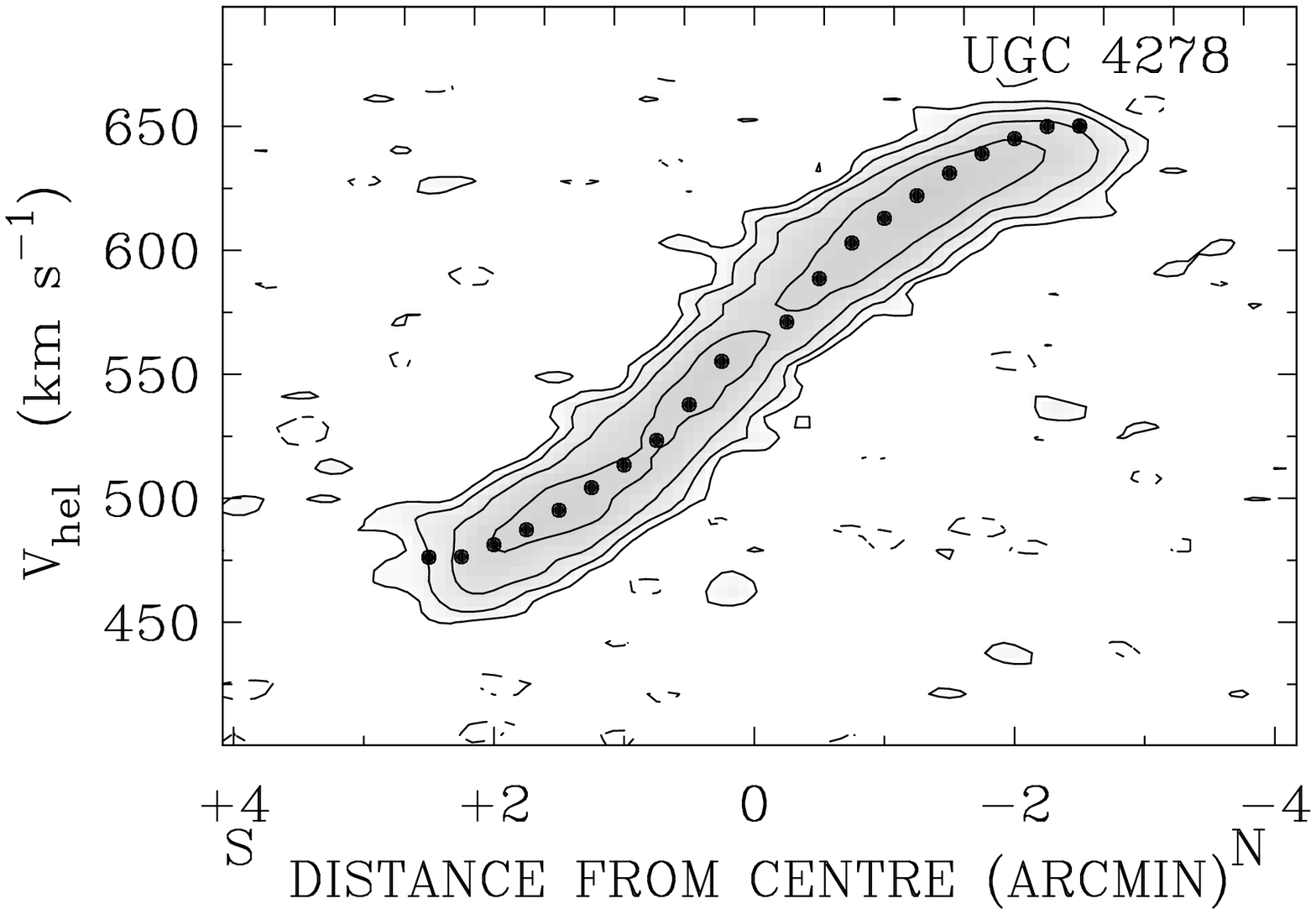}}
\end{minipage}
\hfill
\begin{minipage}[b]{5.7 cm}
\resizebox{5.7cm}{!}{\includegraphics[angle=-90]{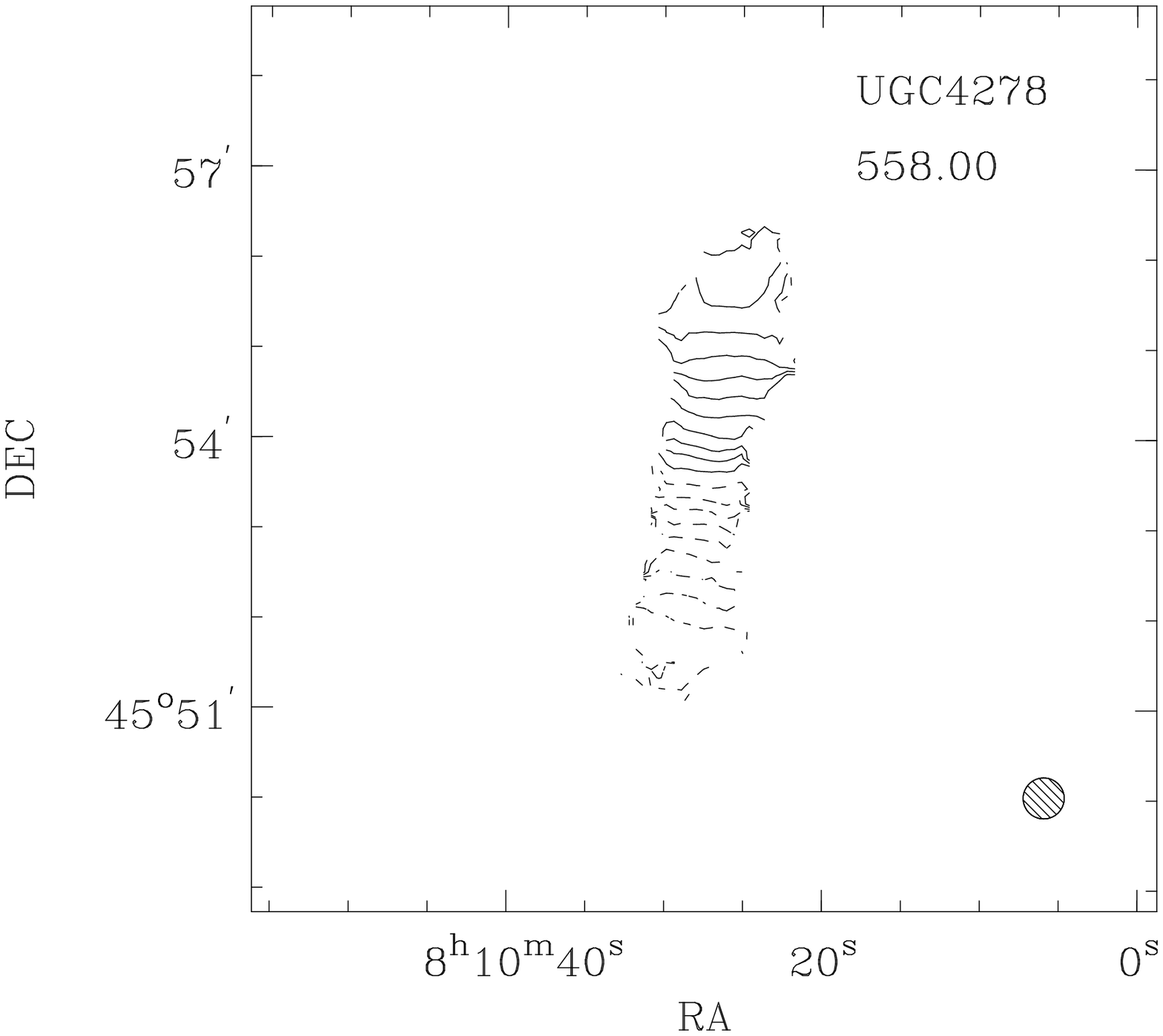}}
\end{minipage}
\hfill
\begin{minipage}[b]{5.7 cm}
\resizebox{5.85cm}{!}{\includegraphics[angle=-90]{U4278.tplfit.ps}}
\end{minipage}
\vspace{-1cm}
\caption{
\small Continued
}
\end{figure*}

\begin{figure*}
\addtocounter{figure}{-1}
\begin{minipage}[b]{5.7 cm}
\resizebox{5.7cm}{!}{\includegraphics[angle=-90]{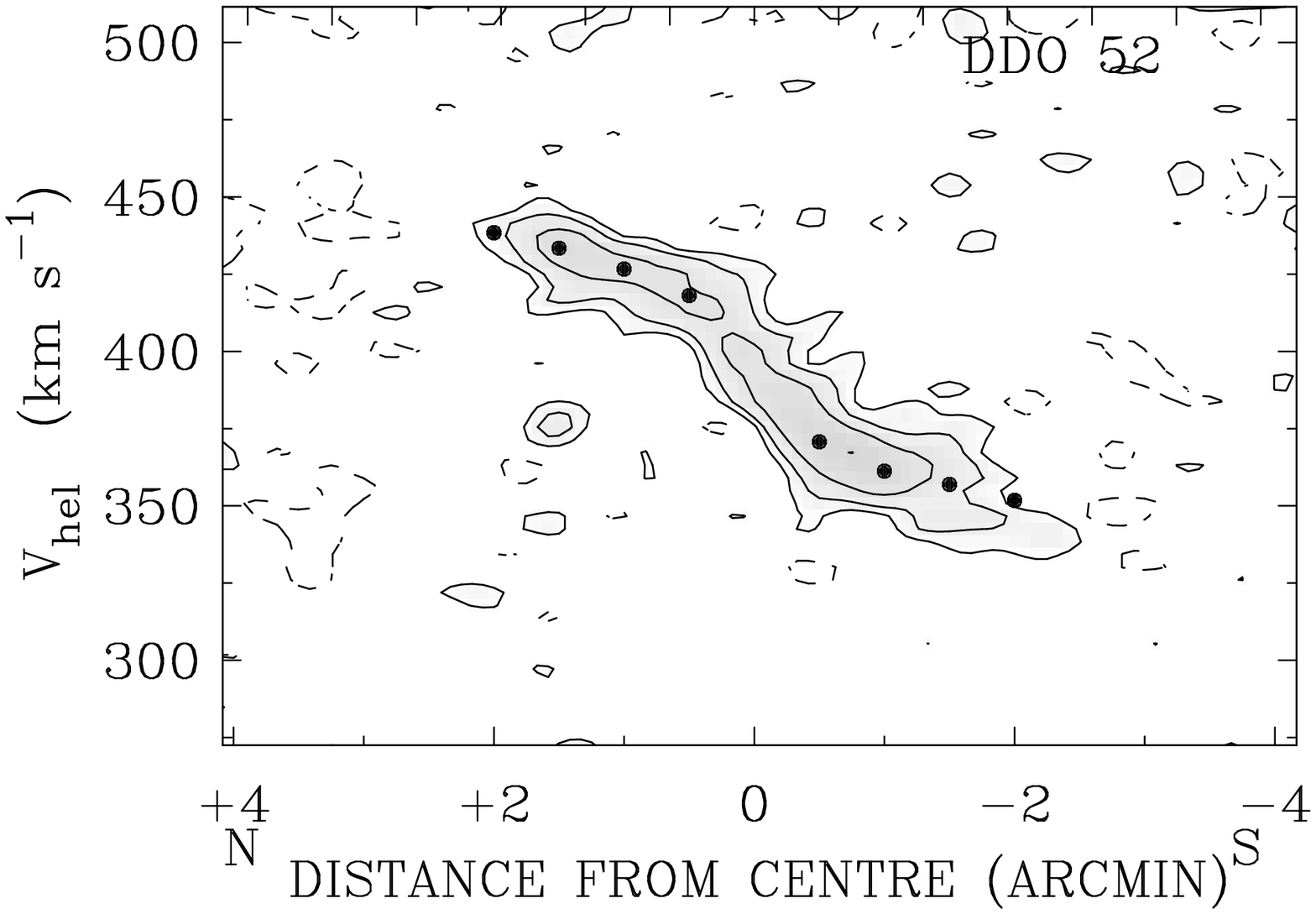}}
\end{minipage}
\hfill
\begin{minipage}[b]{5.7 cm}
\resizebox{5.7cm}{!}{\includegraphics[angle=-90]{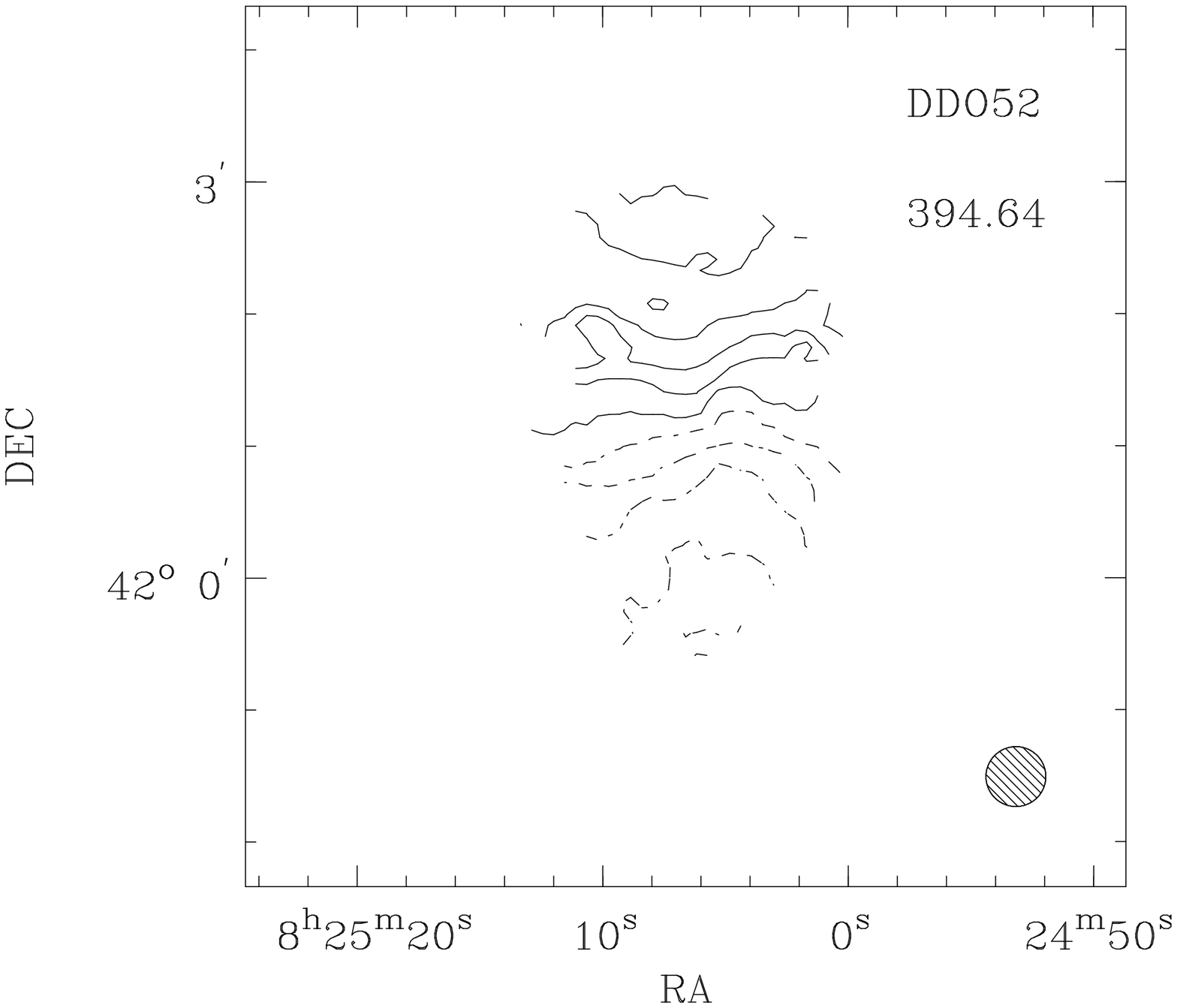}}
\end{minipage}
\hfill
\begin{minipage}[b]{5.7 cm}
\resizebox{5.85cm}{!}{\includegraphics[angle=-90]{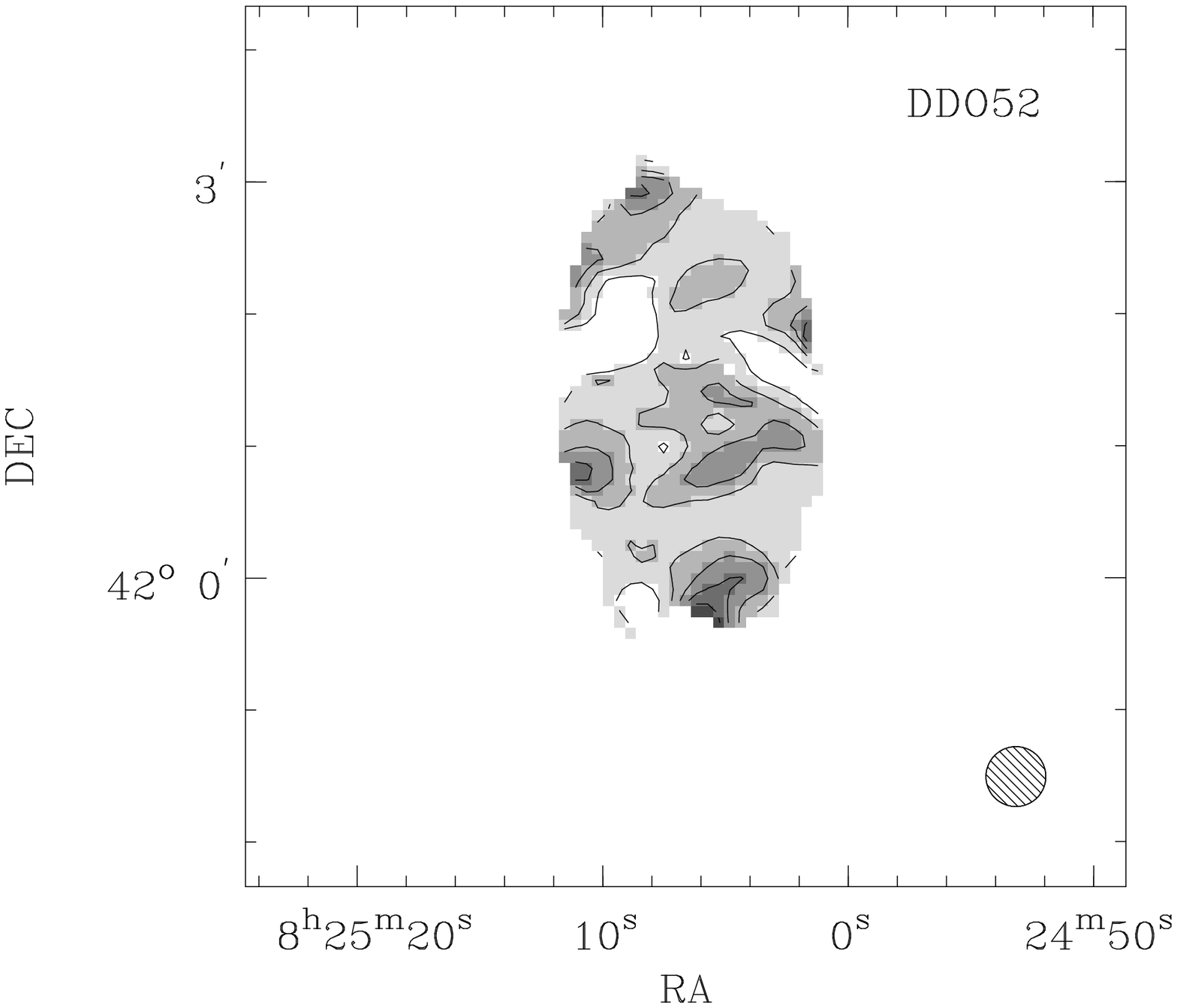}}
\end{minipage}
\begin{minipage}[b]{5.7 cm}
\resizebox{5.7cm}{!}{\includegraphics[angle=-90]{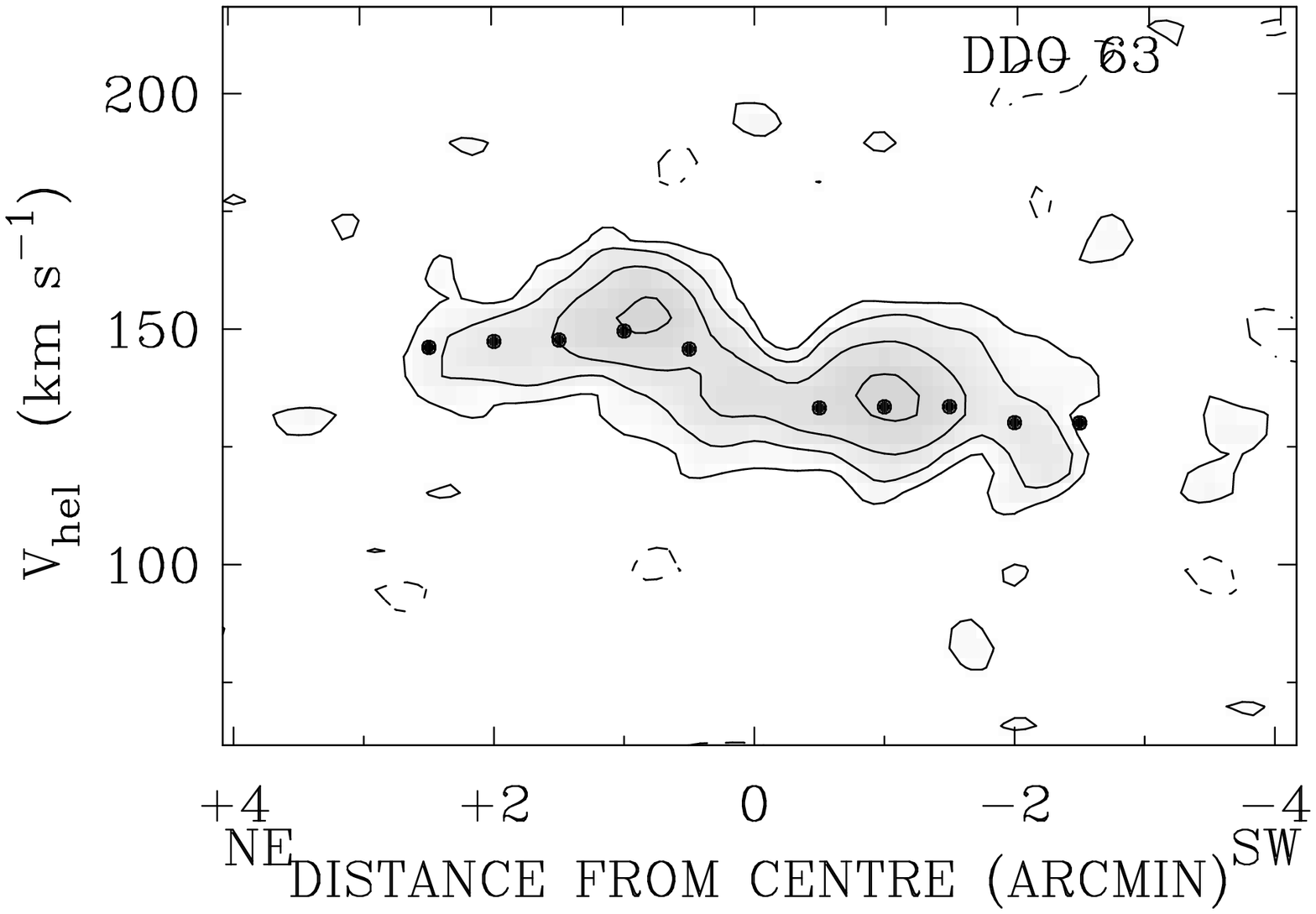}}
\end{minipage}
\hfill
\begin{minipage}[b]{5.7 cm}
\resizebox{5.7cm}{!}{\includegraphics[angle=-90]{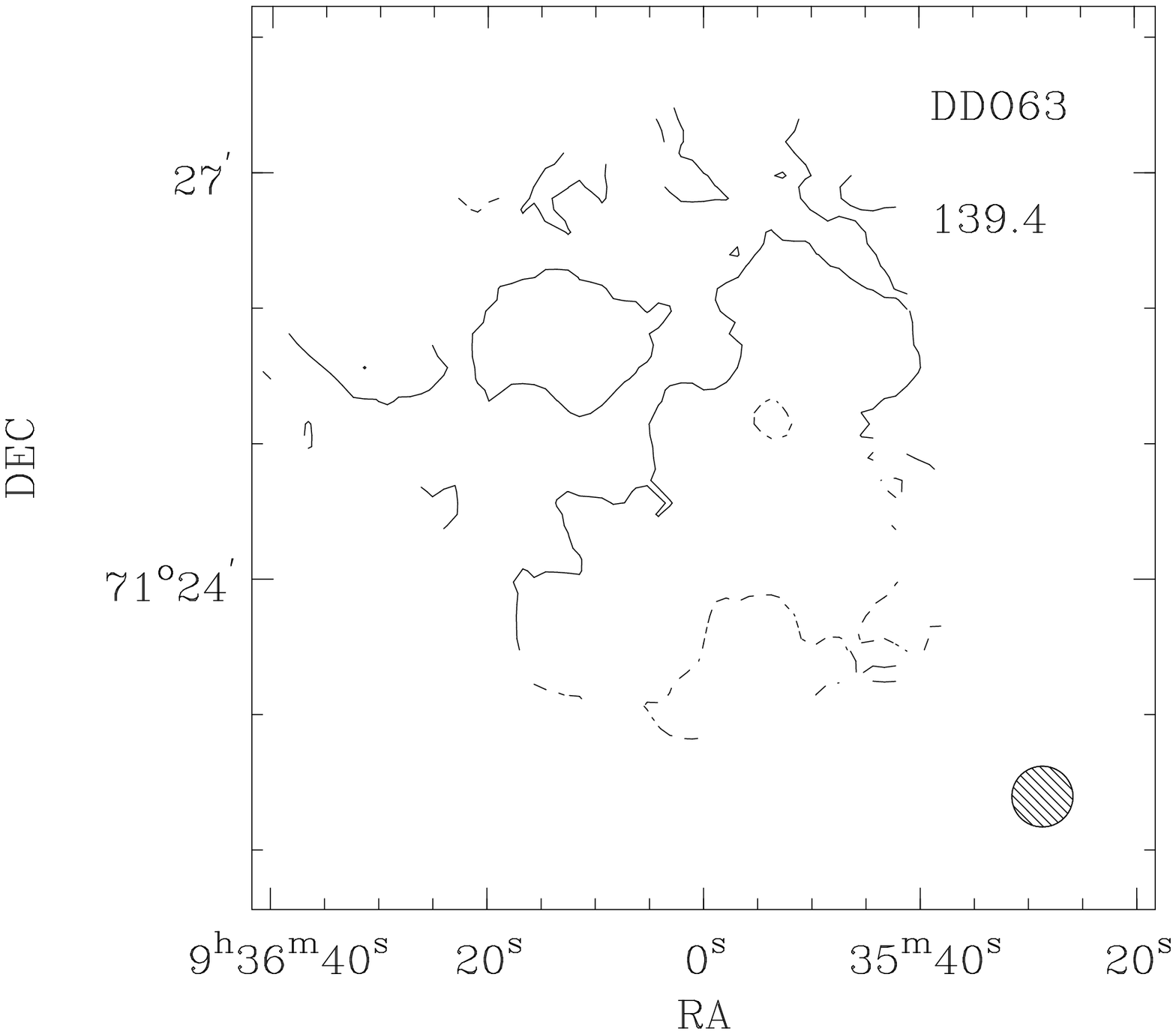}}
\end{minipage}
\hfill
\begin{minipage}[b]{5.7 cm}
\resizebox{5.85cm}{!}{\includegraphics[angle=-90]{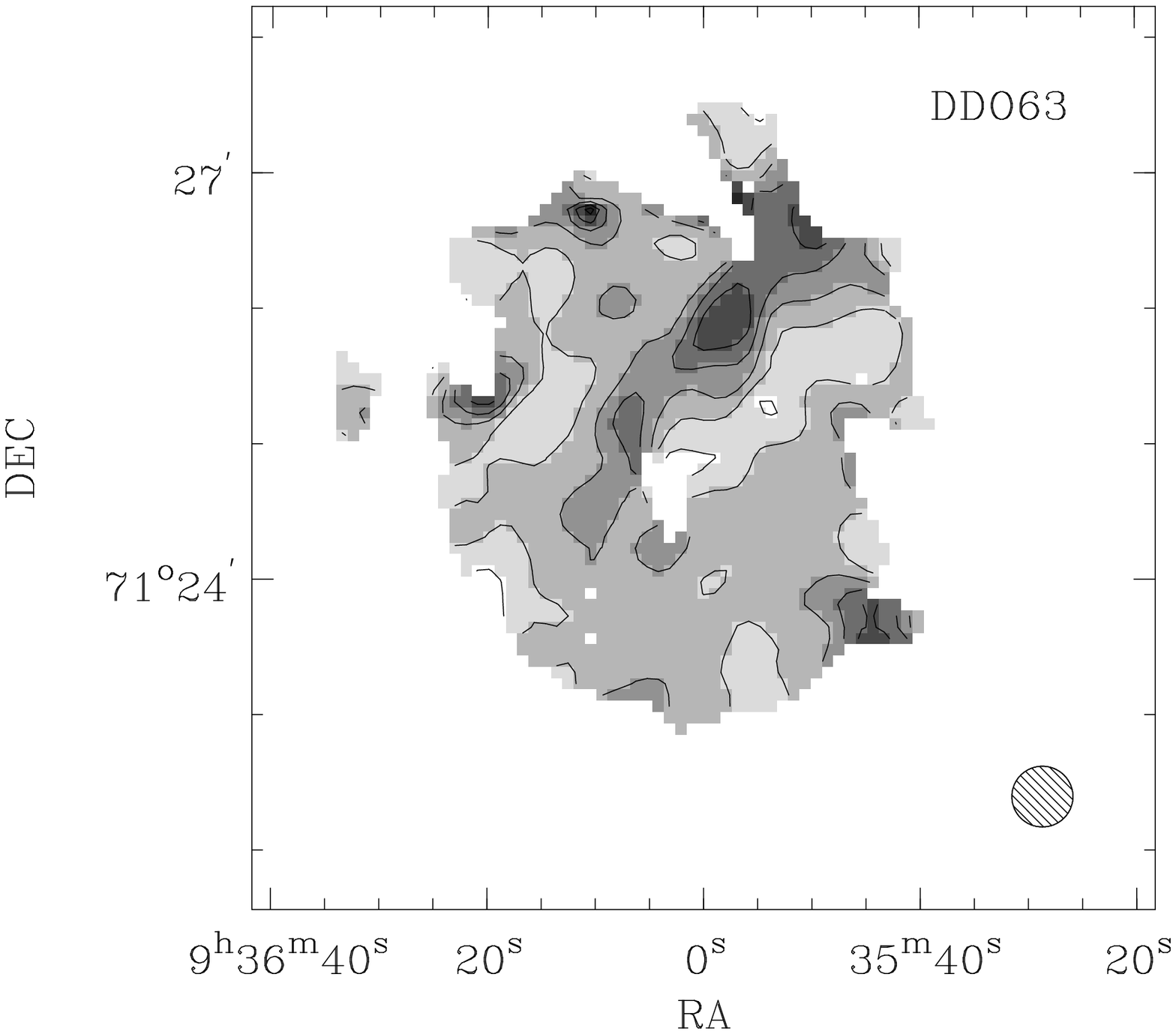}}
\end{minipage}
\begin{minipage}[b]{5.7 cm}
\resizebox{5.7cm}{!}{\includegraphics[angle=-90]{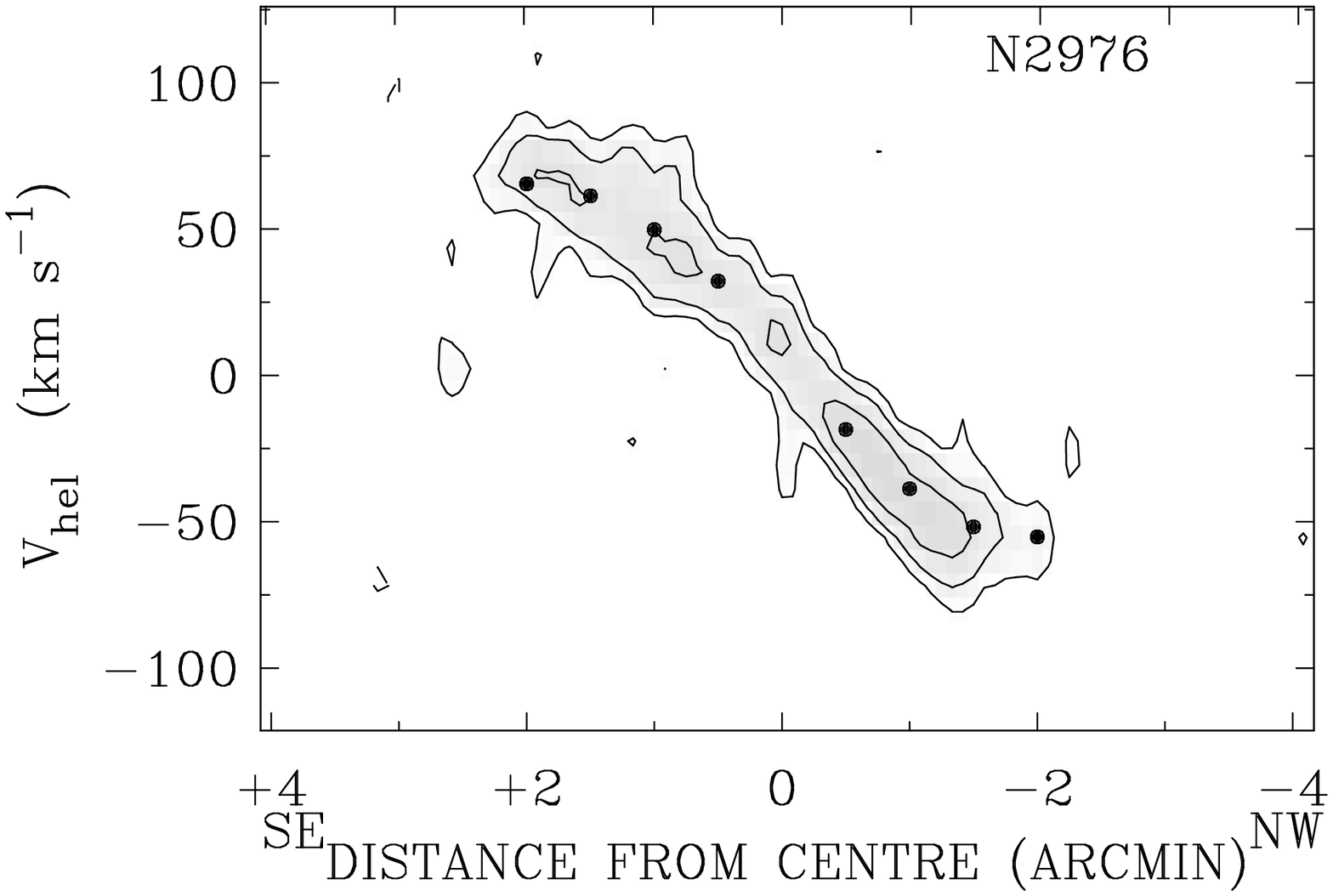}}
\end{minipage}
\hfill
\begin{minipage}[b]{5.7 cm}
\resizebox{5.7cm}{!}{\includegraphics[angle=-90]{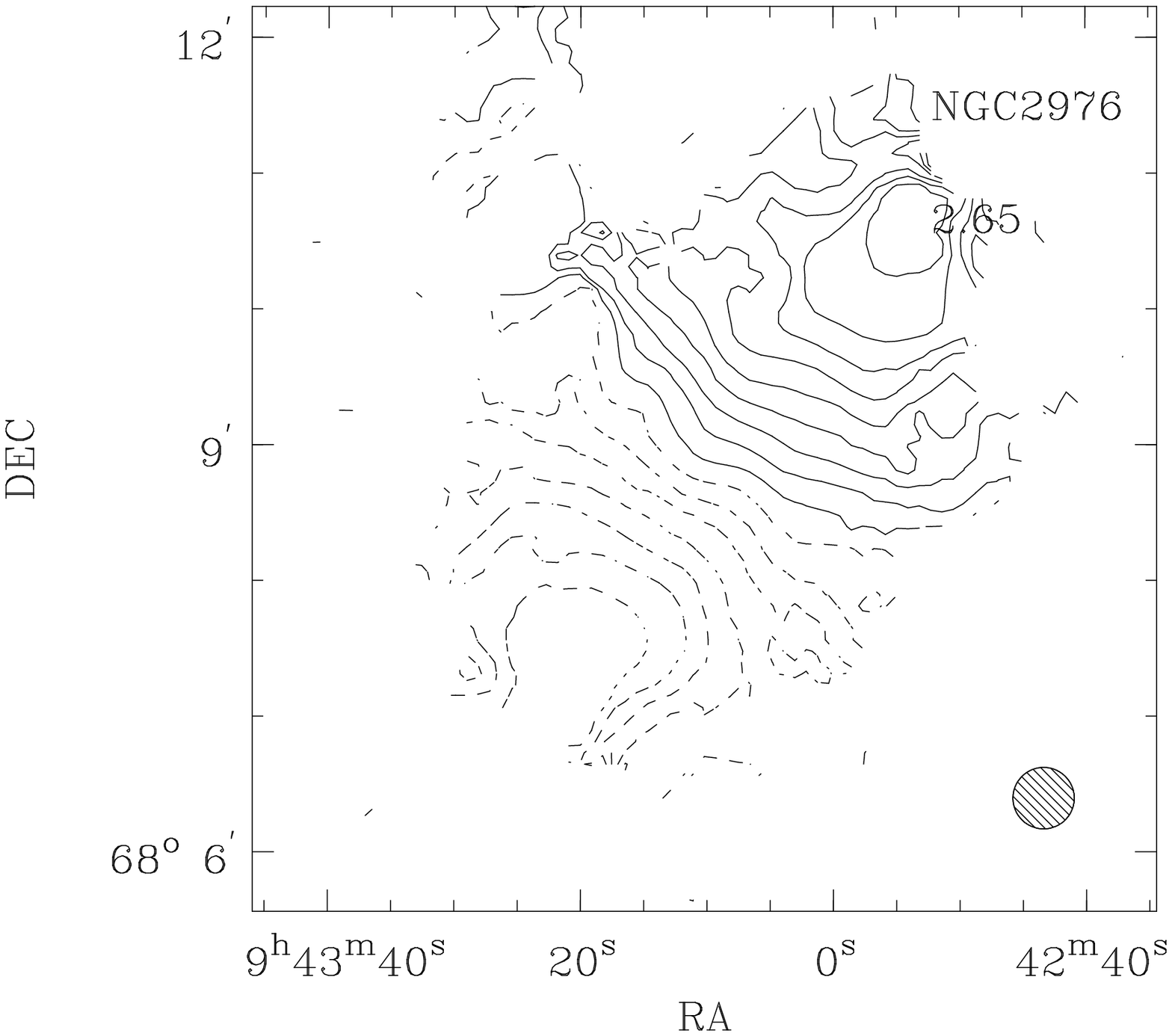}}
\end{minipage}
\hfill
\begin{minipage}[b]{5.7 cm}
\resizebox{5.85cm}{!}{\includegraphics[angle=-90]{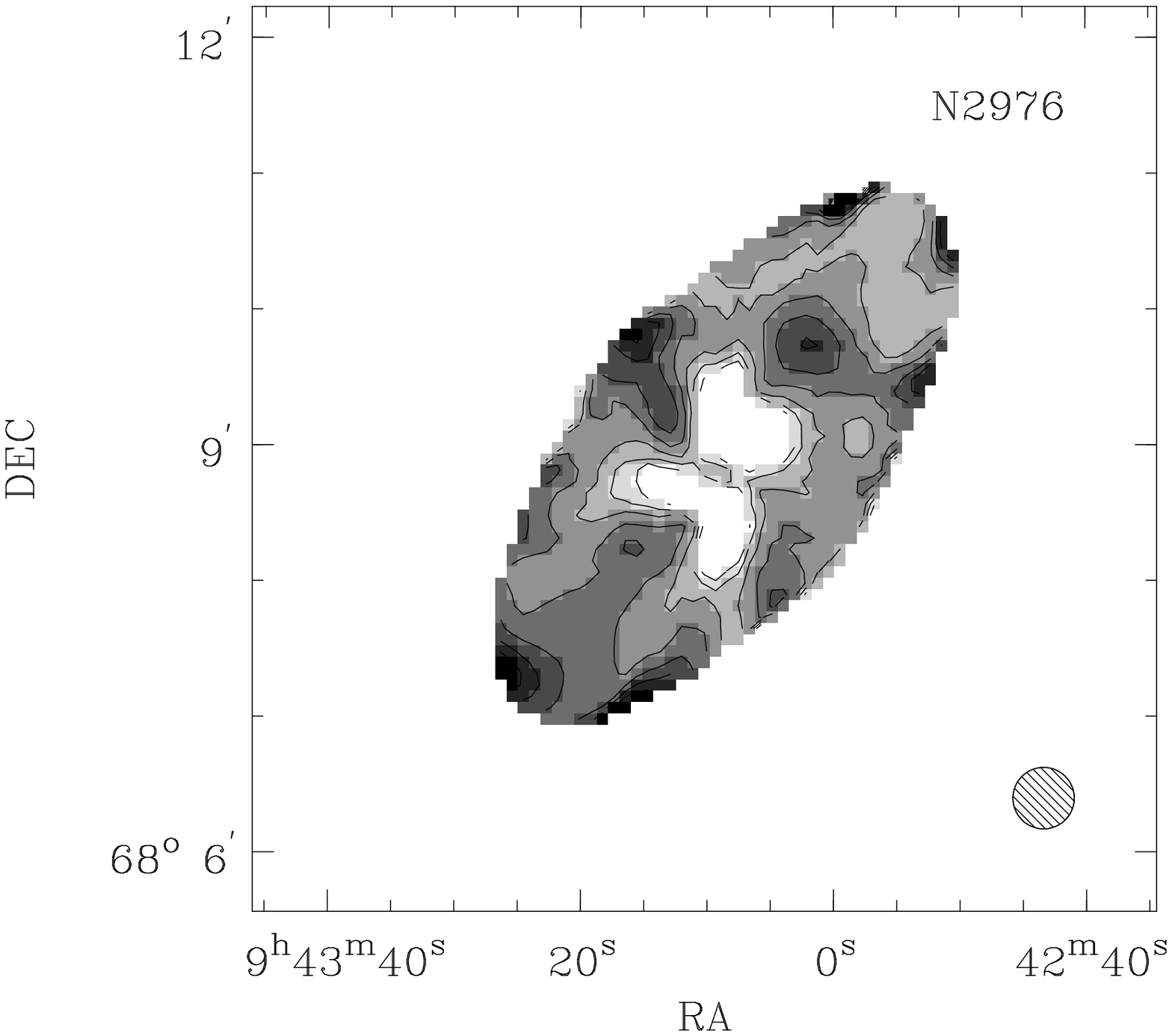}}
\end{minipage}
\begin{minipage}[b]{5.7 cm}
\resizebox{5.7cm}{!}{\includegraphics[angle=-90]{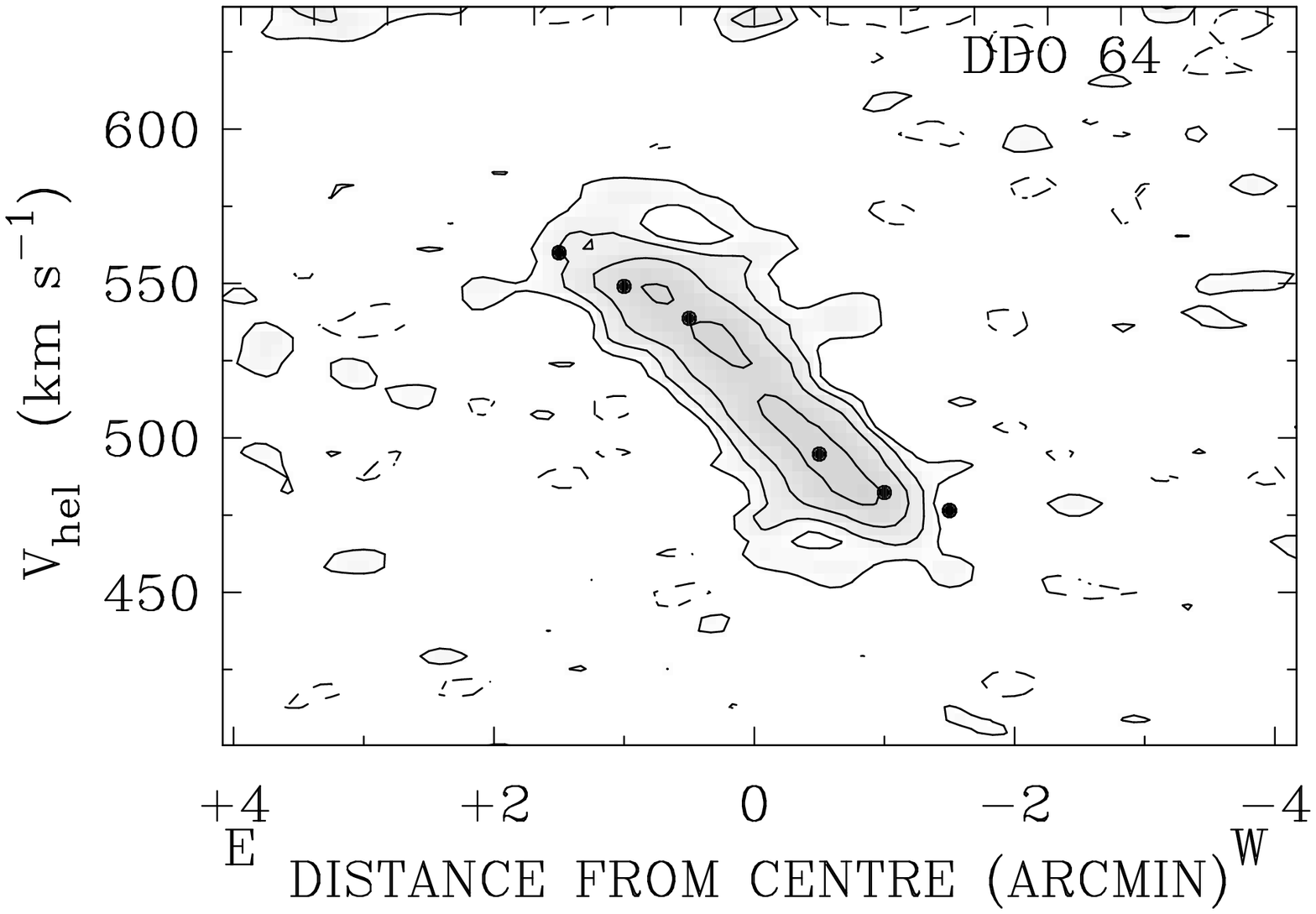}}
\end{minipage}
\hfill
\begin{minipage}[b]{5.7 cm}
\resizebox{5.7cm}{!}{\includegraphics[angle=-90]{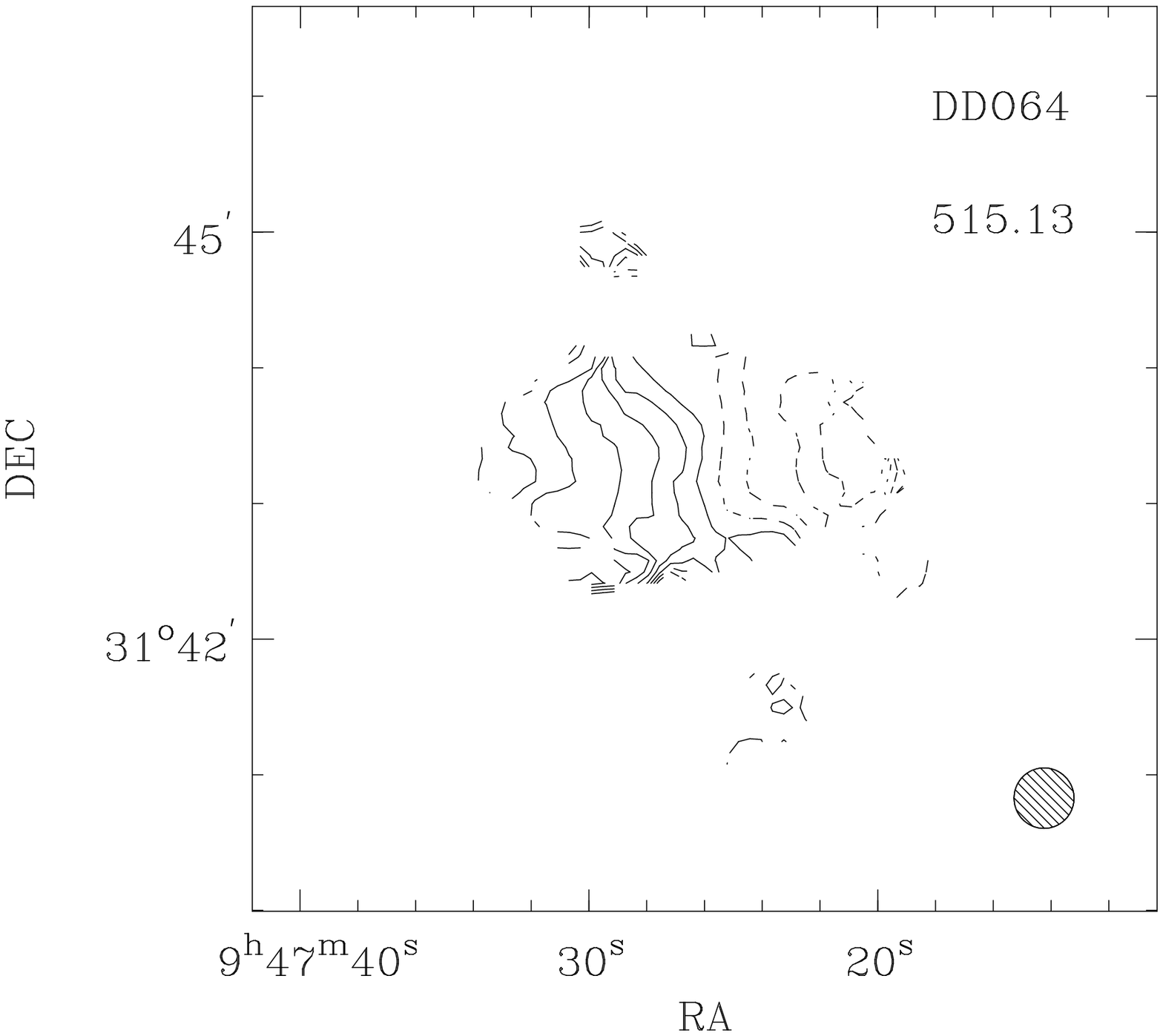}}
\end{minipage}
\hfill
\begin{minipage}[b]{5.7 cm}
\resizebox{5.85cm}{!}{\includegraphics[angle=-90]{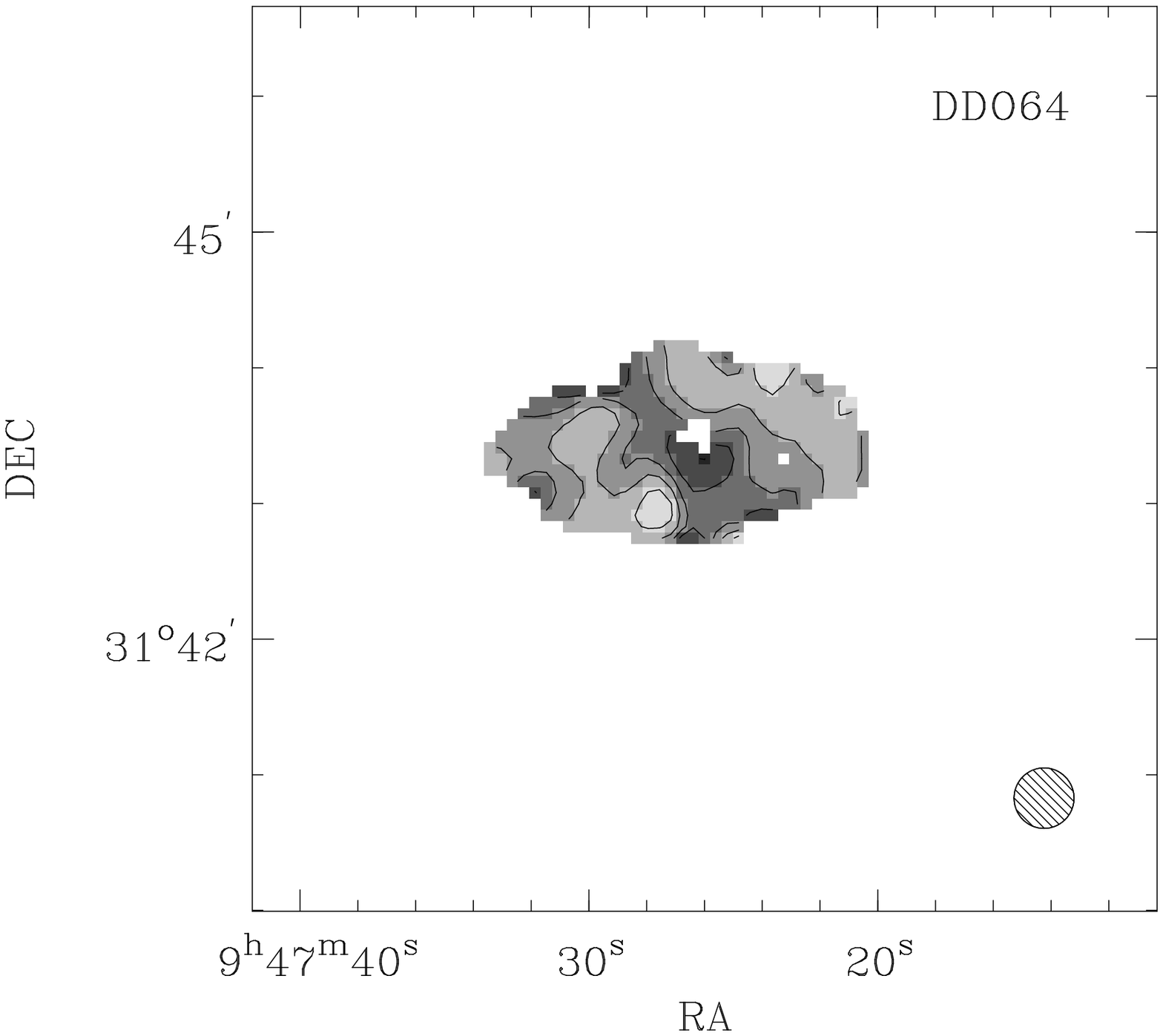}}
\end{minipage}
\caption{
\small Continued
}
\end{figure*}

\begin{figure*}
\addtocounter{figure}{-1}
\begin{minipage}[b]{5.7 cm}
\resizebox{5.7cm}{!}{\includegraphics[angle=-90]{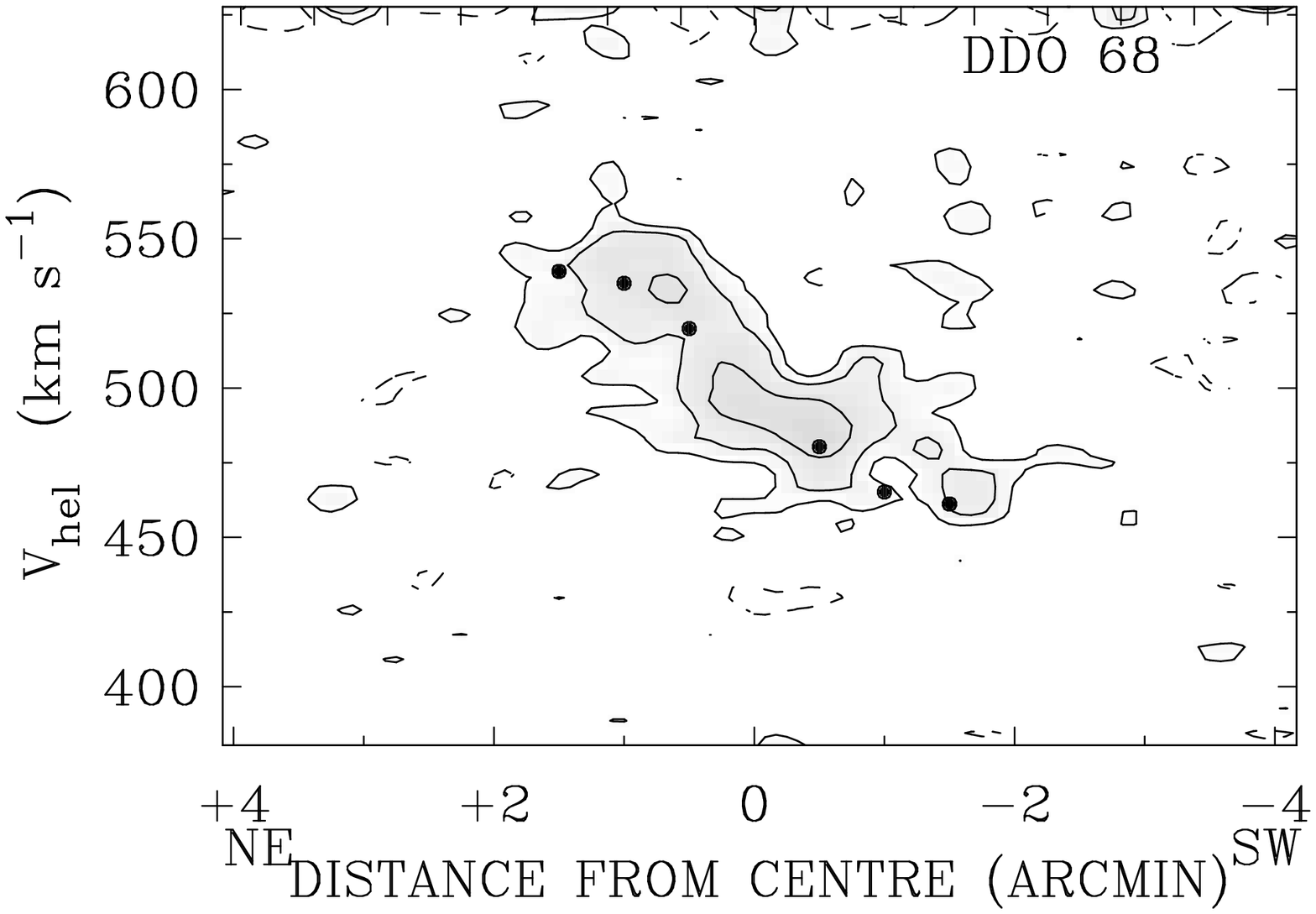}}
\end{minipage}
\hfill
\begin{minipage}[b]{5.7 cm}
\resizebox{5.7cm}{!}{\includegraphics[angle=-90]{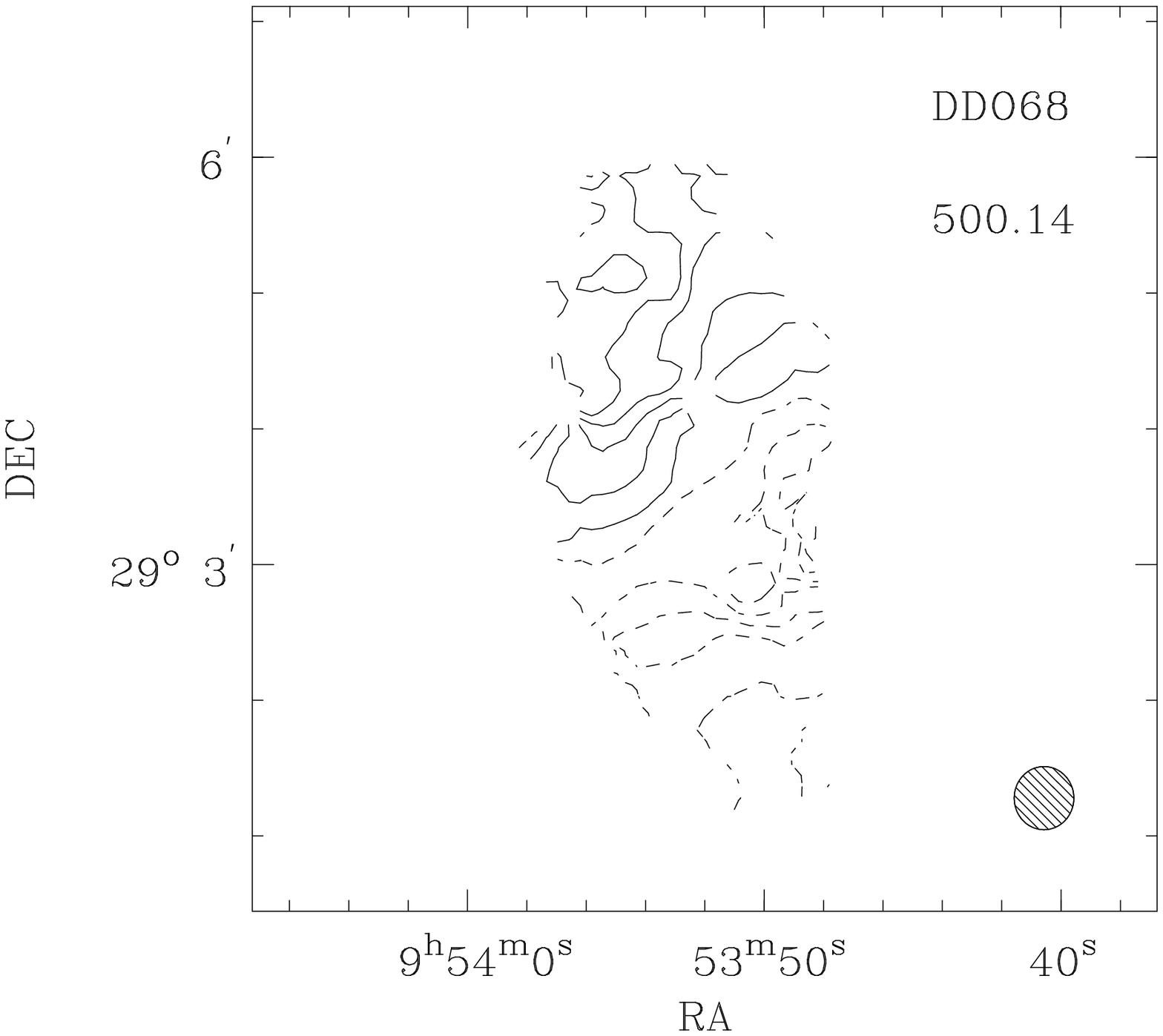}}
\end{minipage}
\hfill
\begin{minipage}[b]{5.7 cm}
\resizebox{5.85cm}{!}{\includegraphics[angle=-90]{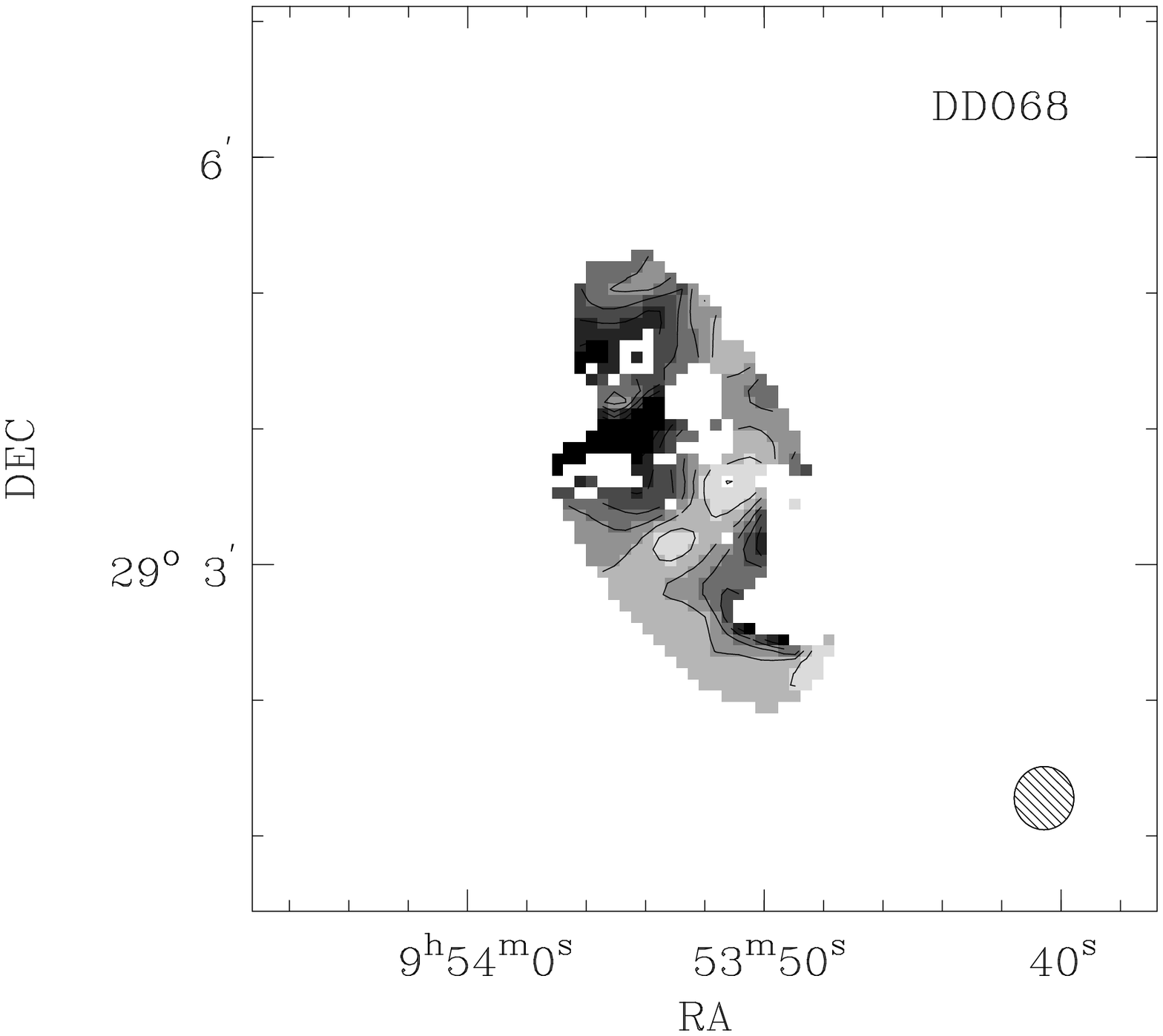}}
\end{minipage}
\begin{minipage}[b]{5.7 cm}
\resizebox{5.7cm}{!}{\includegraphics[angle=-90]{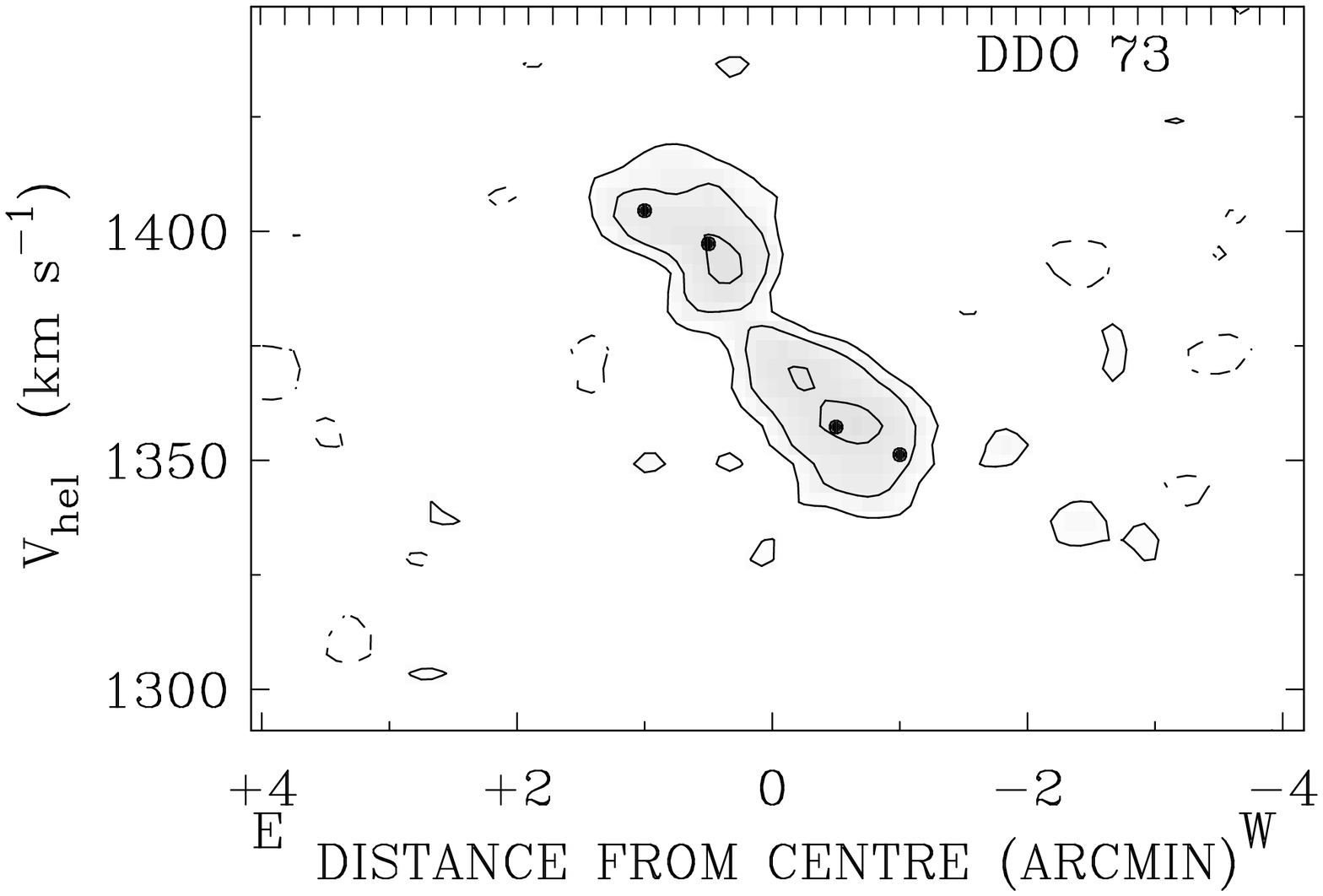}}
\end{minipage}
\hfill
\begin{minipage}[b]{5.7 cm}
\resizebox{5.7cm}{!}{\includegraphics[angle=-90]{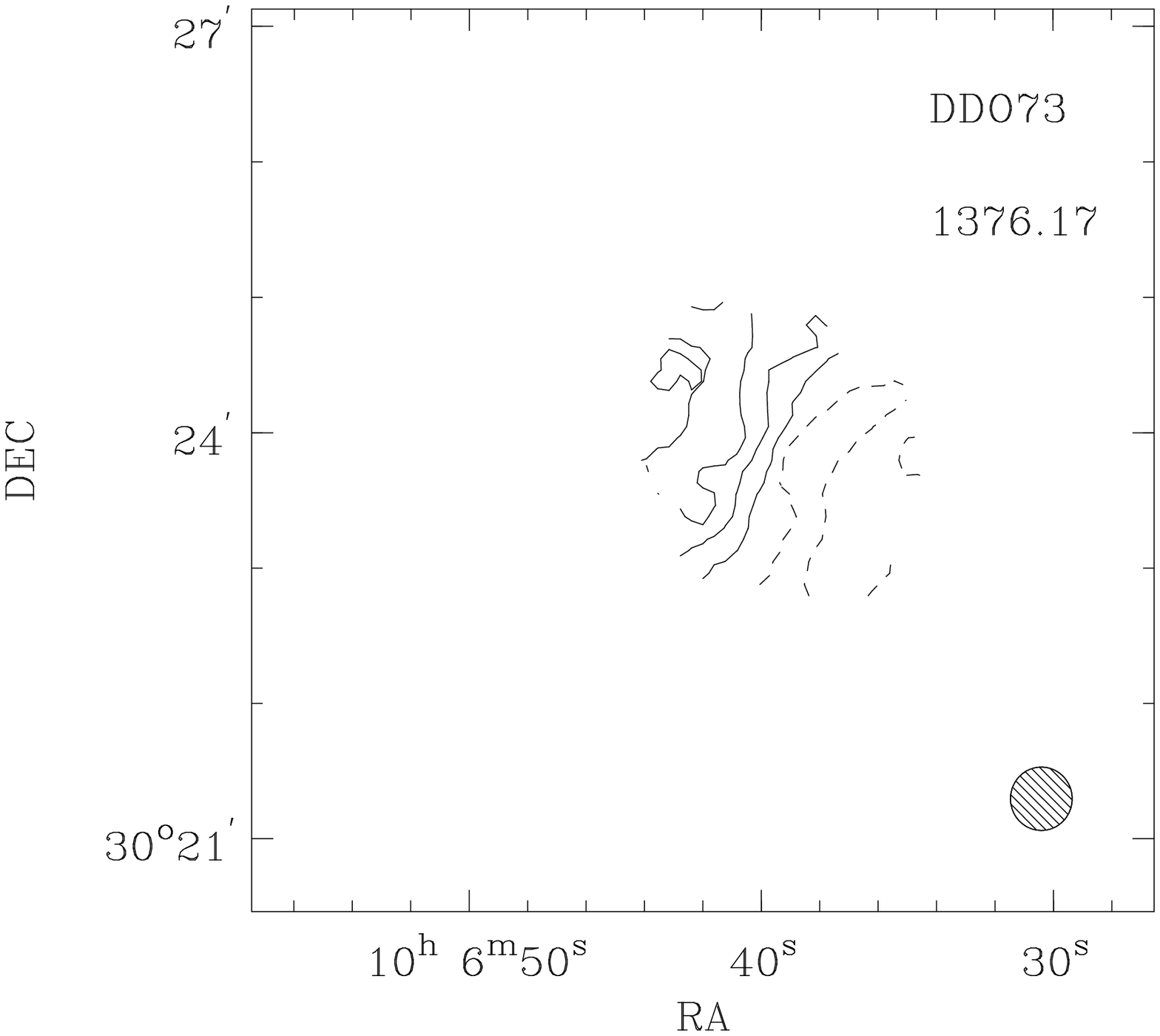}}
\end{minipage}
\hfill
\begin{minipage}[b]{5.7 cm}
\resizebox{5.85cm}{!}{\includegraphics[angle=-90]{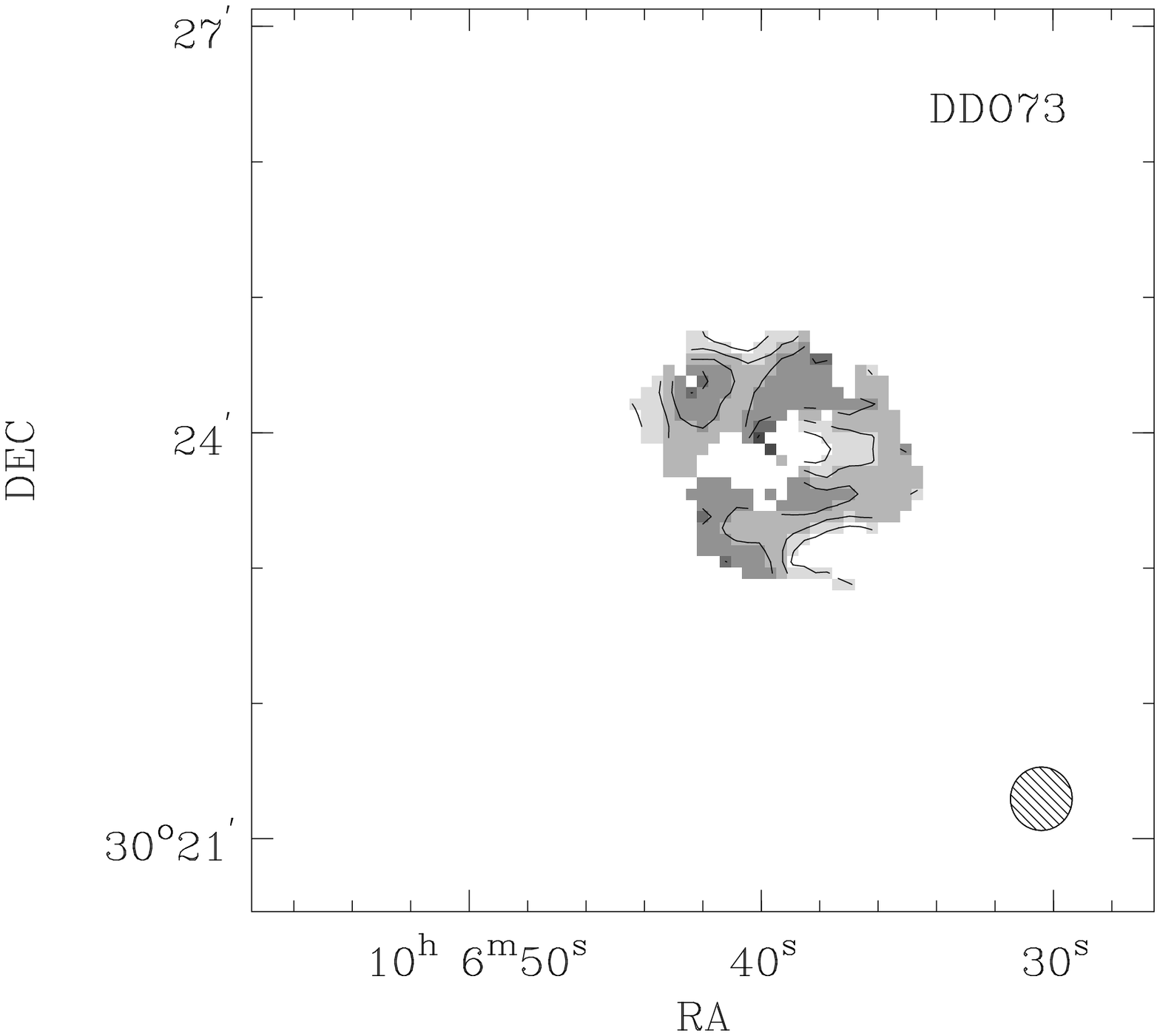}}
\end{minipage}
\begin{minipage}[b]{5.7 cm}
\resizebox{5.7cm}{!}{\includegraphics[angle=-90]{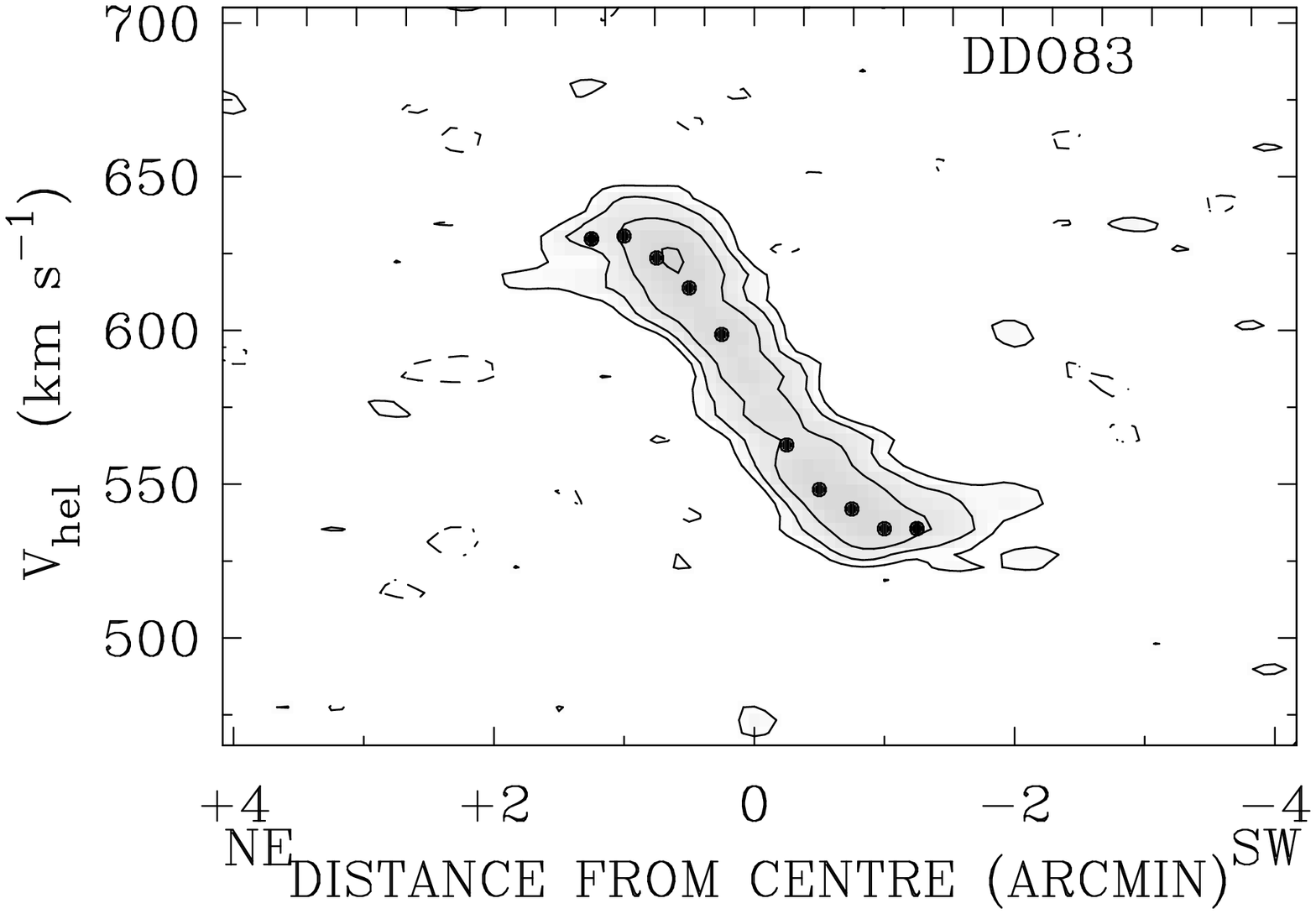}}
\end{minipage}
\hfill
\begin{minipage}[b]{5.7 cm}
\resizebox{5.7cm}{!}{\includegraphics[angle=-90]{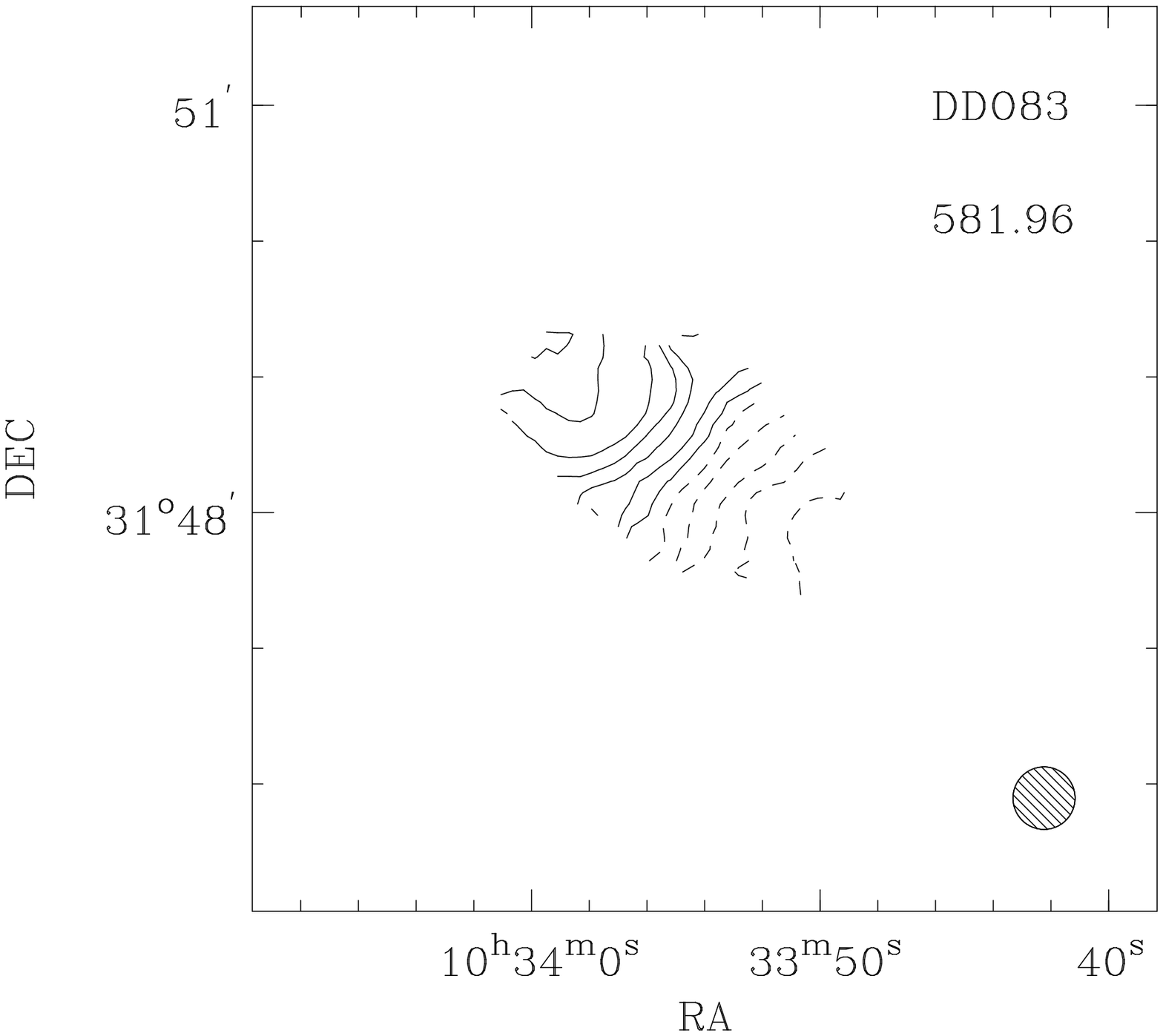}}
\end{minipage}
\hfill
\begin{minipage}[b]{5.7 cm}
\resizebox{5.85cm}{!}{\includegraphics[angle=-90]{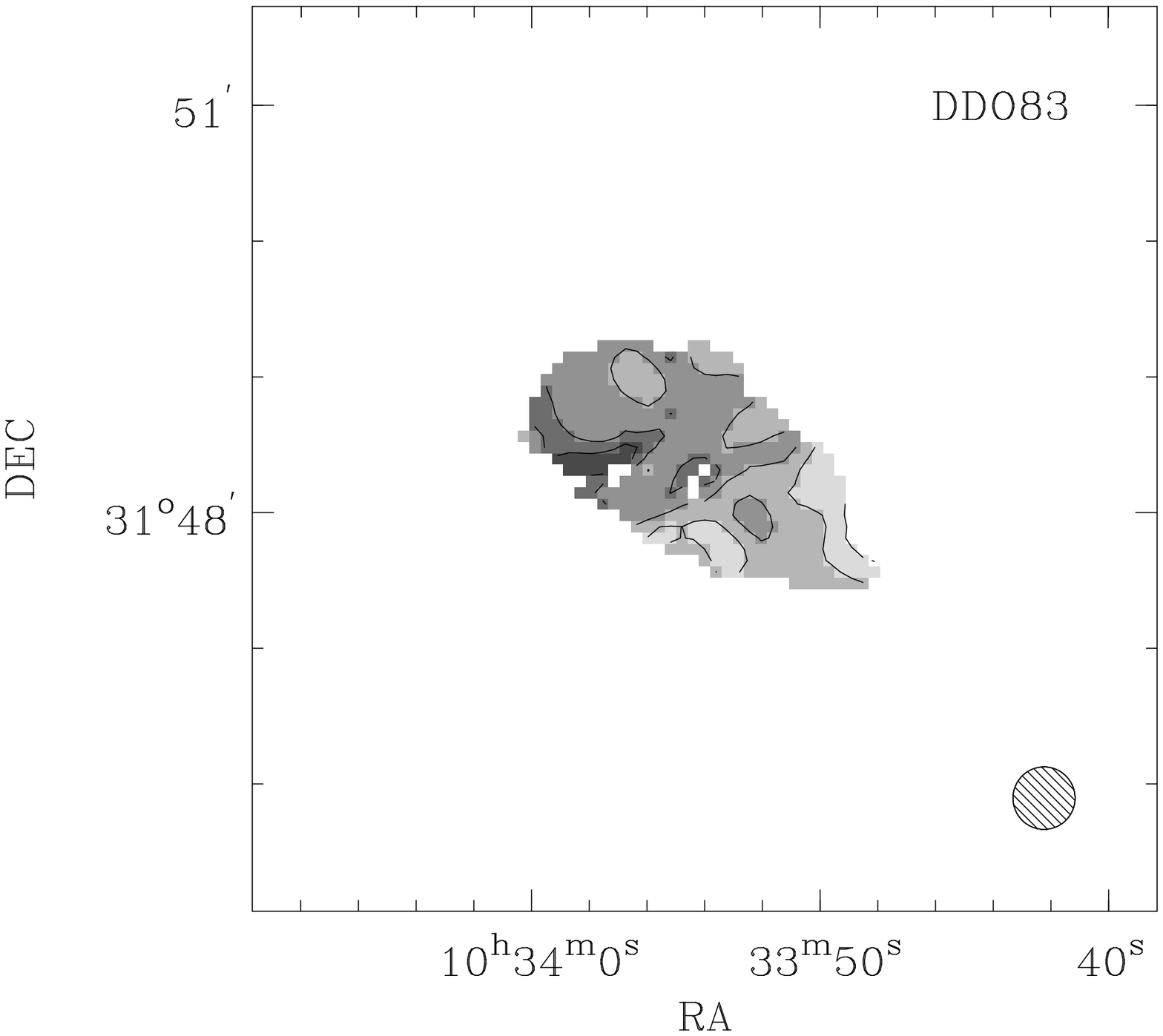}}
\end{minipage}
\begin{minipage}[b]{5.7 cm}
\resizebox{5.7cm}{!}{\includegraphics[angle=-90]{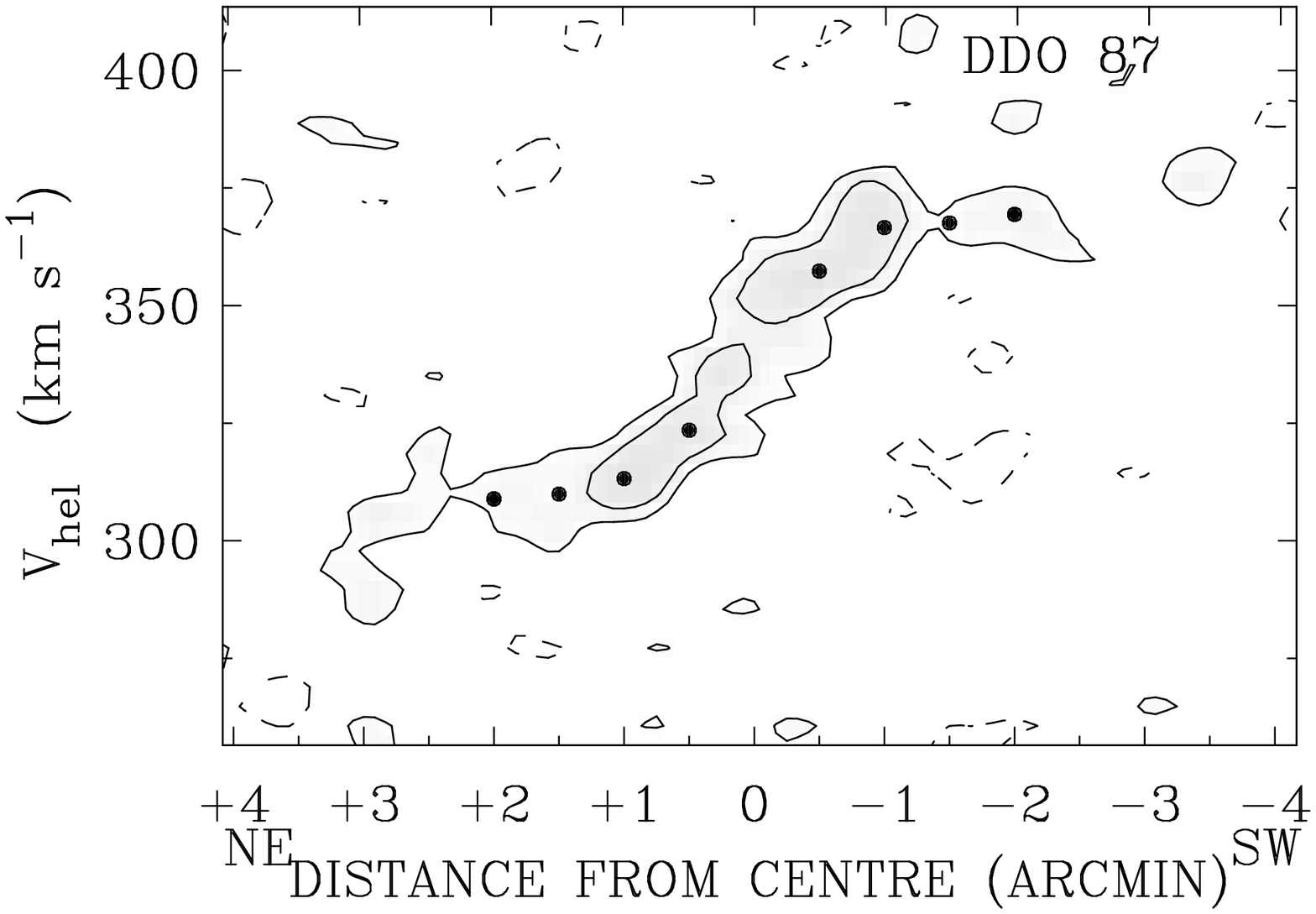}}
\end{minipage}
\hfill
\begin{minipage}[b]{5.7 cm}
\resizebox{5.7cm}{!}{\includegraphics[angle=-90]{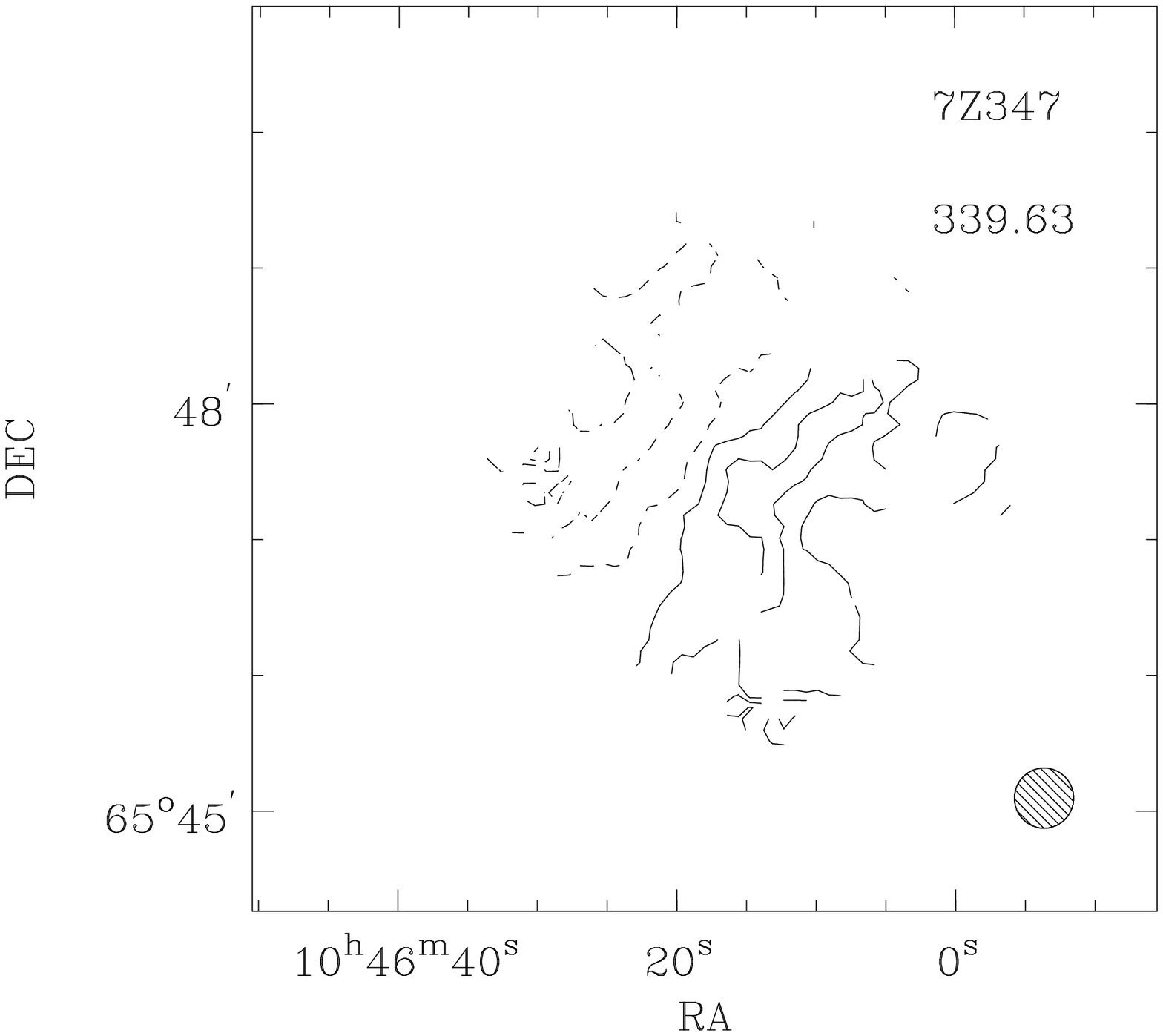}}
\end{minipage}
\hfill
\begin{minipage}[b]{5.7 cm}
\resizebox{5.85cm}{!}{\includegraphics[angle=-90]{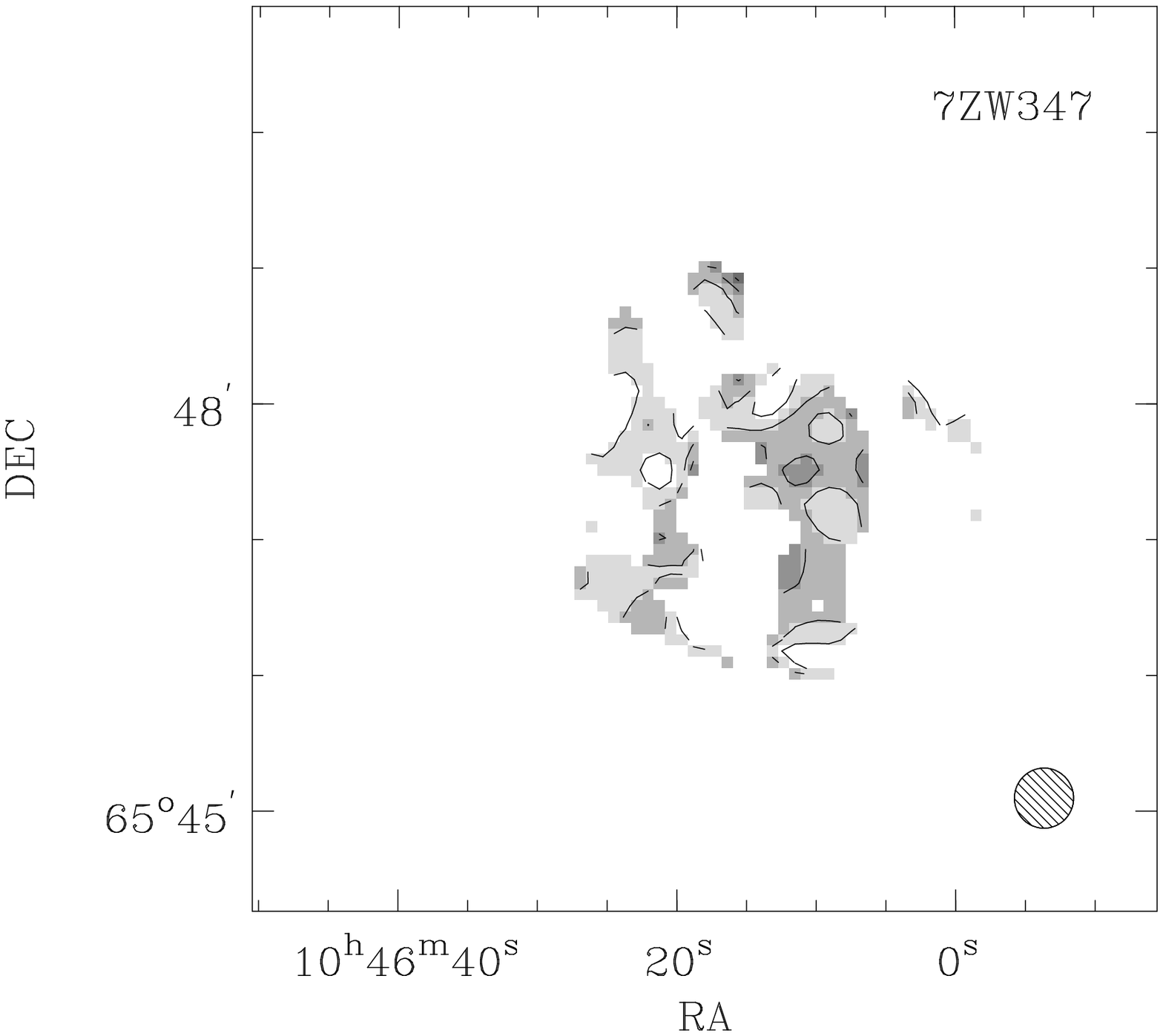}}
\end{minipage}
\caption{
\small Continued
}
\end{figure*}

\begin{figure*}
\addtocounter{figure}{-1}
\begin{minipage}[b]{5.7 cm}
\resizebox{5.7cm}{!}{\includegraphics[angle=-90]{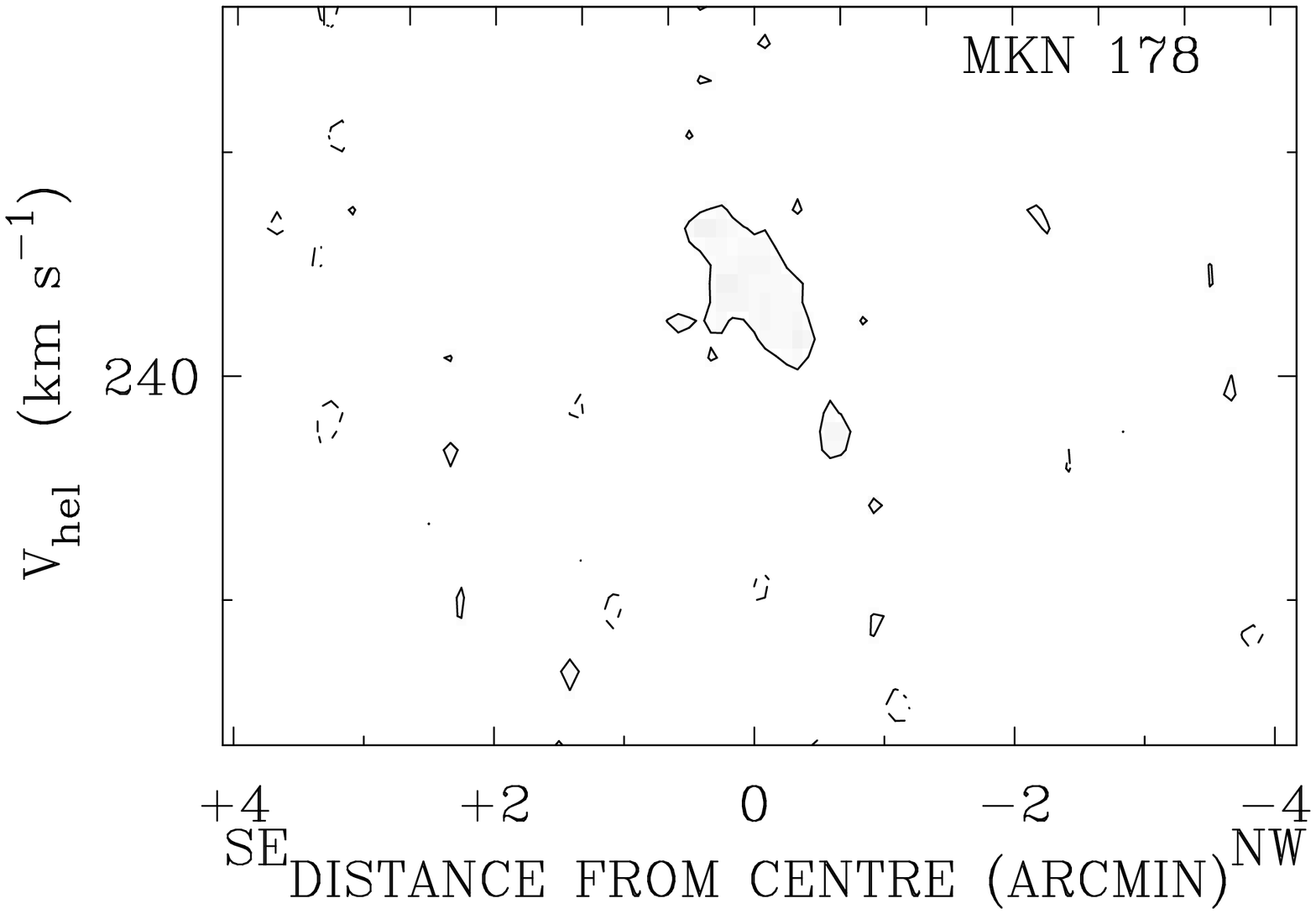}}
\end{minipage}
\hfill
\begin{minipage}[b]{5.7 cm}
\resizebox{5.7cm}{!}{\includegraphics[angle=-90]{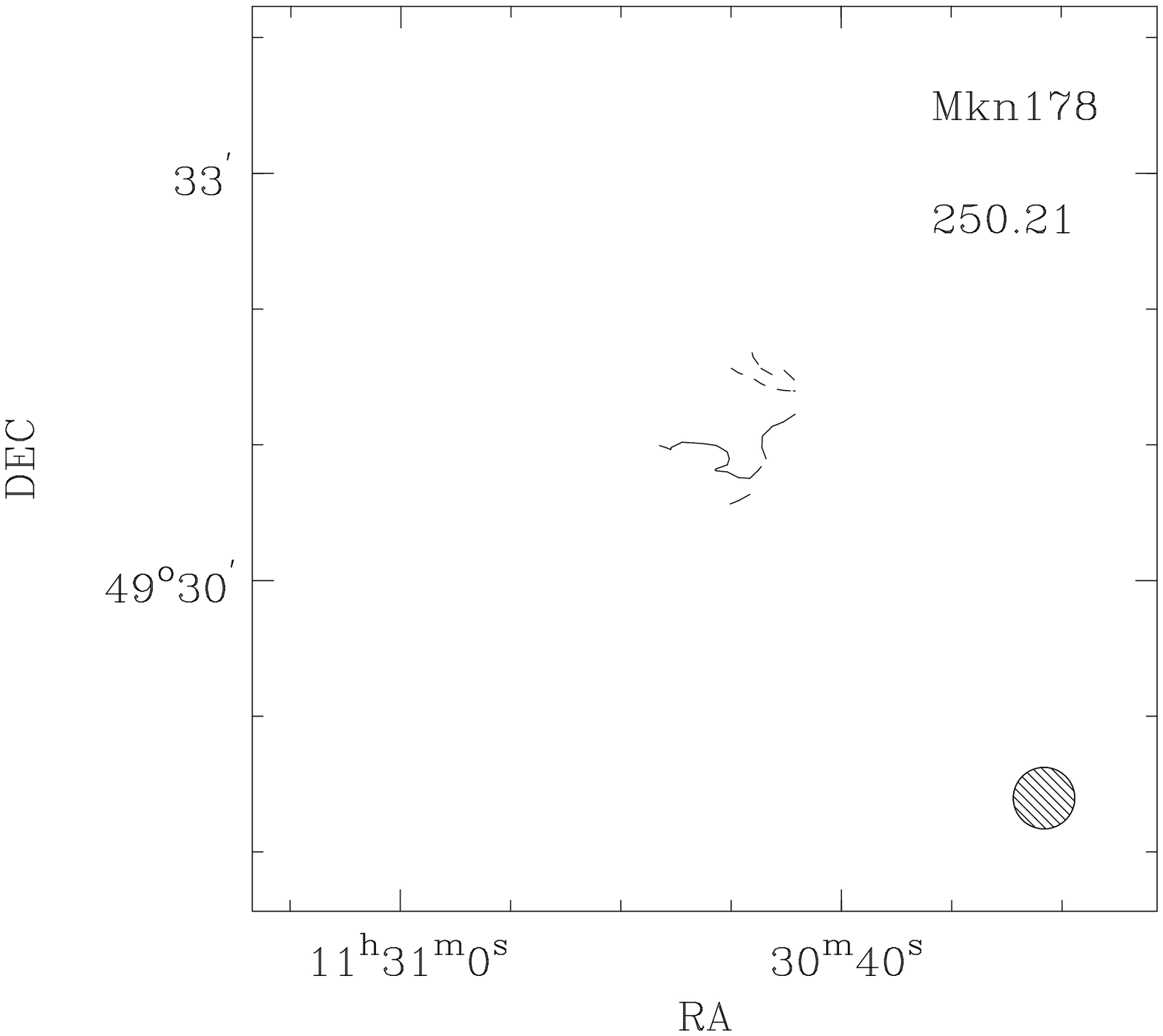}}
\end{minipage}
\hfill
\begin{minipage}[b]{5.7 cm}
\resizebox{5.85cm}{!}{\includegraphics[angle=-90]{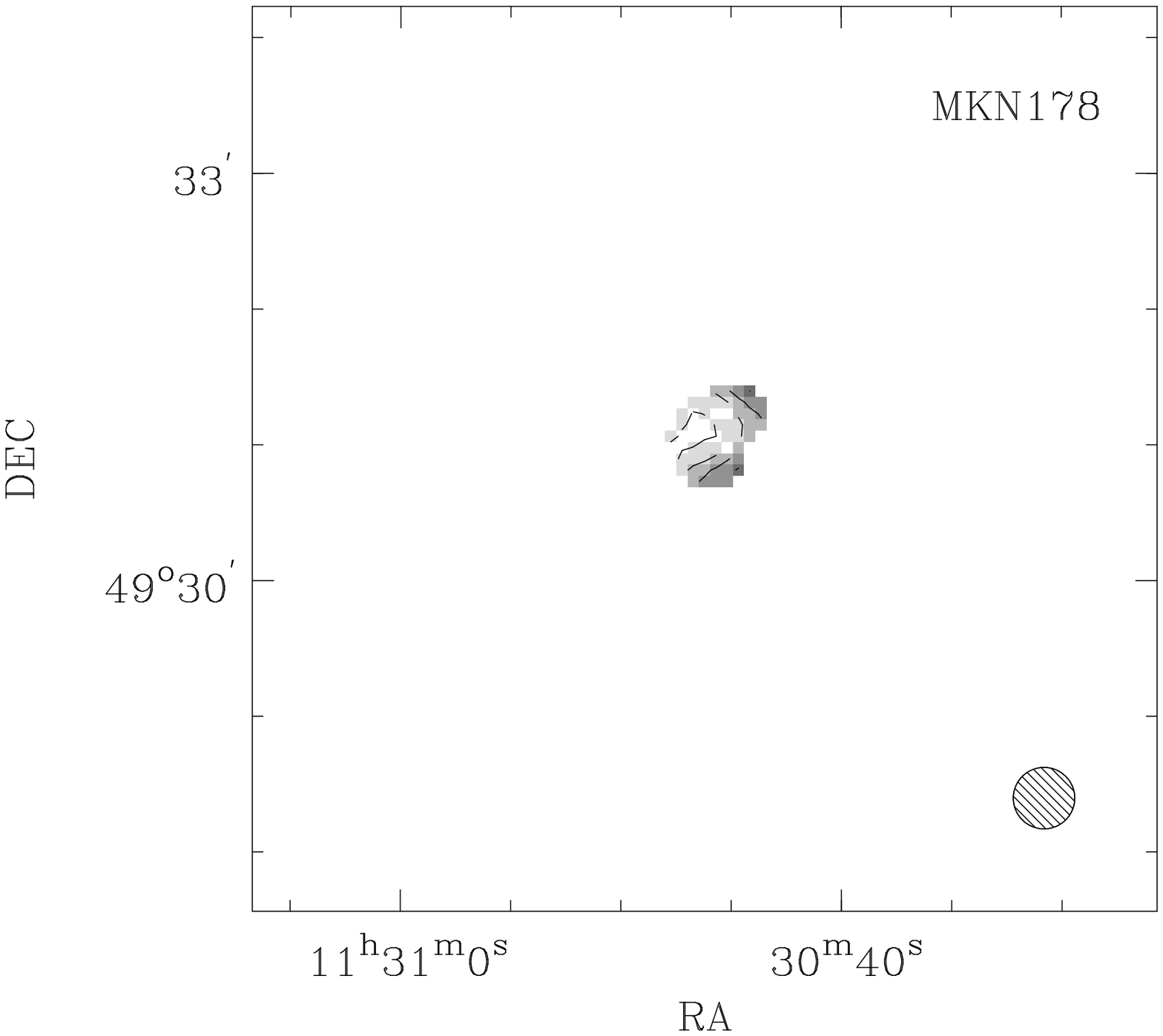}}
\end{minipage}
\begin{minipage}[b]{5.7 cm}
\resizebox{5.7cm}{!}{\includegraphics[angle=-90]{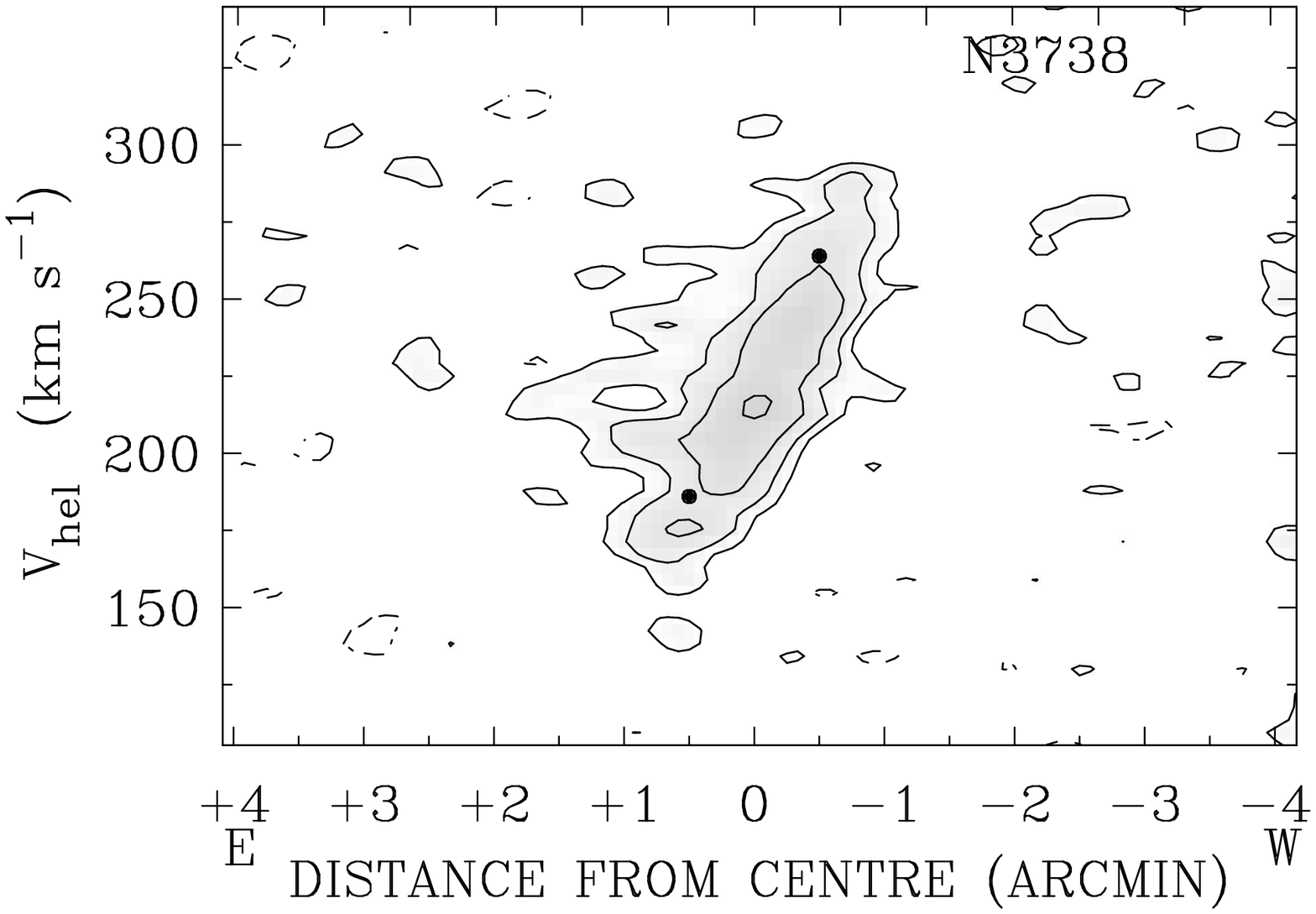}}
\end{minipage}
\hfill
\begin{minipage}[b]{5.7 cm}
\resizebox{5.7cm}{!}{\includegraphics[angle=-90]{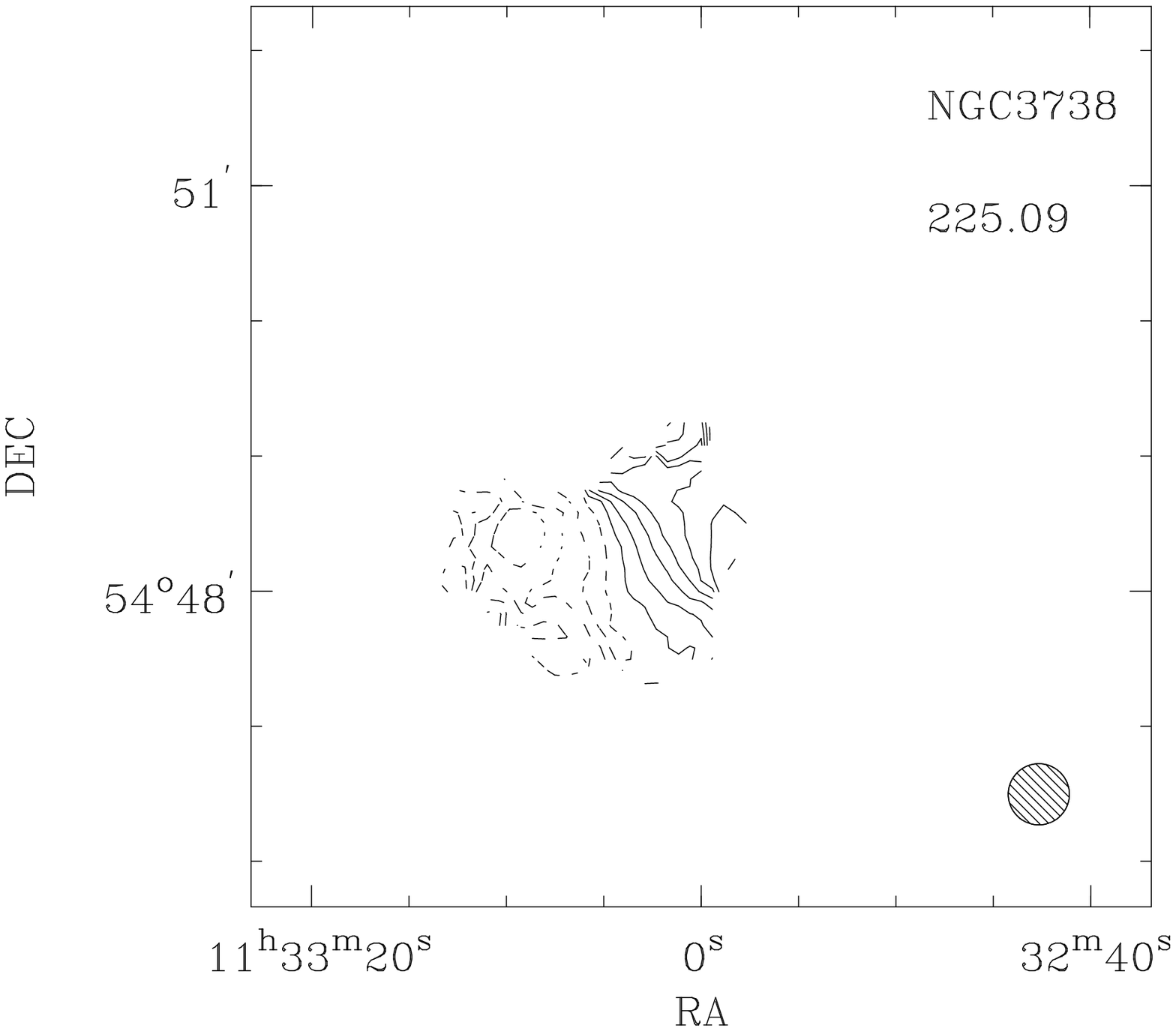}}
\end{minipage}
\hfill
\begin{minipage}[b]{5.7 cm}
\resizebox{5.85cm}{!}{\includegraphics[angle=-90]{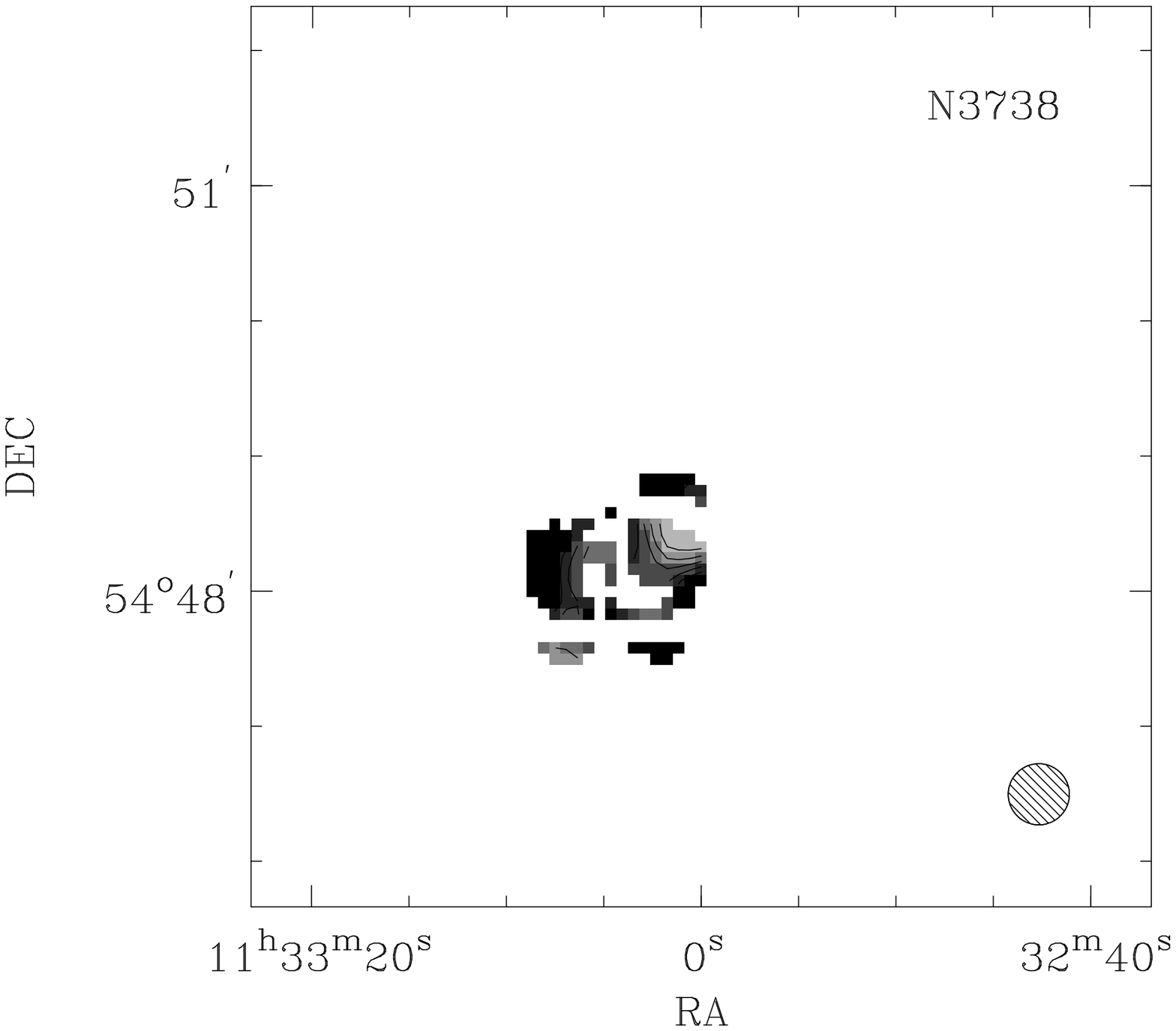}}
\end{minipage}
\begin{minipage}[b]{5.7 cm}
\resizebox{5.7cm}{!}{\includegraphics[angle=-90]{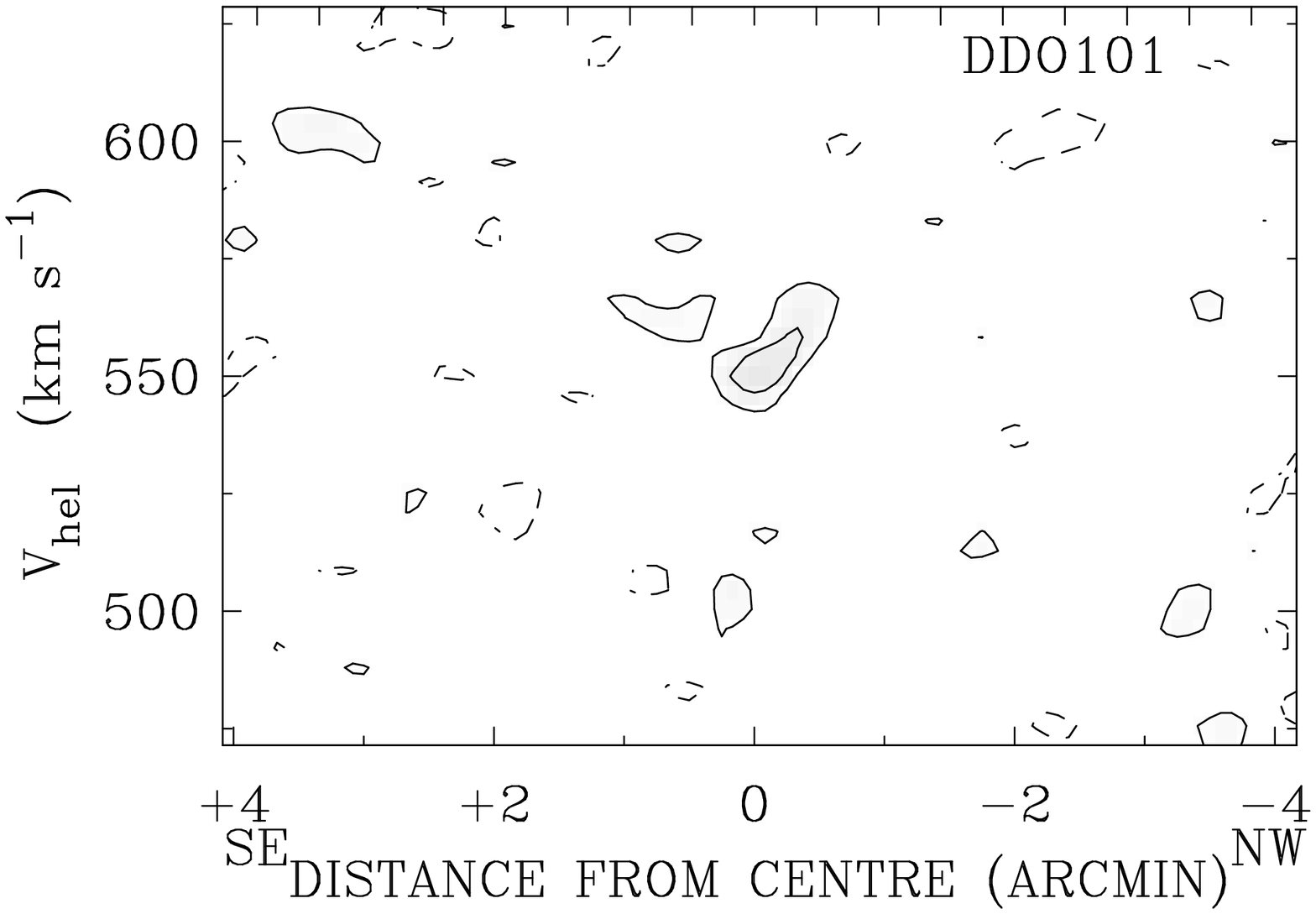}}
\end{minipage}
\hfill
\begin{minipage}[b]{5.7 cm}
\resizebox{5.7cm}{!}{\includegraphics[angle=-90]{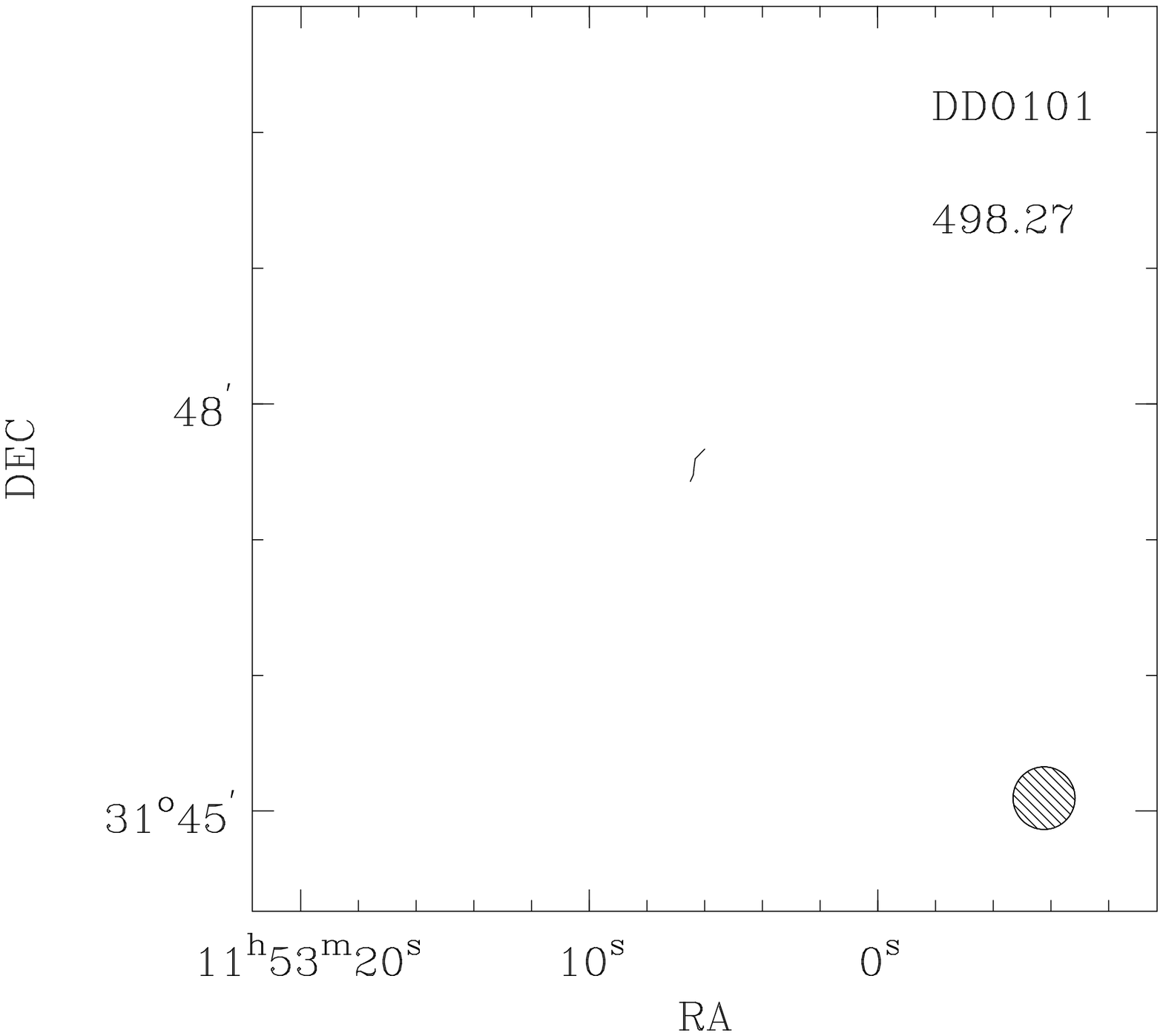}}
\end{minipage}
\hfill
\begin{minipage}[b]{5.7 cm}
\resizebox{5.85cm}{!}{\includegraphics[angle=-90]{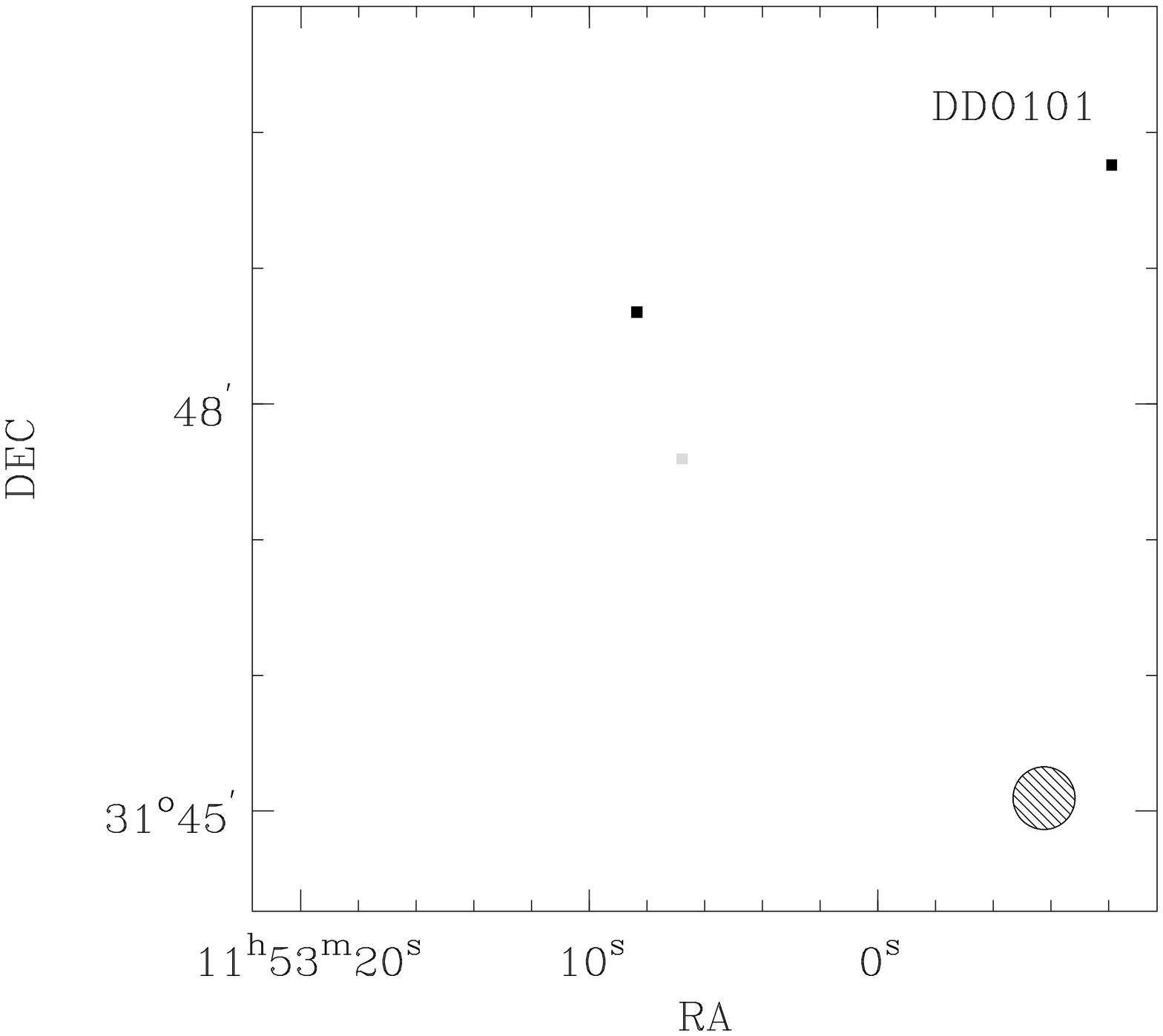}}
\end{minipage}
\begin{minipage}[b]{5.7 cm}
\resizebox{5.7cm}{!}{\includegraphics[angle=-90]{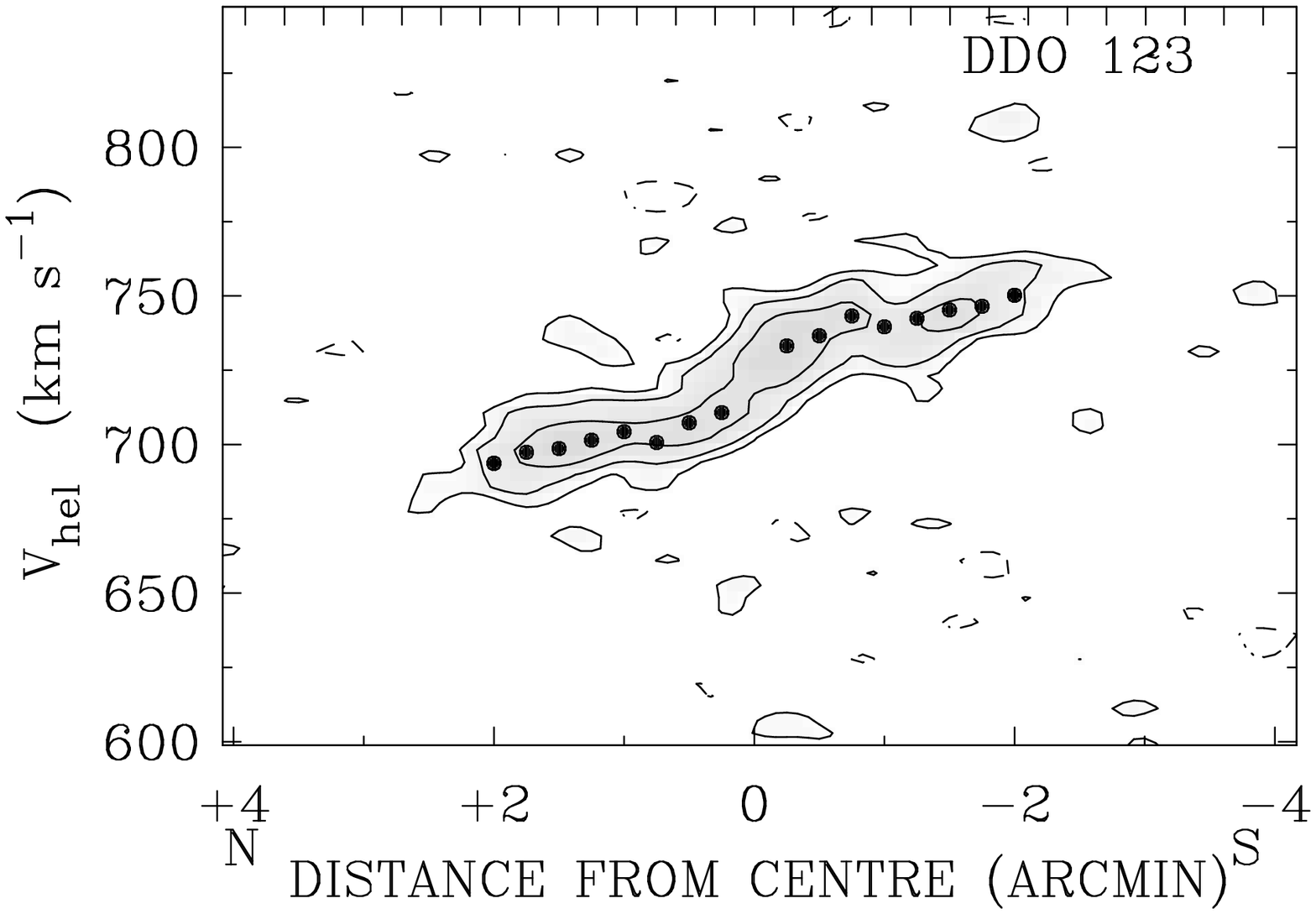}}
\end{minipage}
\hfill
\begin{minipage}[b]{5.7 cm}
\resizebox{5.7cm}{!}{\includegraphics[angle=-90]{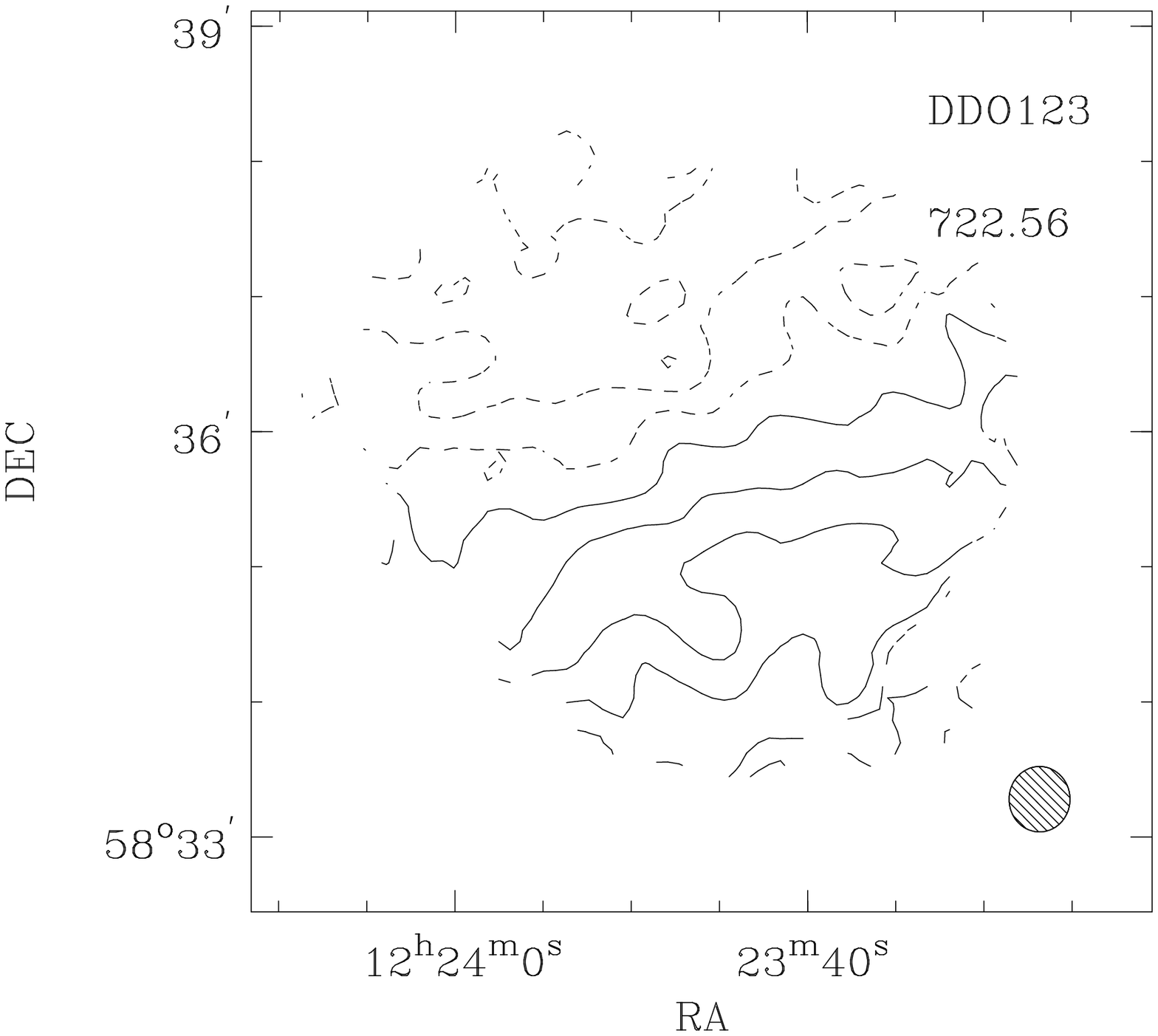}}
\end{minipage}
\hfill
\begin{minipage}[b]{5.7 cm}
\resizebox{5.85cm}{!}{\includegraphics[angle=-90]{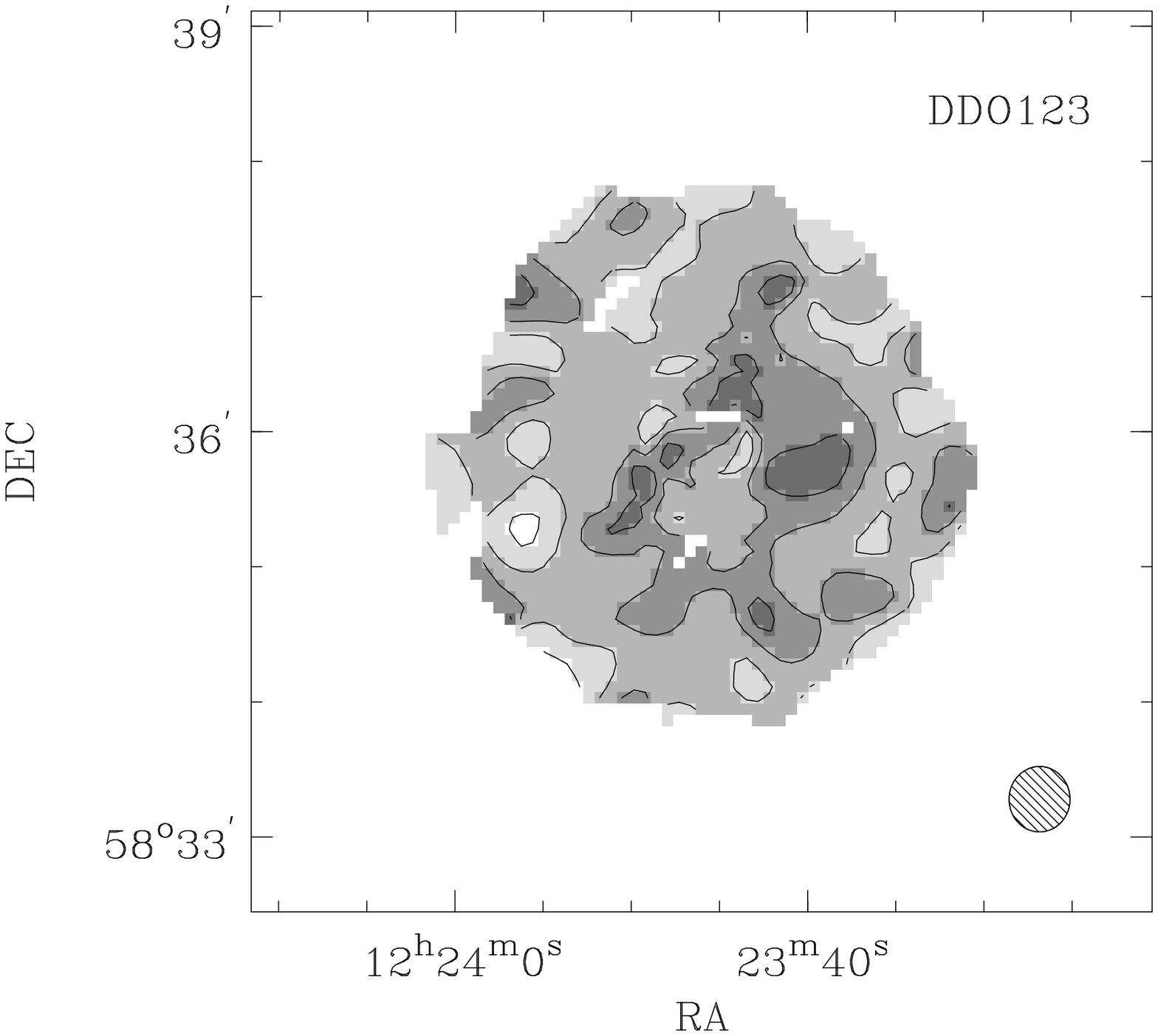}}
\end{minipage}
\caption{
\small Continued
}
\end{figure*}

\begin{figure*}
\addtocounter{figure}{-1}
\begin{minipage}[b]{5.7 cm}
\resizebox{5.7cm}{!}{\includegraphics[angle=-90]{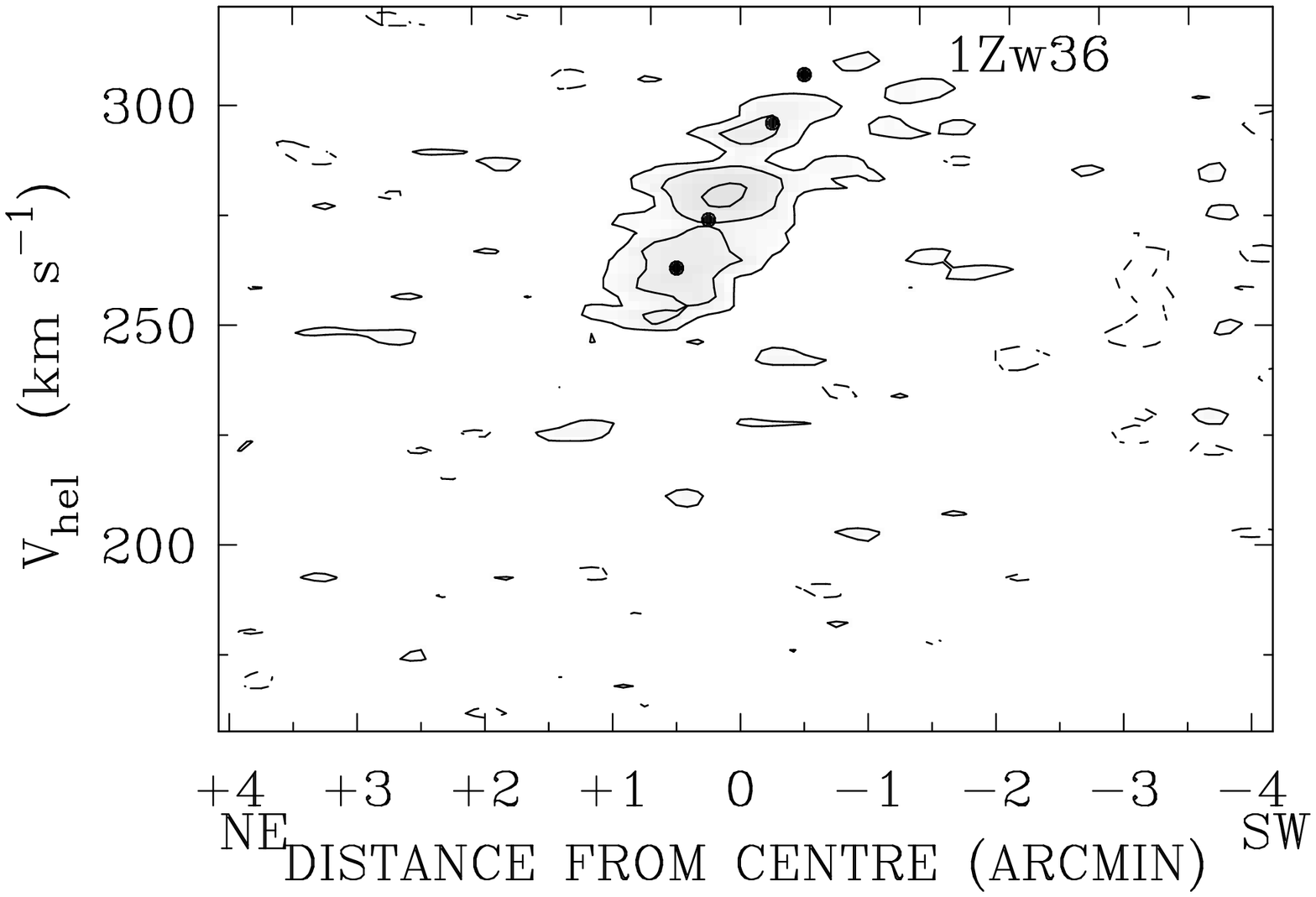}}
\end{minipage}
\hfill
\begin{minipage}[b]{5.7 cm}
\resizebox{5.7cm}{!}{\includegraphics[angle=-90]{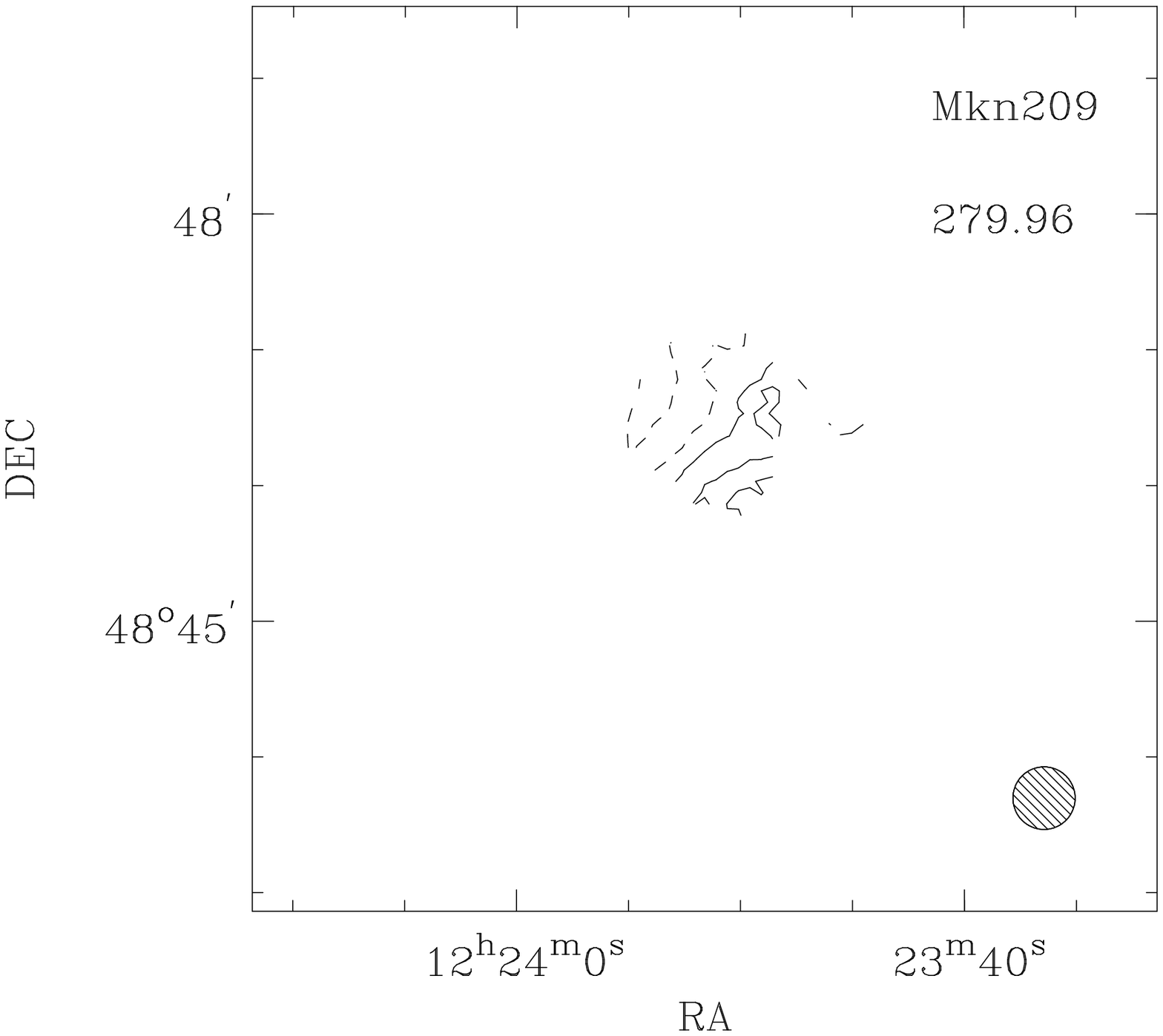}}
\end{minipage}
\hfill
\begin{minipage}[b]{5.7 cm}
\resizebox{5.85cm}{!}{\includegraphics[angle=-90]{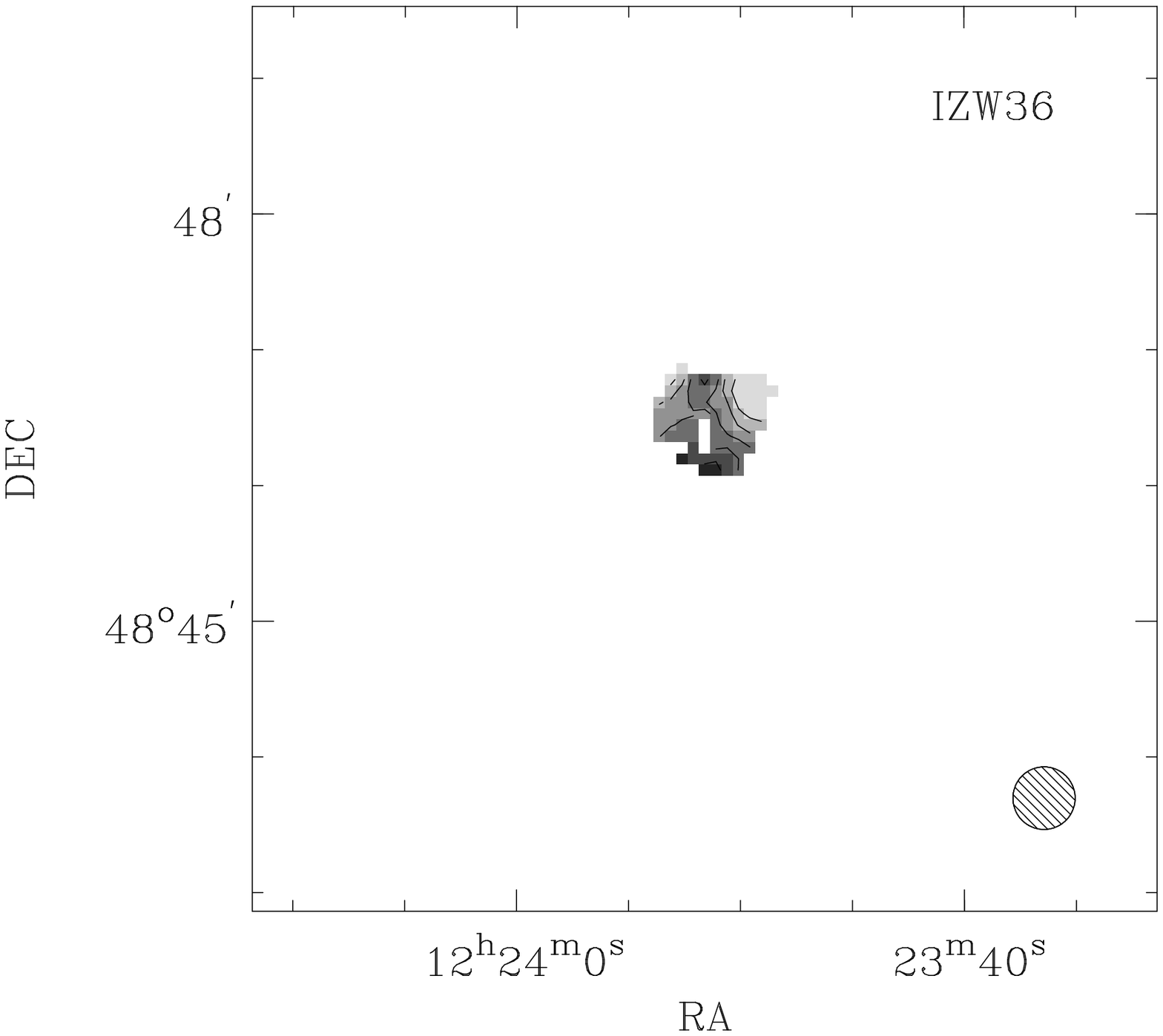}}
\end{minipage}
\begin{minipage}[b]{5.7 cm}
\resizebox{5.7cm}{!}{\includegraphics[angle=-90]{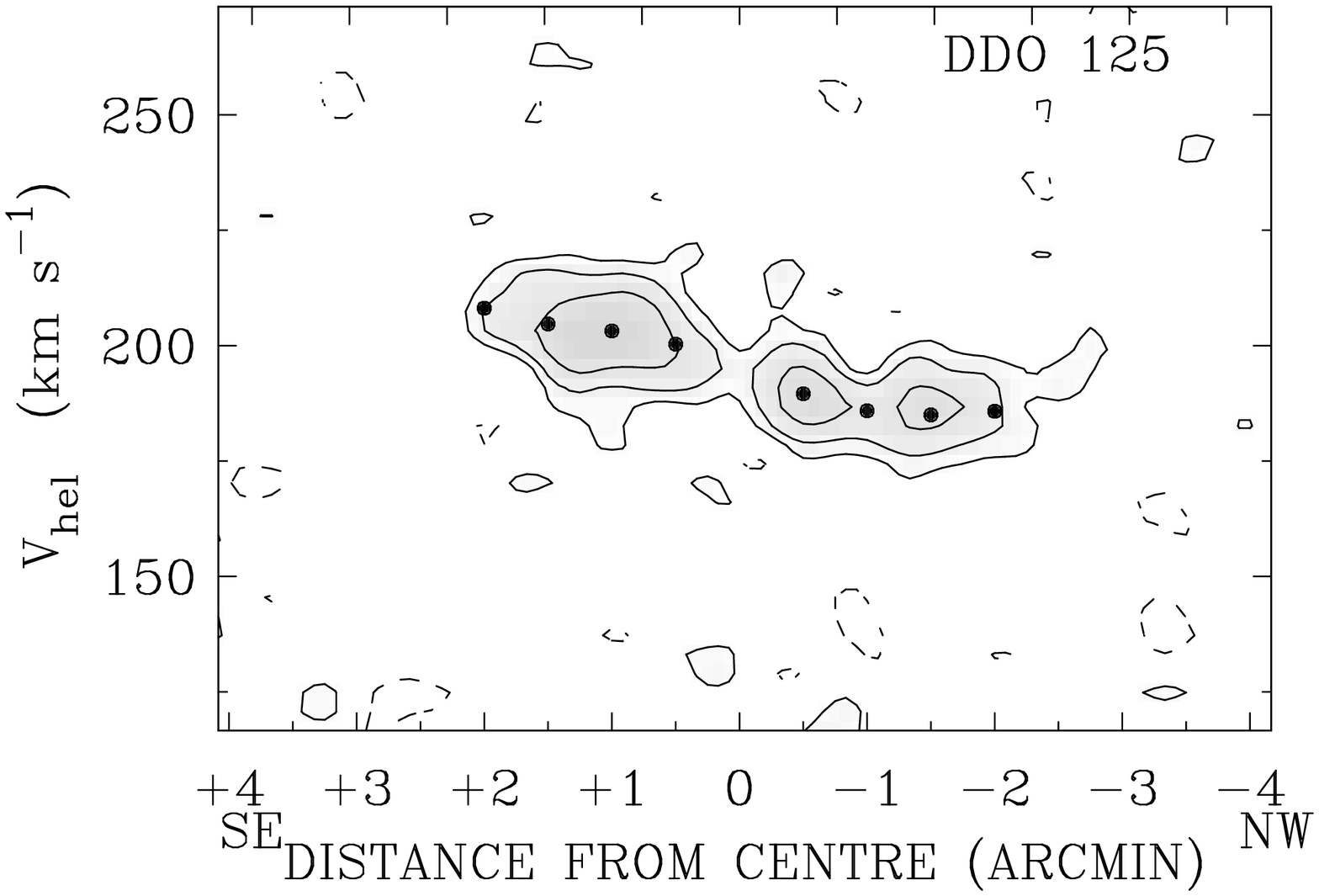}}
\end{minipage}
\hfill
\begin{minipage}[b]{5.7 cm}
\resizebox{5.7cm}{!}{\includegraphics[angle=-90]{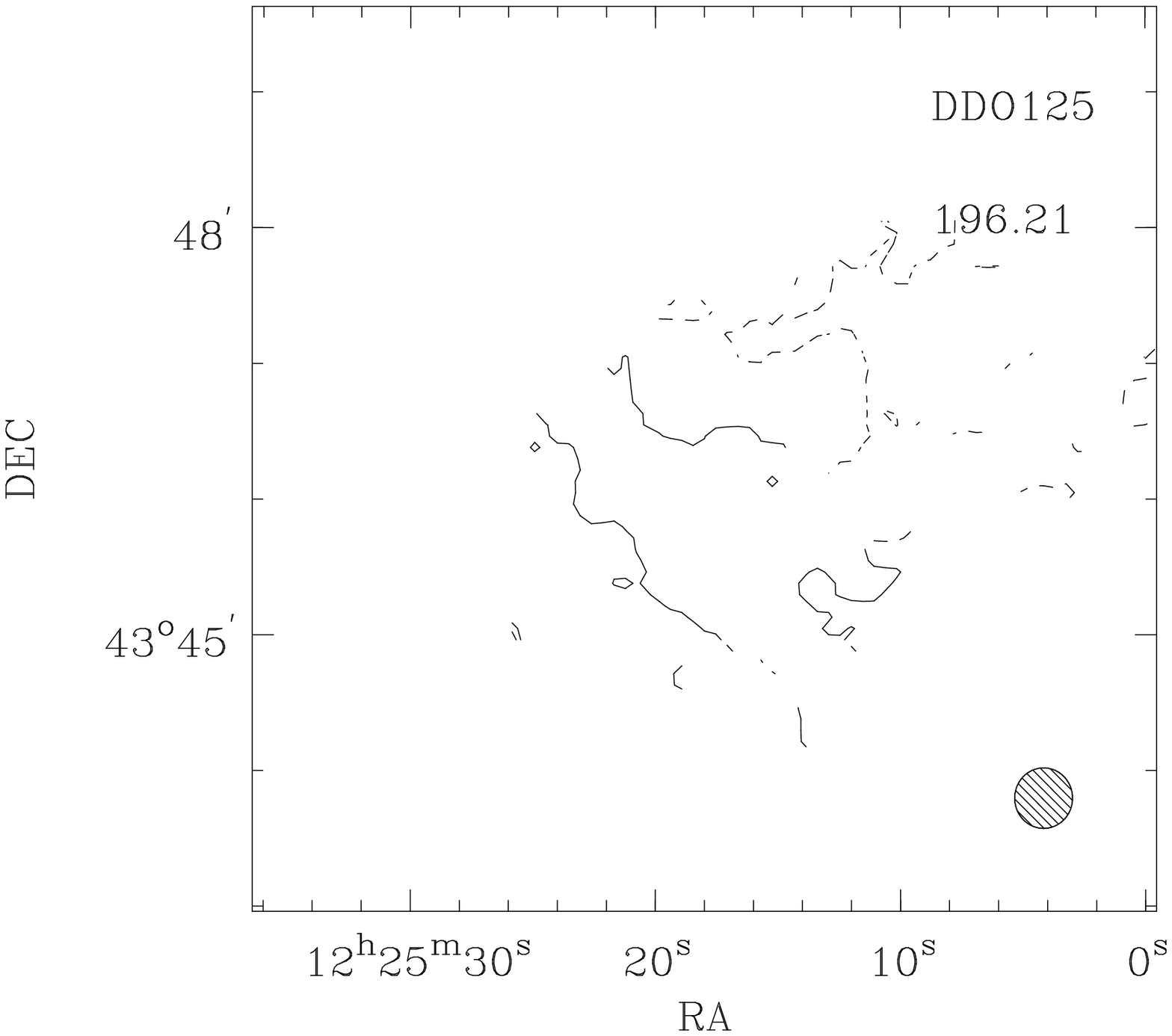}}
\end{minipage}
\hfill
\begin{minipage}[b]{5.7 cm}
\resizebox{5.85cm}{!}{\includegraphics[angle=-90]{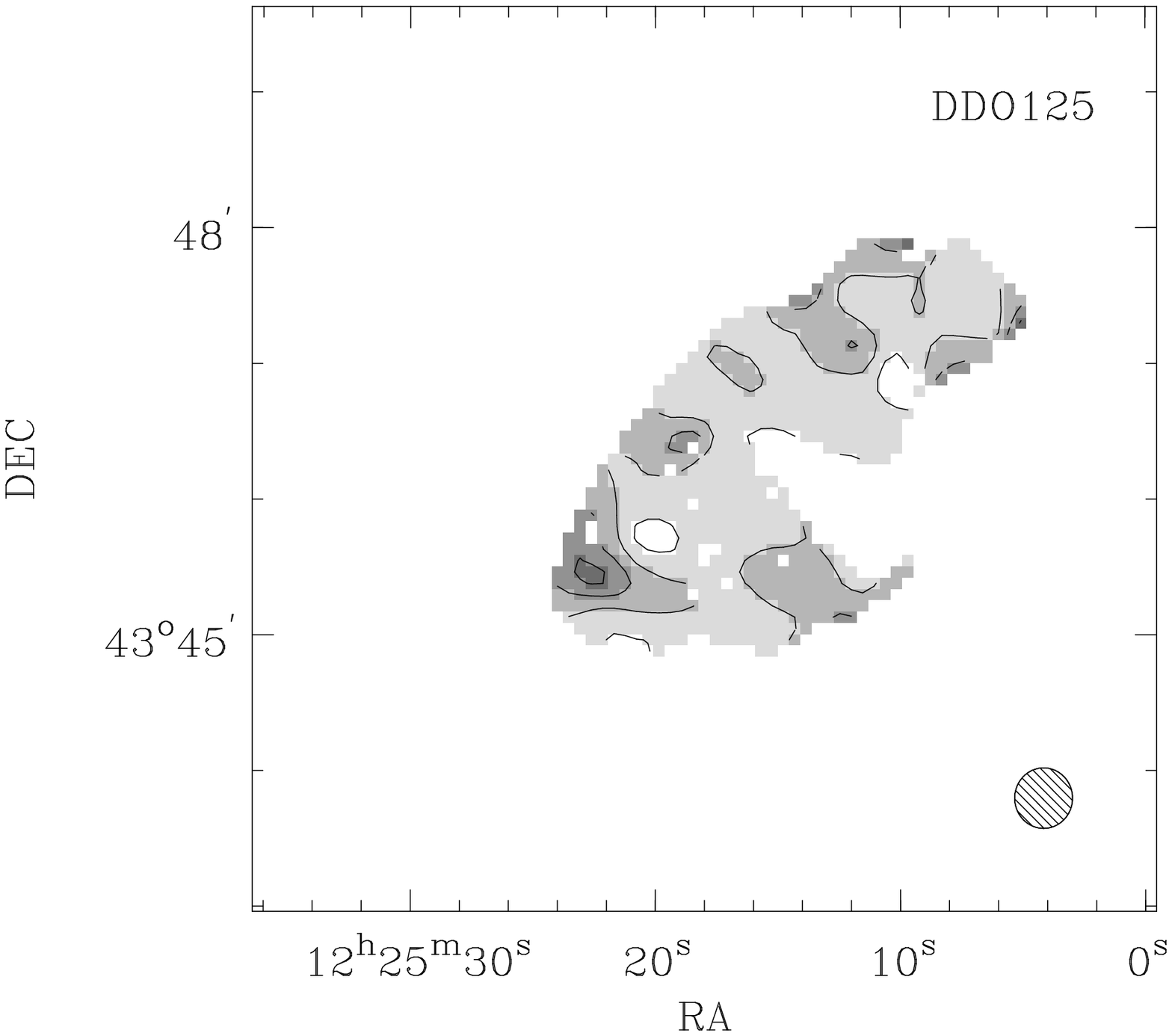}}
\end{minipage}
\begin{minipage}[b]{5.7 cm}
\resizebox{5.7cm}{!}{\includegraphics[angle=-90]{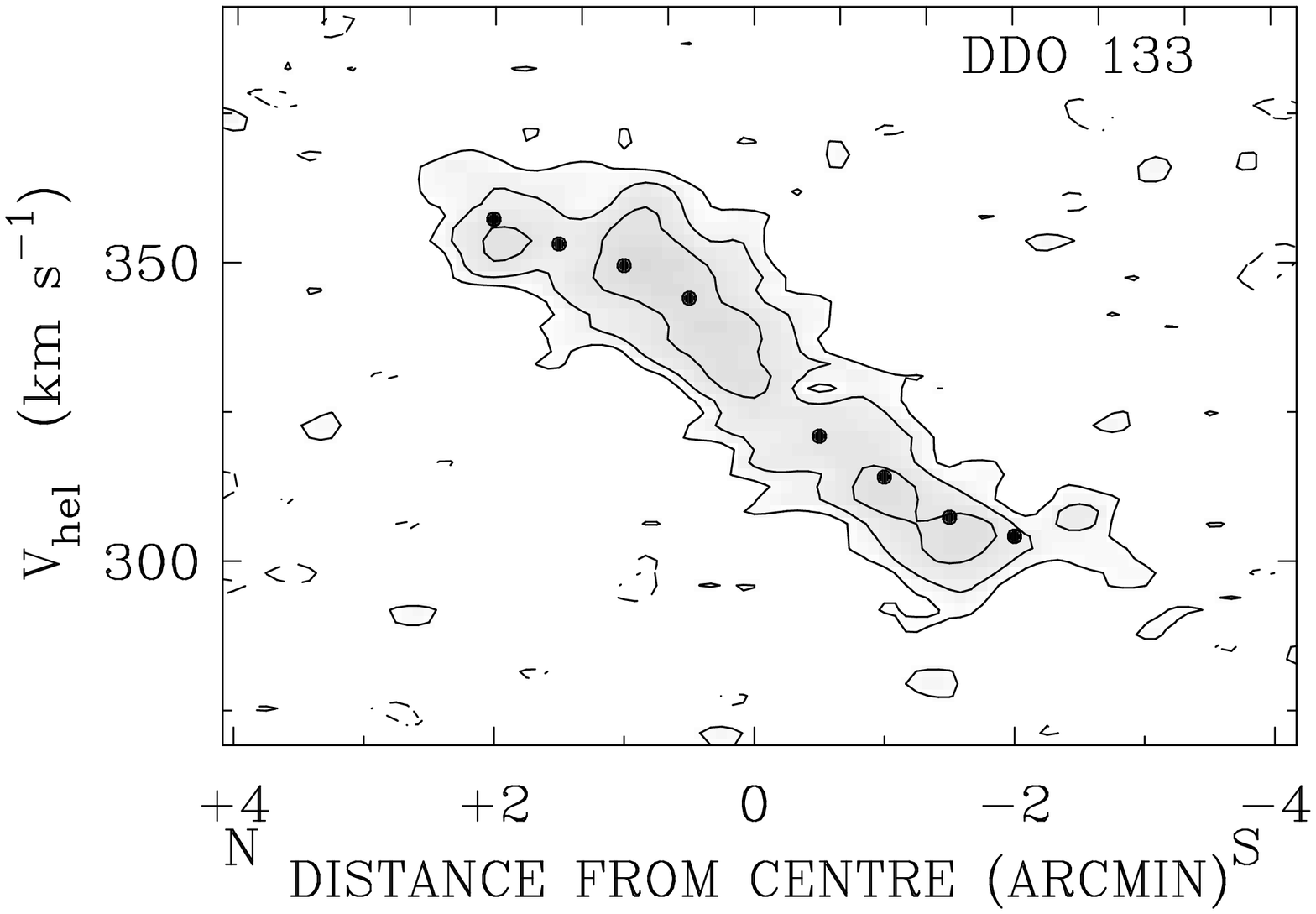}}
\end{minipage}
\hfill
\begin{minipage}[b]{5.7 cm}
\resizebox{5.7cm}{!}{\includegraphics[angle=-90]{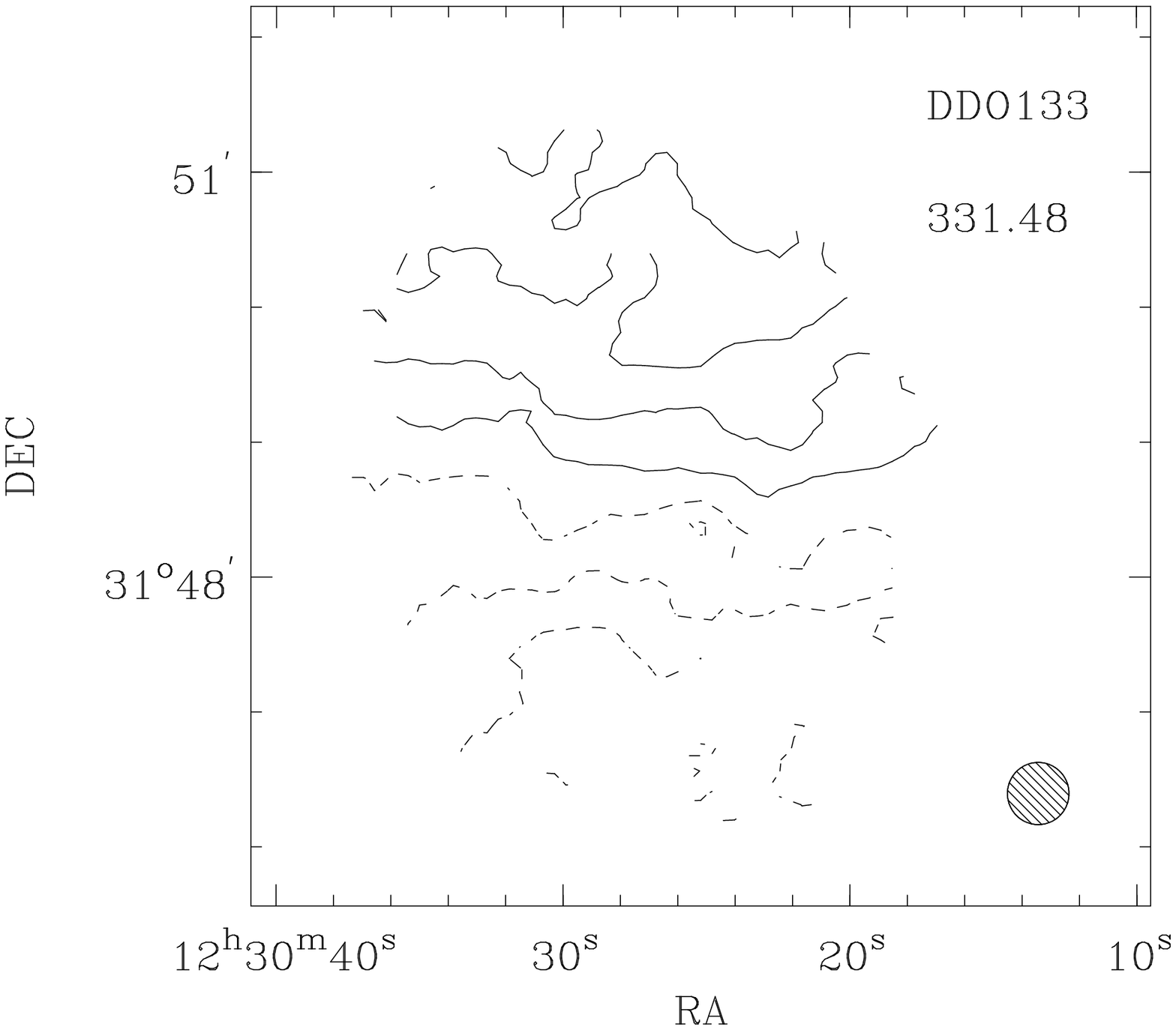}}
\end{minipage}
\hfill
\begin{minipage}[b]{5.7 cm}
\resizebox{5.85cm}{!}{\includegraphics[angle=-90]{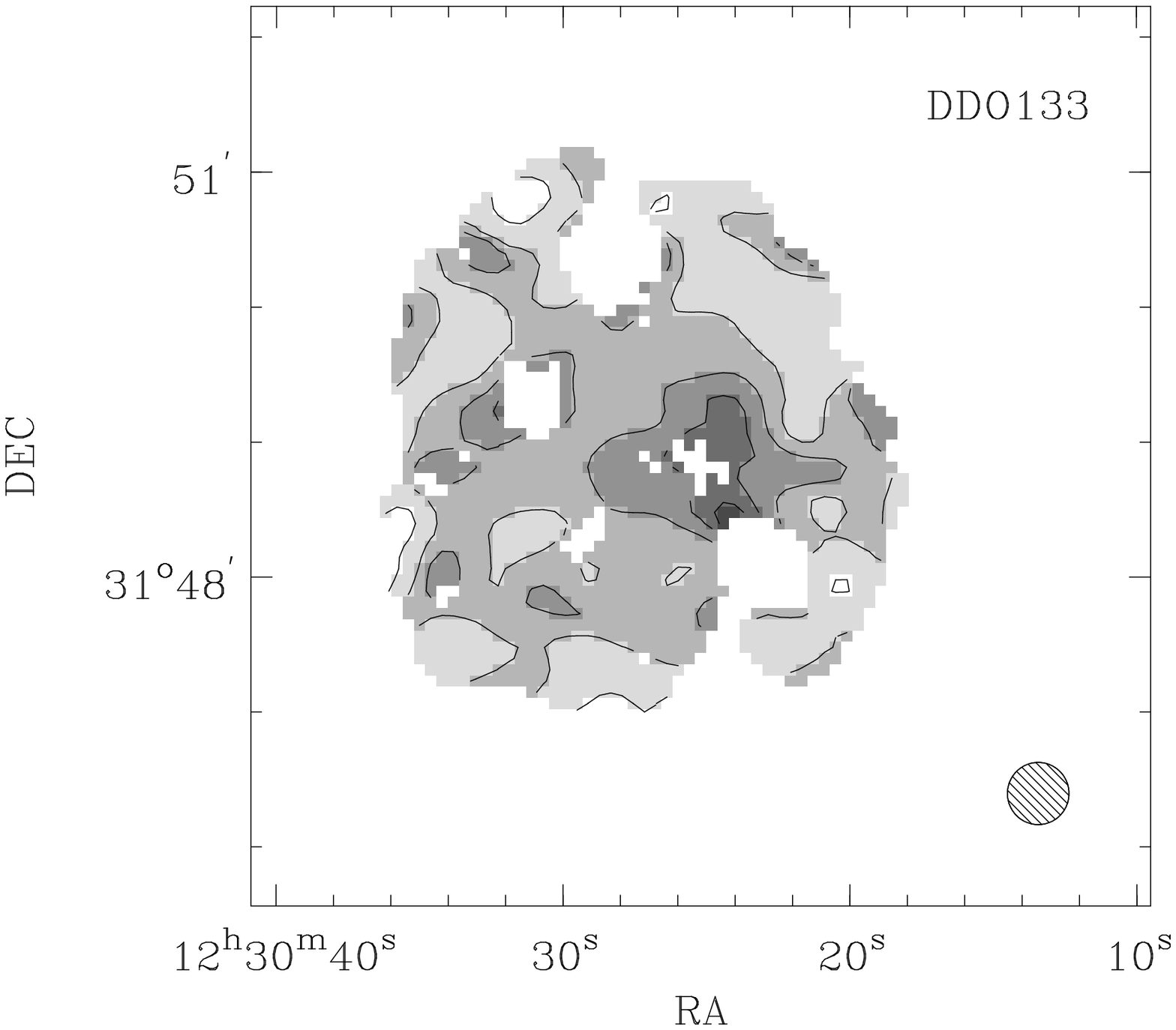}}
\end{minipage}
\begin{minipage}[b]{5.7 cm}
\resizebox{5.7cm}{!}{\includegraphics[angle=-90]{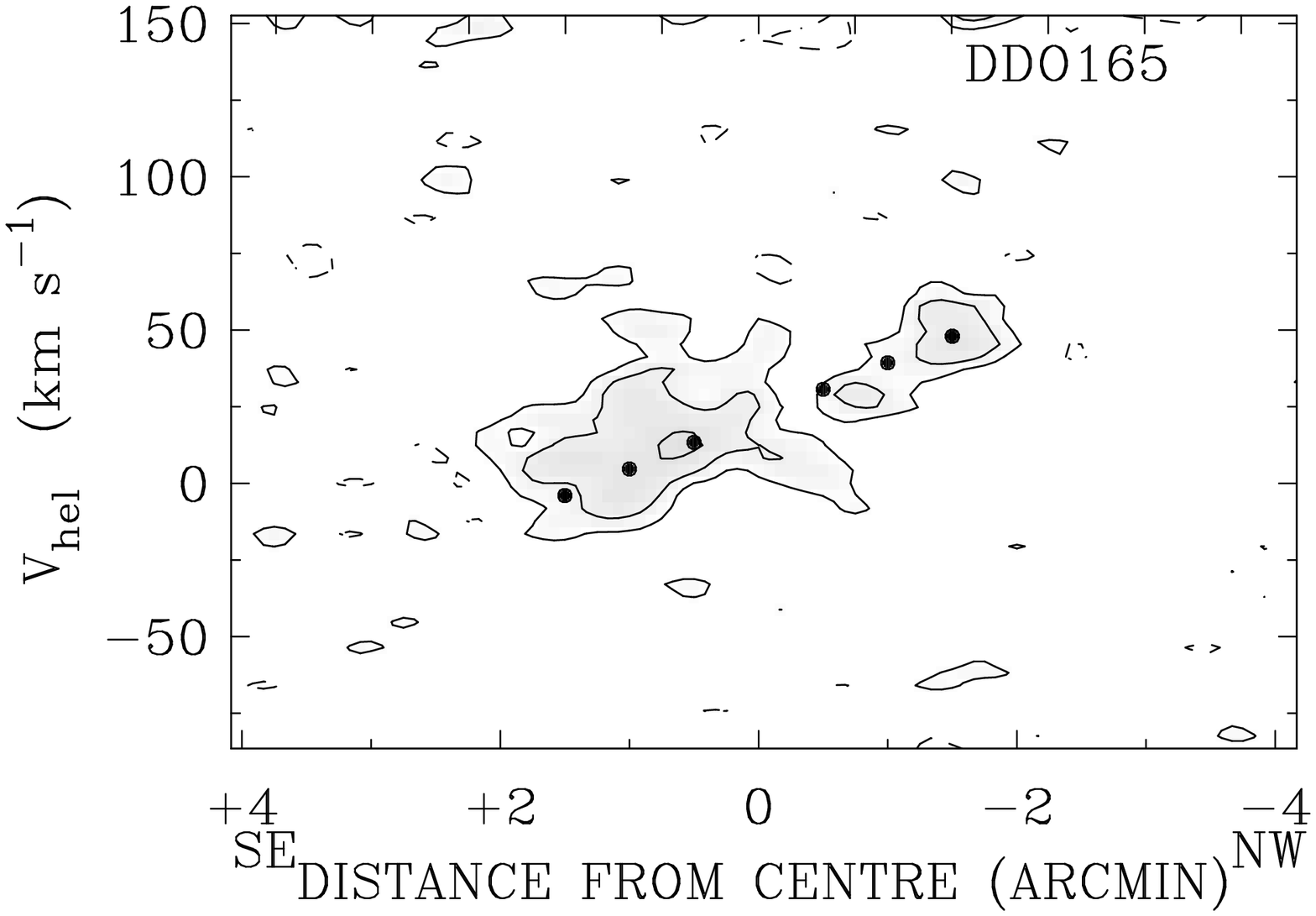}}
\end{minipage}
\hfill
\begin{minipage}[b]{5.7 cm}
\resizebox{5.7cm}{!}{\includegraphics[angle=-90]{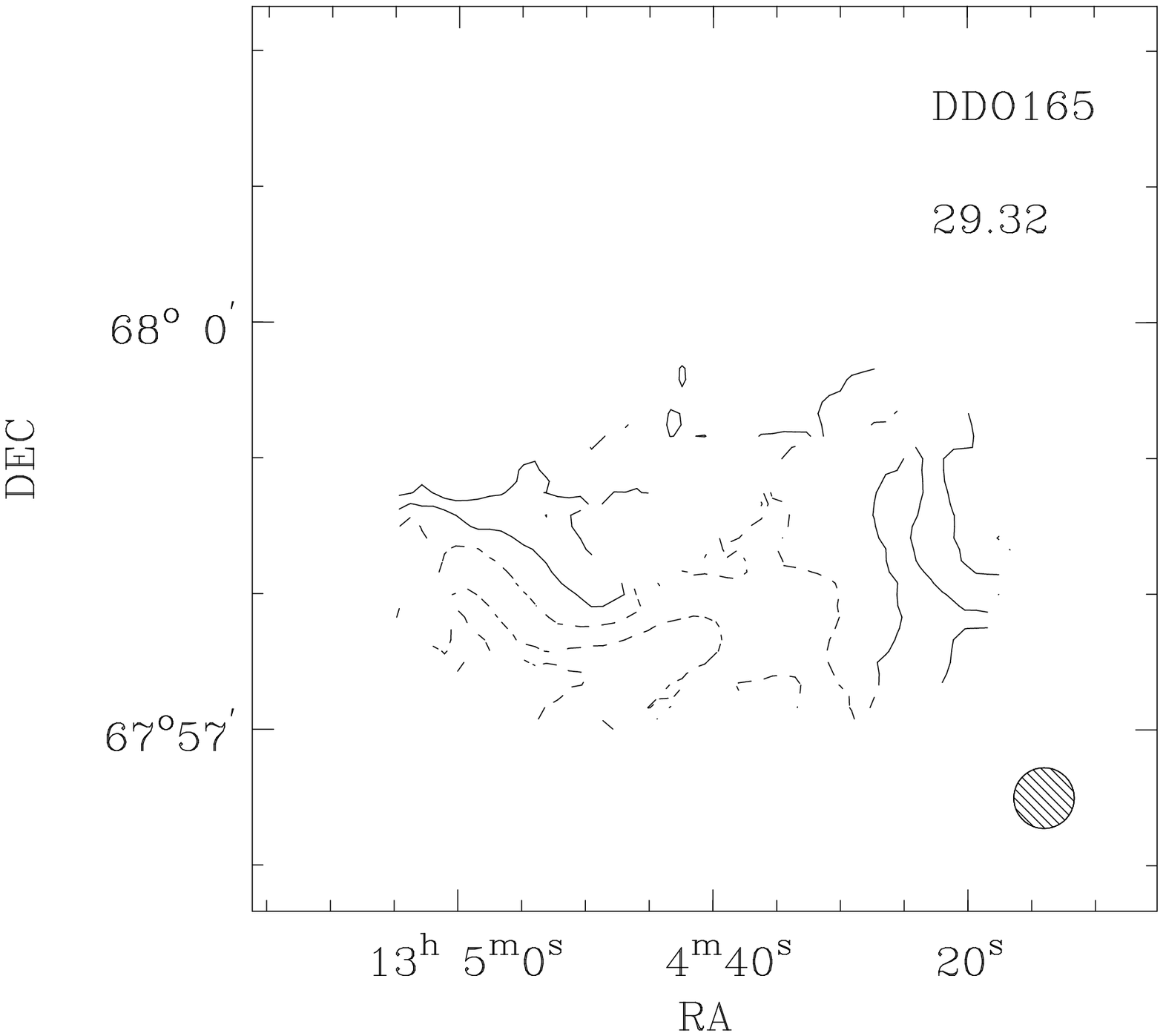}}
\end{minipage}
\hfill
\begin{minipage}[b]{5.7 cm}
\resizebox{5.85cm}{!}{\includegraphics[angle=-90]{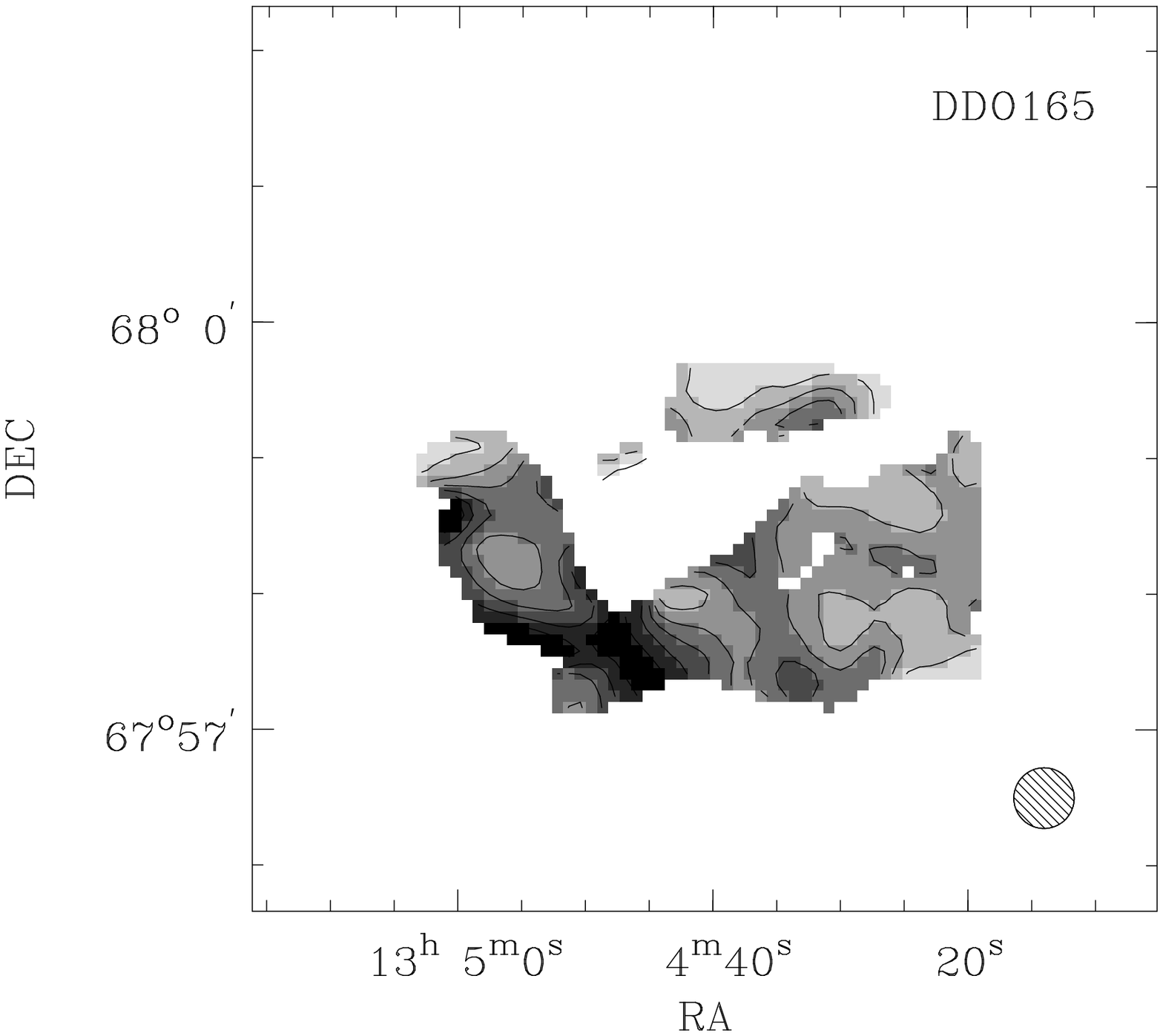}}
\end{minipage}
\caption{
\small Continued
}
\end{figure*}

\begin{figure*}
\addtocounter{figure}{-1}
\begin{minipage}[b]{5.7 cm}
\resizebox{5.7cm}{!}{\includegraphics[angle=-90]{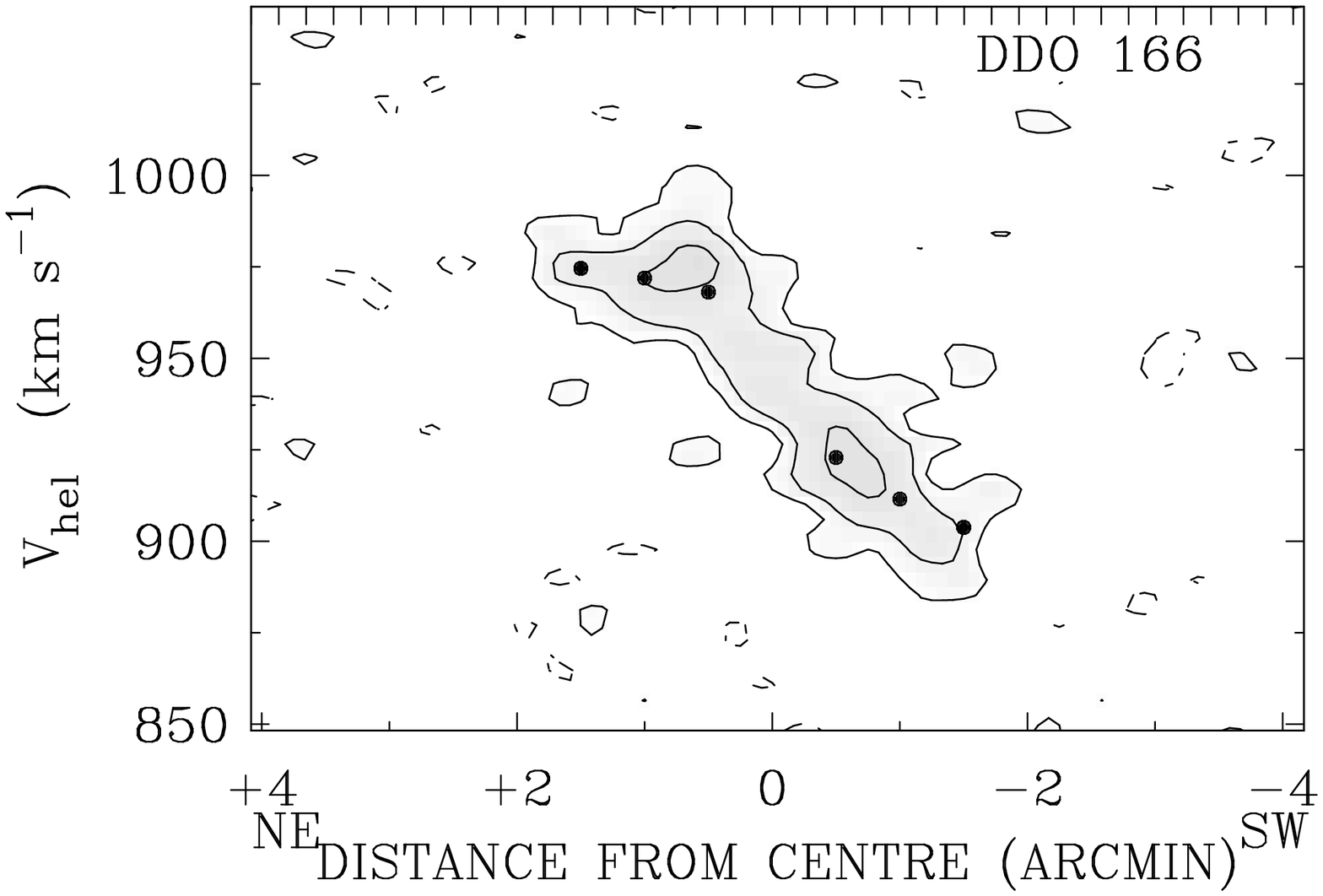}}
\end{minipage}
\hfill
\begin{minipage}[b]{5.7 cm}
\resizebox{5.7cm}{!}{\includegraphics[angle=-90]{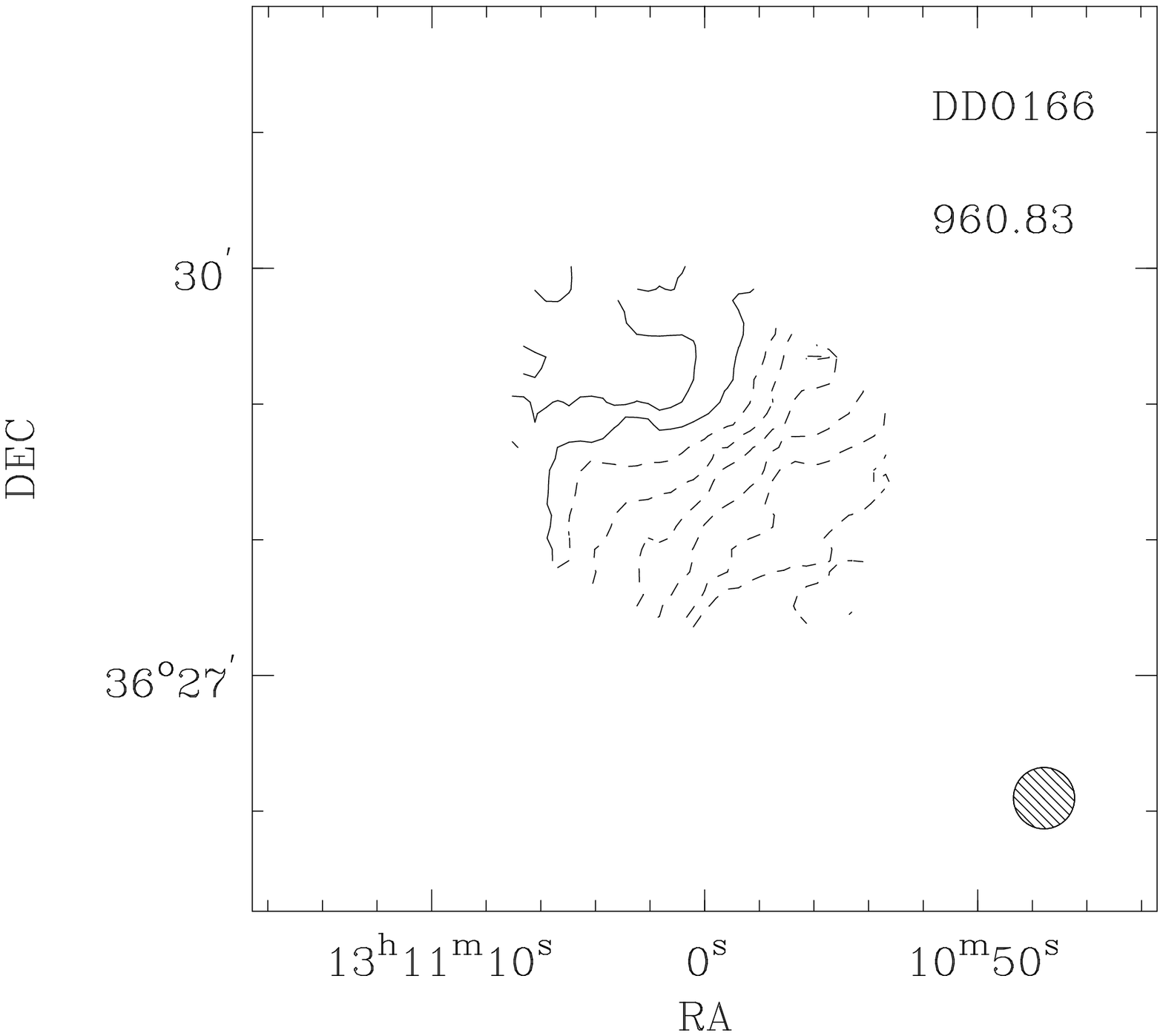}}
\end{minipage}
\hfill
\begin{minipage}[b]{5.7 cm}
\resizebox{5.85cm}{!}{\includegraphics[angle=-90]{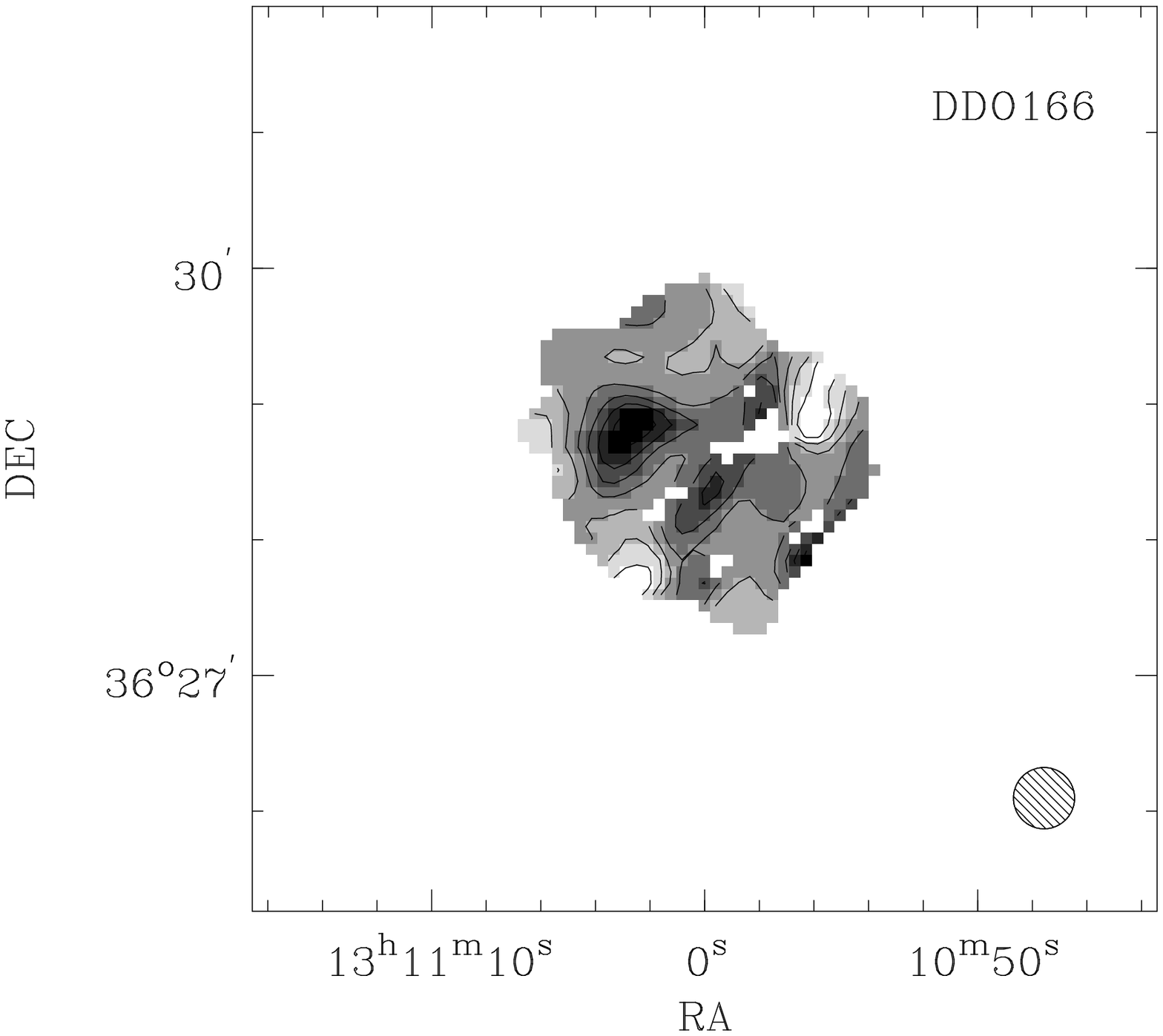}}
\end{minipage}
\begin{minipage}[b]{5.7 cm}
\resizebox{5.7cm}{!}{\includegraphics[angle=-90]{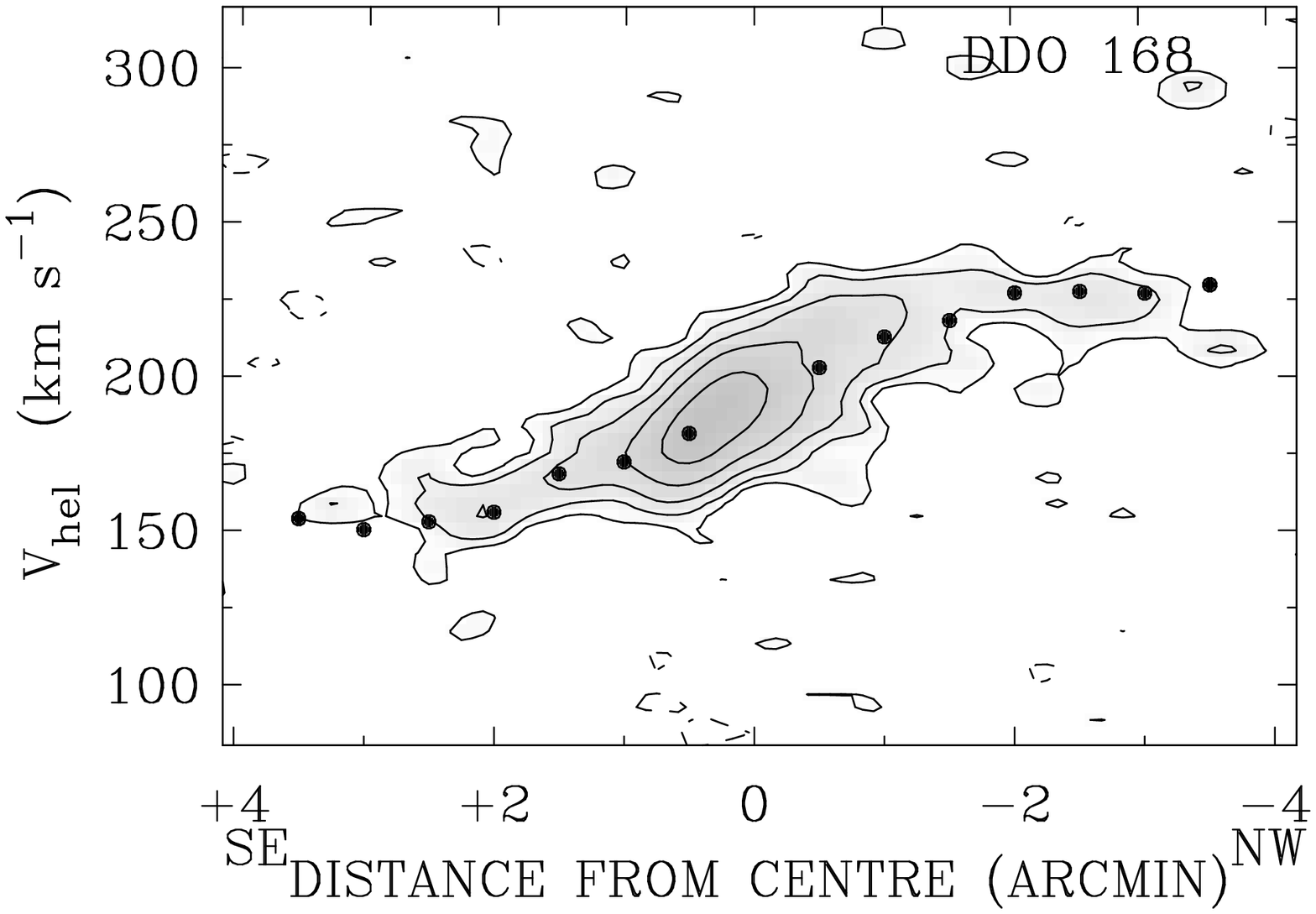}}
\end{minipage}
\hfill
\begin{minipage}[b]{5.7 cm}
\resizebox{5.7cm}{!}{\includegraphics[angle=-90]{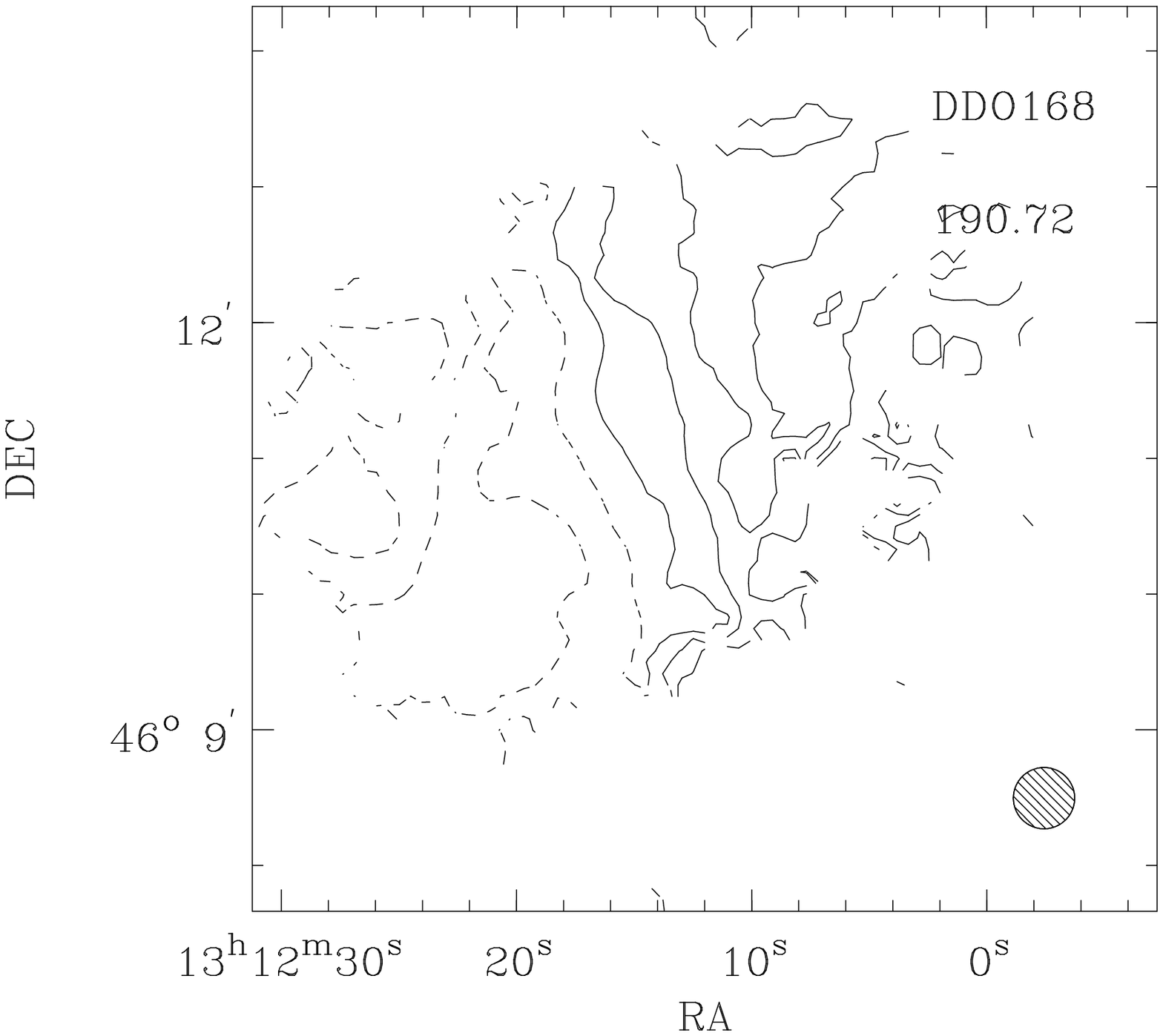}}
\end{minipage}
\hfill
\begin{minipage}[b]{5.7 cm}
\resizebox{5.85cm}{!}{\includegraphics[angle=-90]{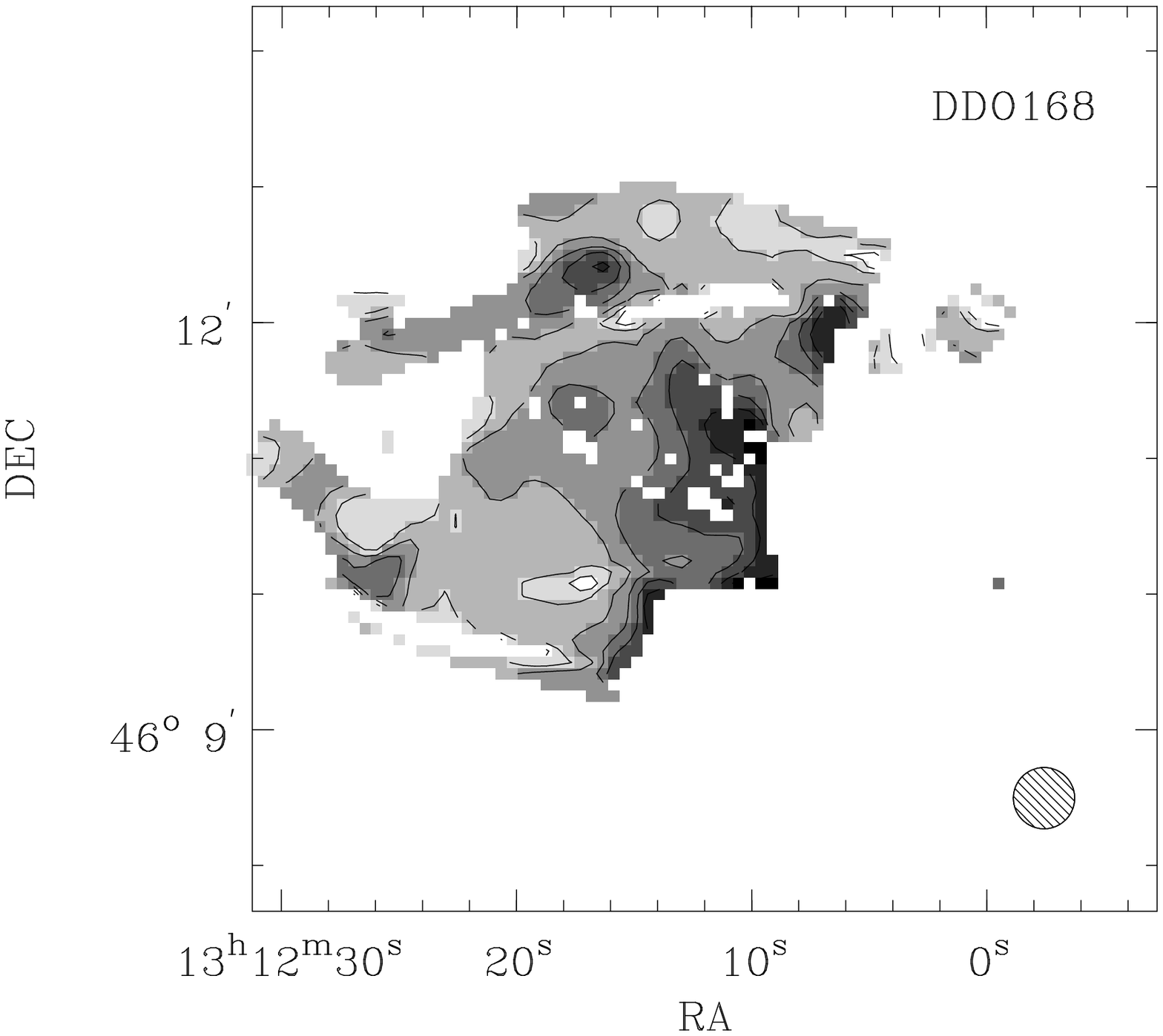}}
\end{minipage}
\begin{minipage}[b]{5.7 cm}
\resizebox{5.7cm}{!}{\includegraphics[angle=-90]{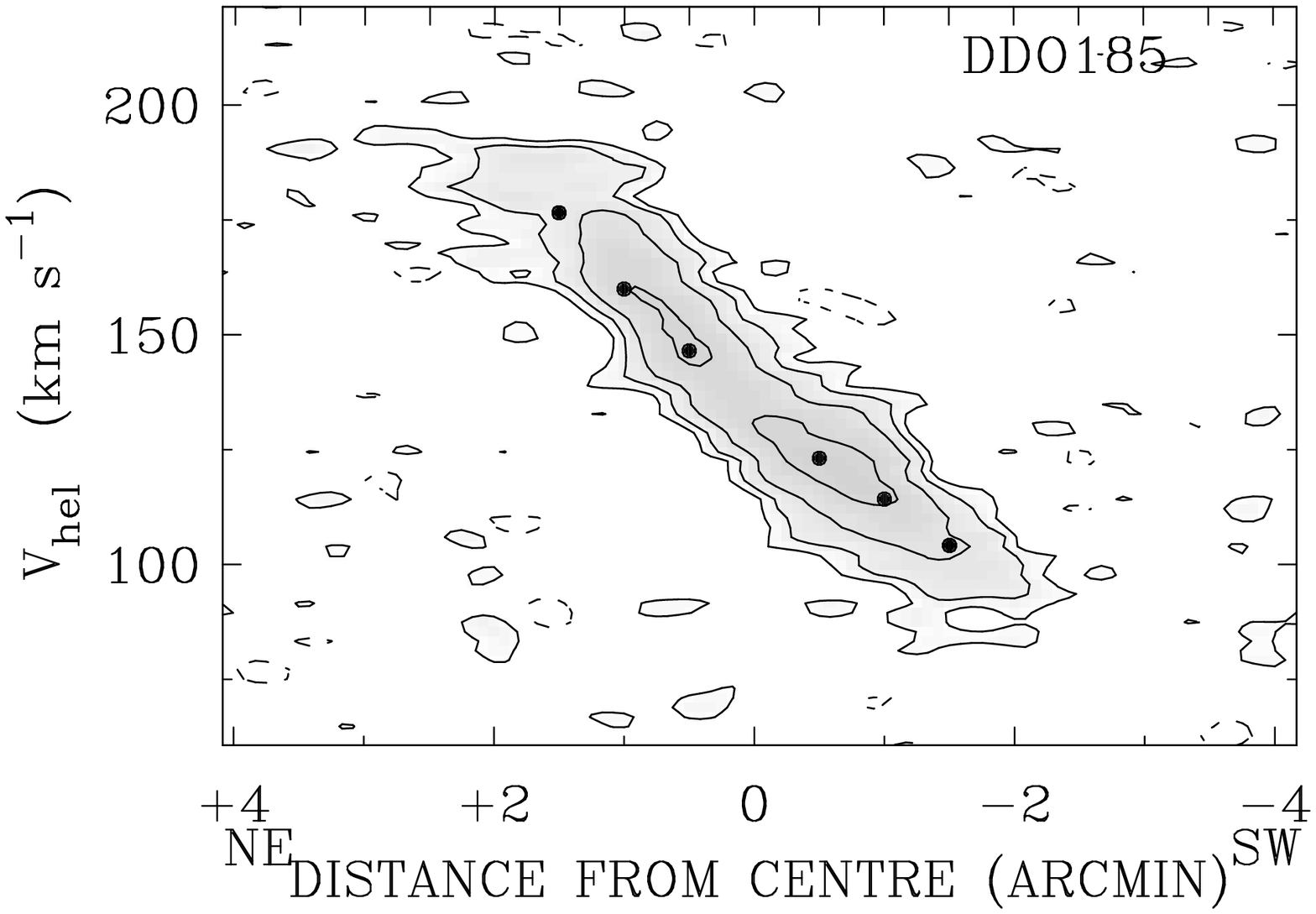}}
\end{minipage}
\hfill
\begin{minipage}[b]{5.7 cm}
\resizebox{5.7cm}{!}{\includegraphics[angle=-90]{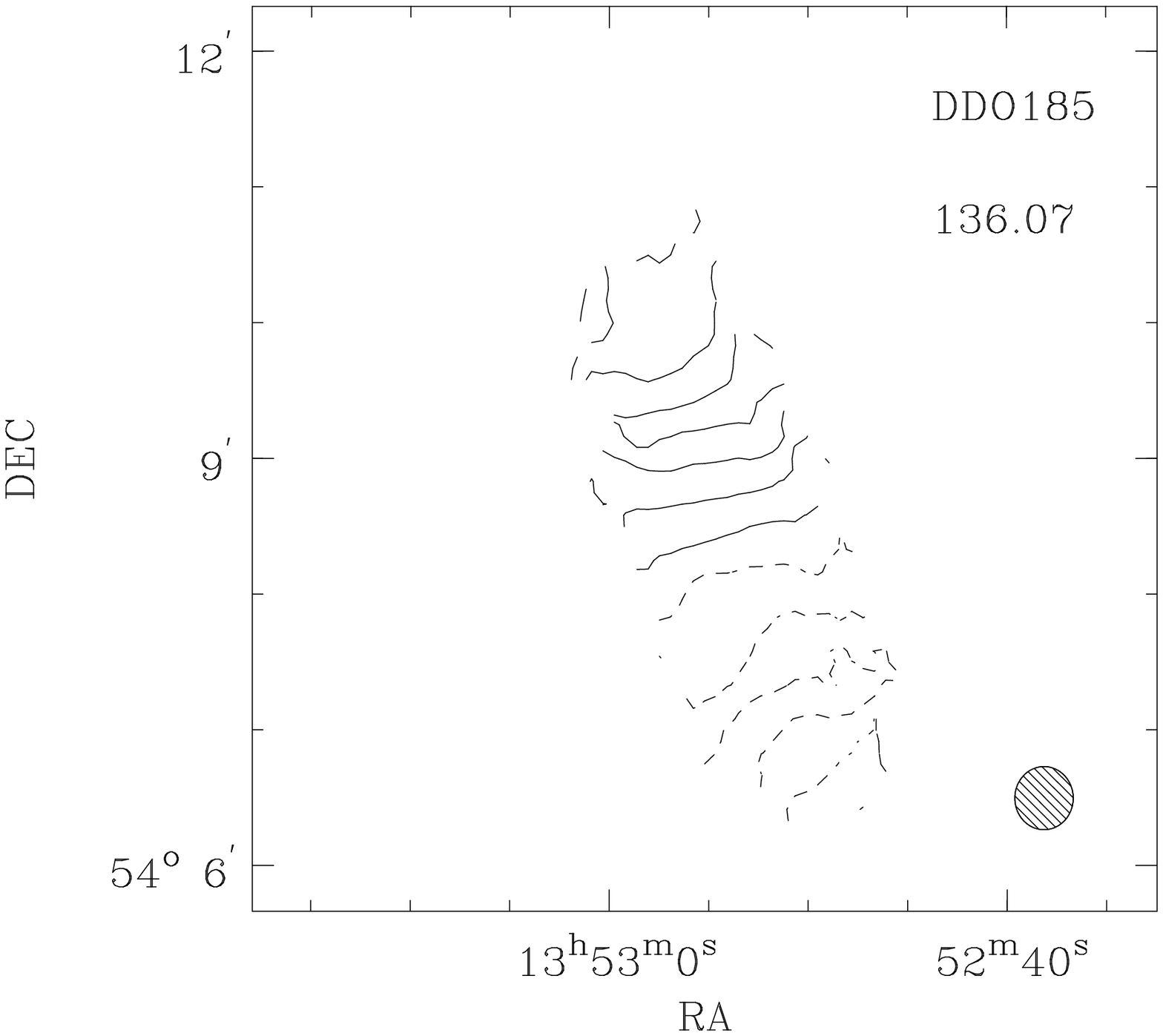}}
\end{minipage}
\hfill
\begin{minipage}[b]{5.7 cm}
\resizebox{5.85cm}{!}{\includegraphics[angle=-90]{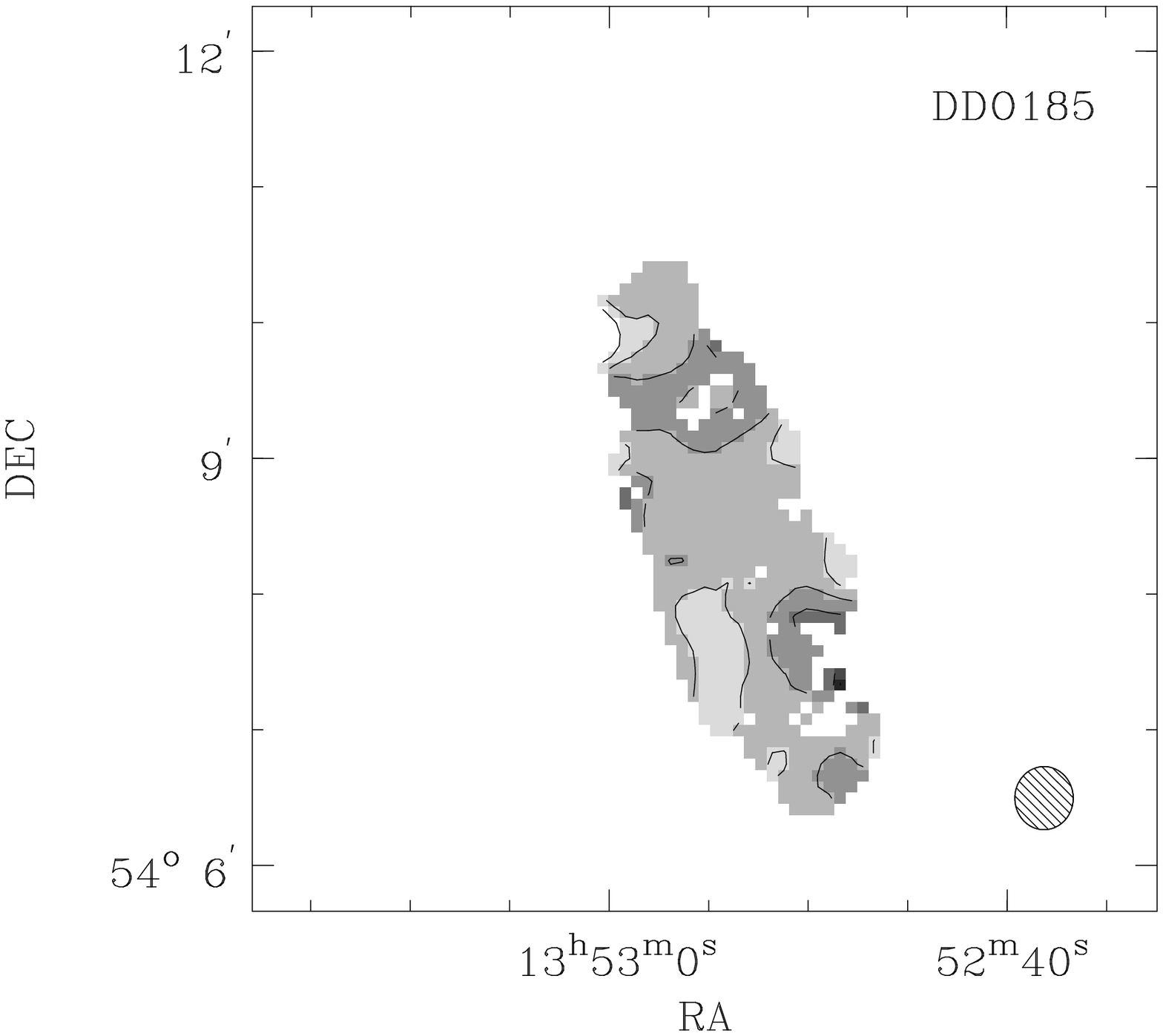}}
\end{minipage}
\begin{minipage}[b]{5.7 cm}
\resizebox{5.7cm}{!}{\includegraphics[angle=-90]{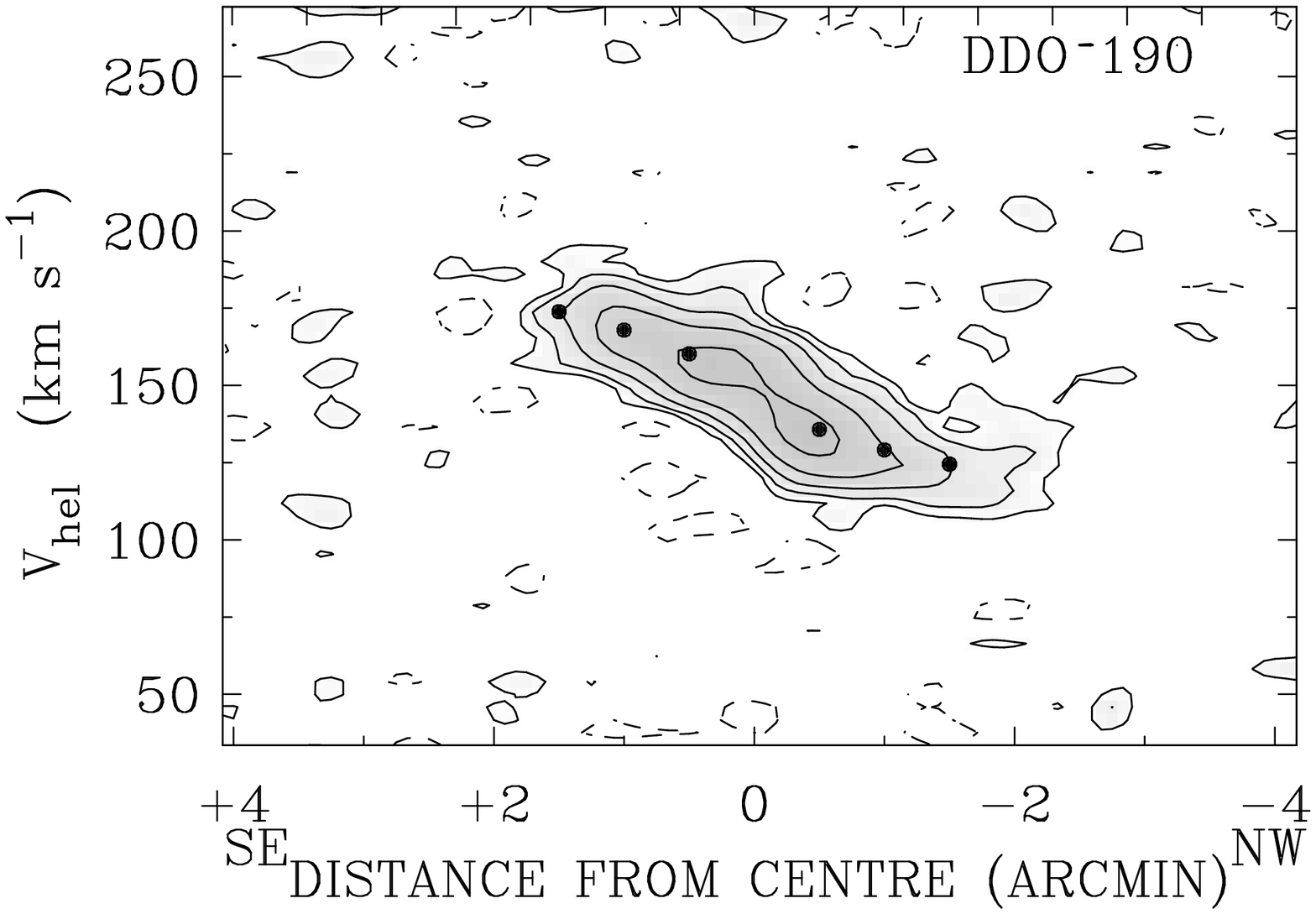}}
\end{minipage}
\hfill
\begin{minipage}[b]{5.7 cm}
\resizebox{5.7cm}{!}{\includegraphics[angle=-90]{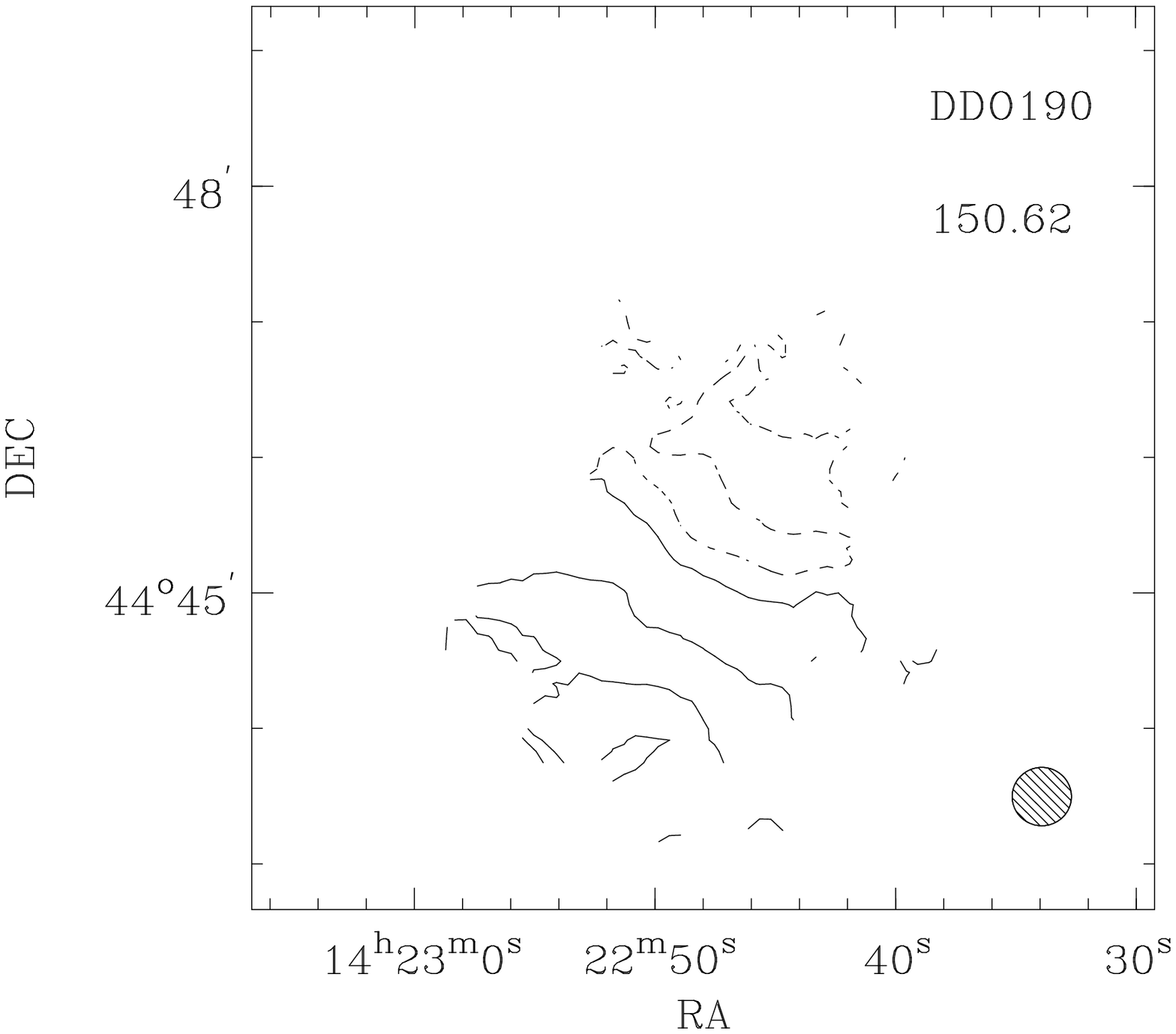}}
\end{minipage}
\hfill
\begin{minipage}[b]{5.7 cm}
\resizebox{5.85cm}{!}{\includegraphics[angle=-90]{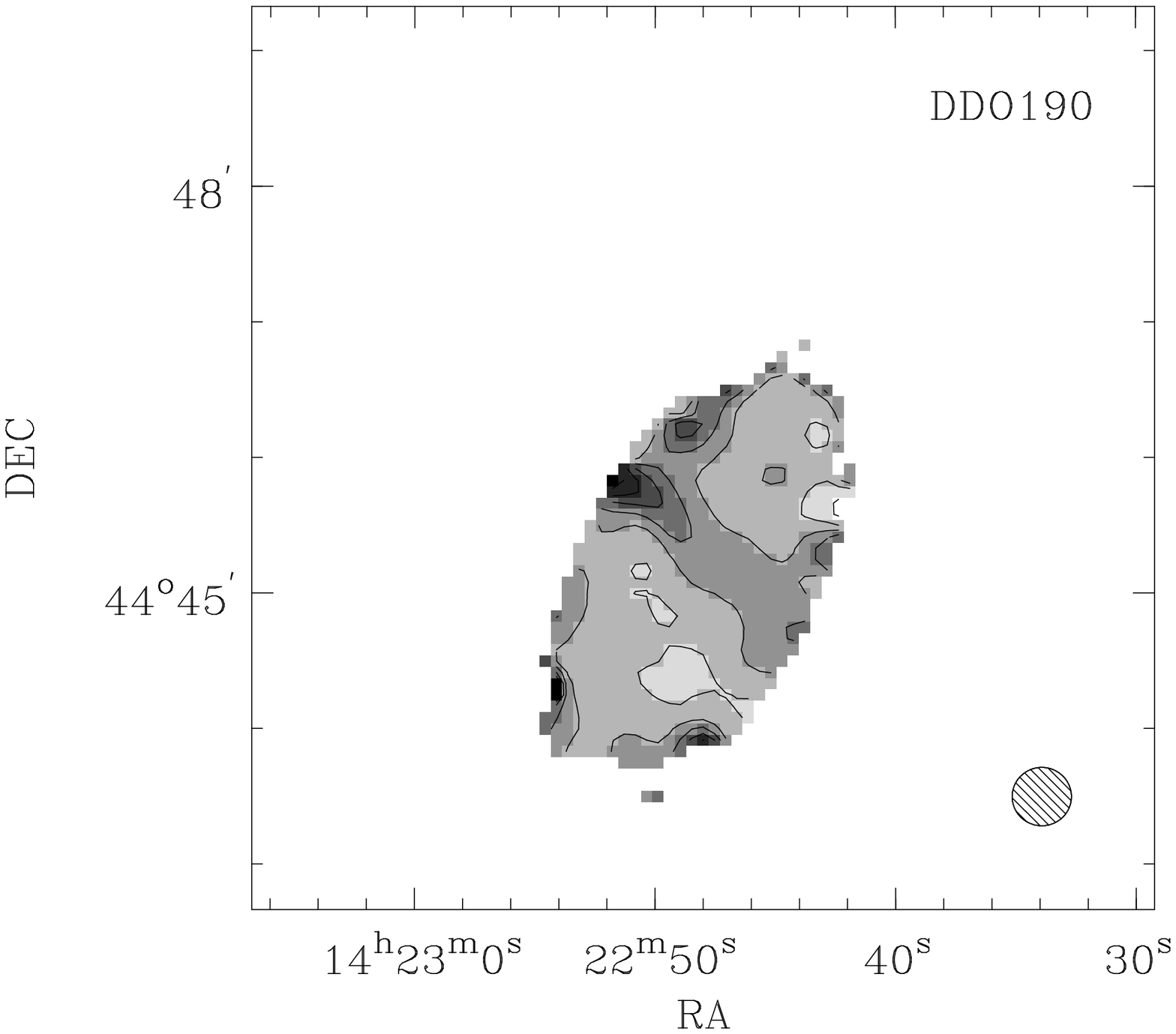}}
\end{minipage}
\caption{
\small Continued
}
\end{figure*}

\begin{figure*}
\addtocounter{figure}{-1}
\begin{minipage}[b]{5.7 cm}
\resizebox{5.7cm}{!}{\includegraphics[angle=-90]{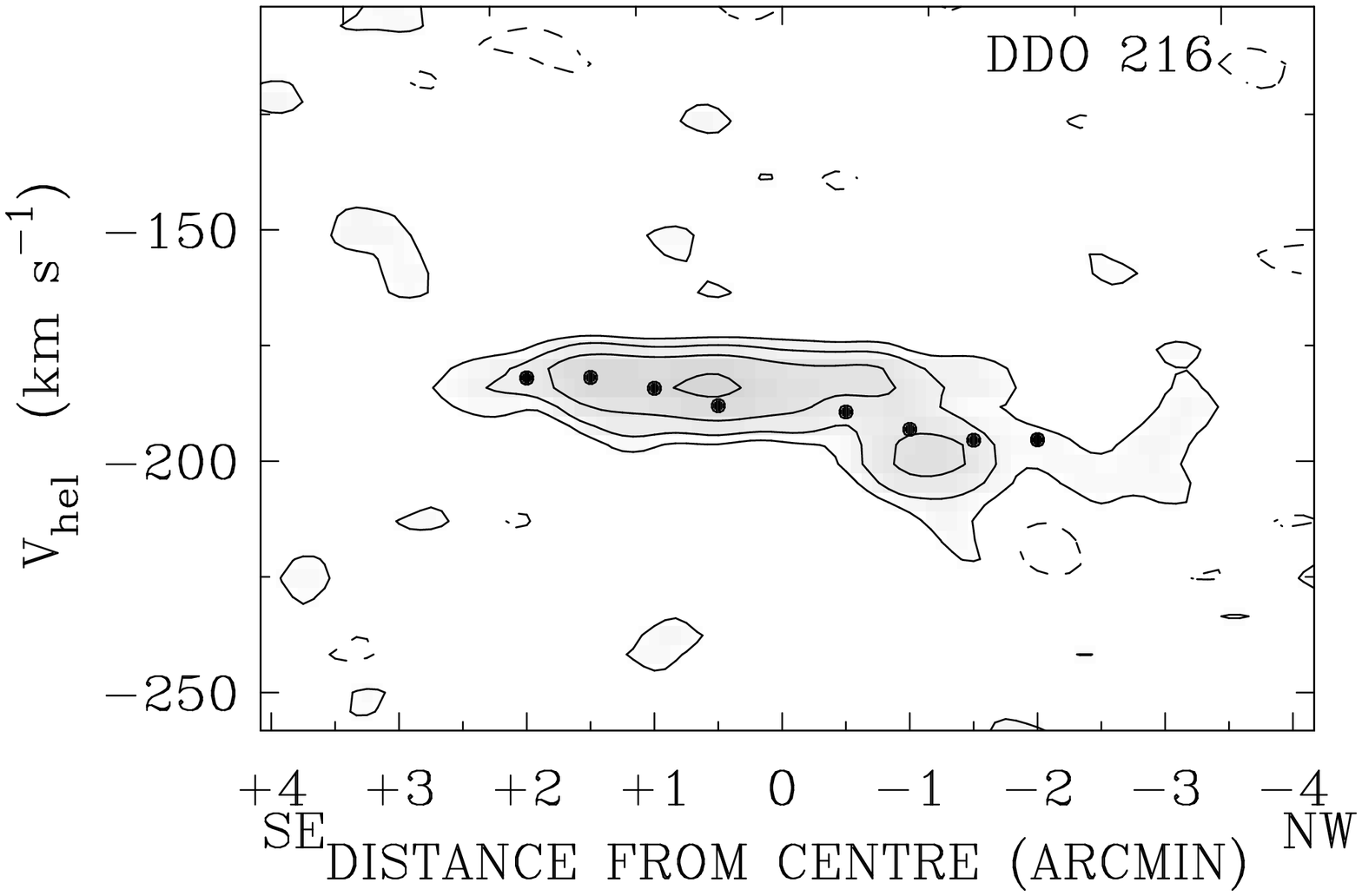}}
\end{minipage}
\hfill
\begin{minipage}[b]{5.7 cm}
\resizebox{5.7cm}{!}{\includegraphics[angle=-90]{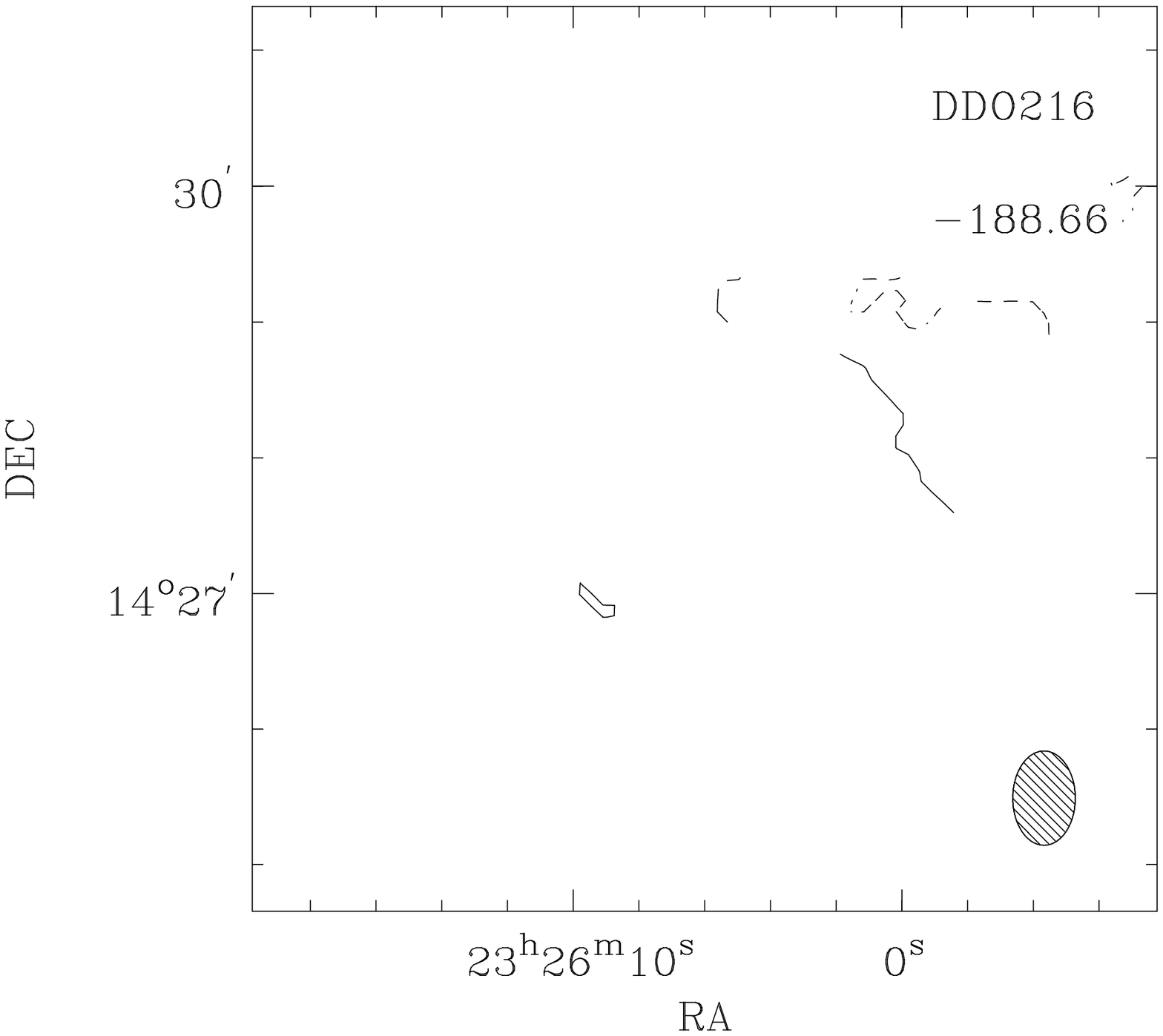}}
\end{minipage}
\hfill
\begin{minipage}[b]{5.7 cm}
\resizebox{5.85cm}{!}{\includegraphics[angle=-90]{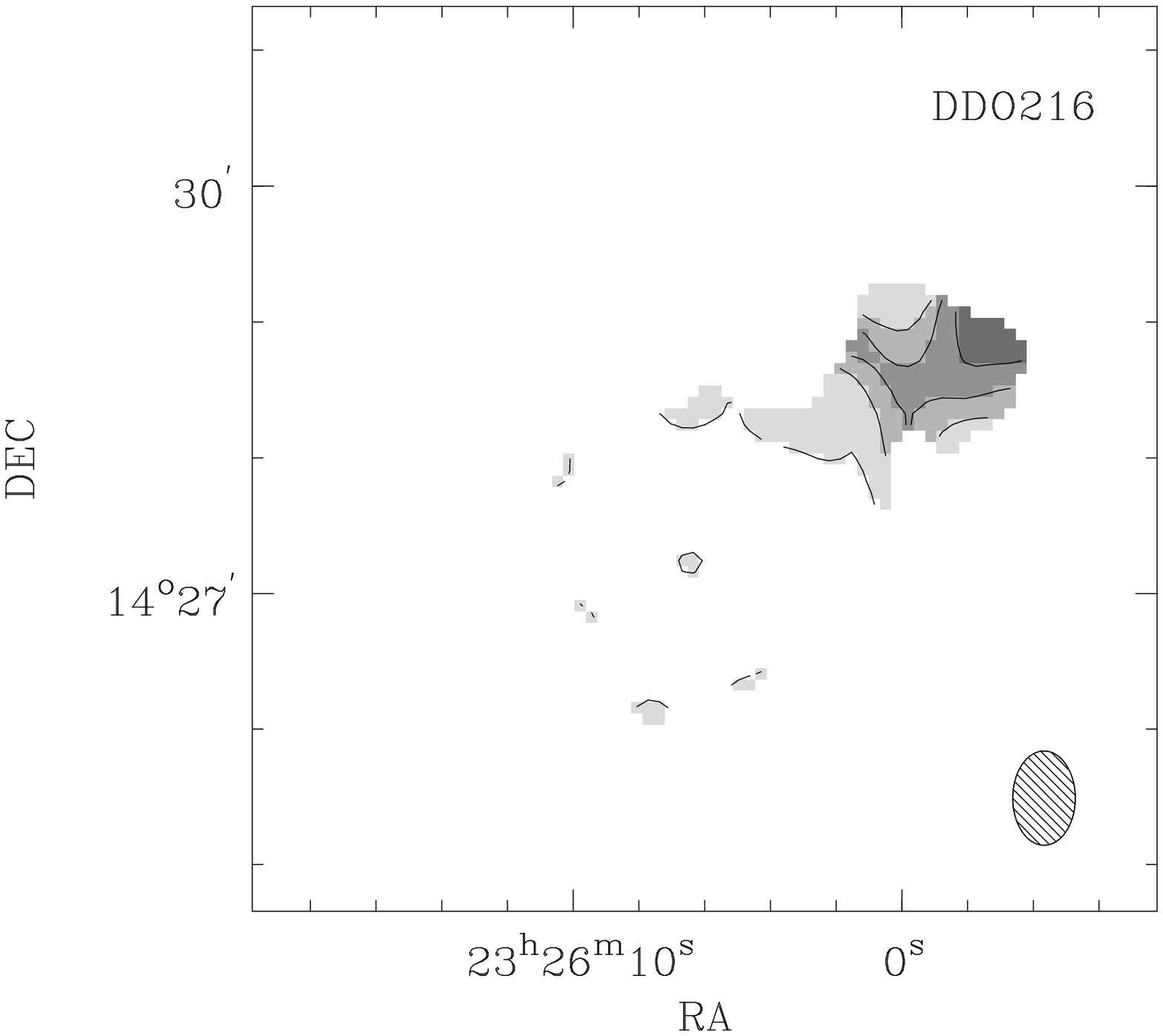}}
\end{minipage}
\begin{minipage}[b]{5.7 cm}
\resizebox{5.7cm}{!}{\includegraphics[angle=-90]{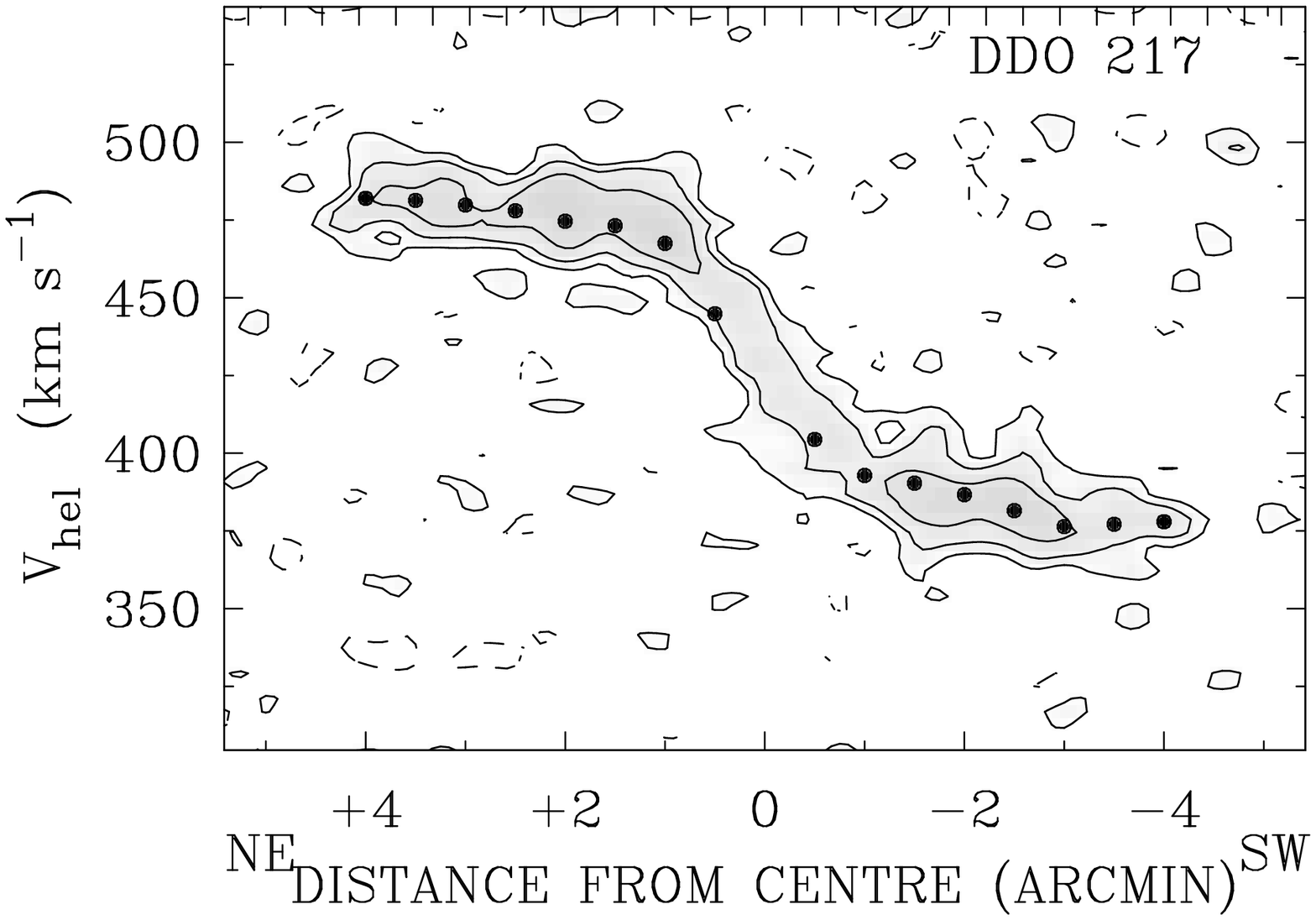}}
\end{minipage}
\hfill
\begin{minipage}[b]{5.7 cm}
\resizebox{5.7cm}{!}{\includegraphics[angle=-90]{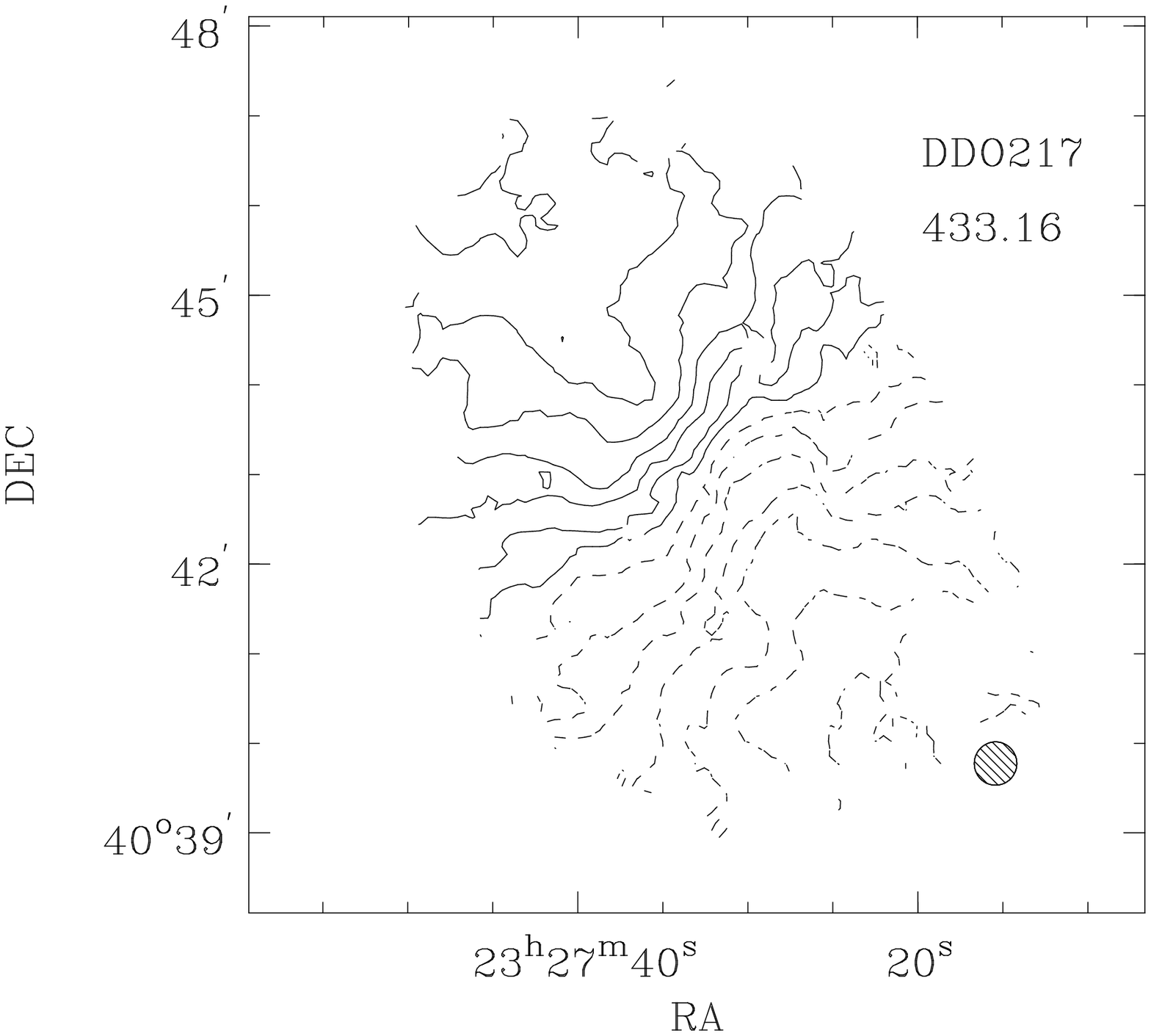}}
\end{minipage}
\hfill
\begin{minipage}[b]{5.7 cm}
\resizebox{5.85cm}{!}{\includegraphics[angle=-90]{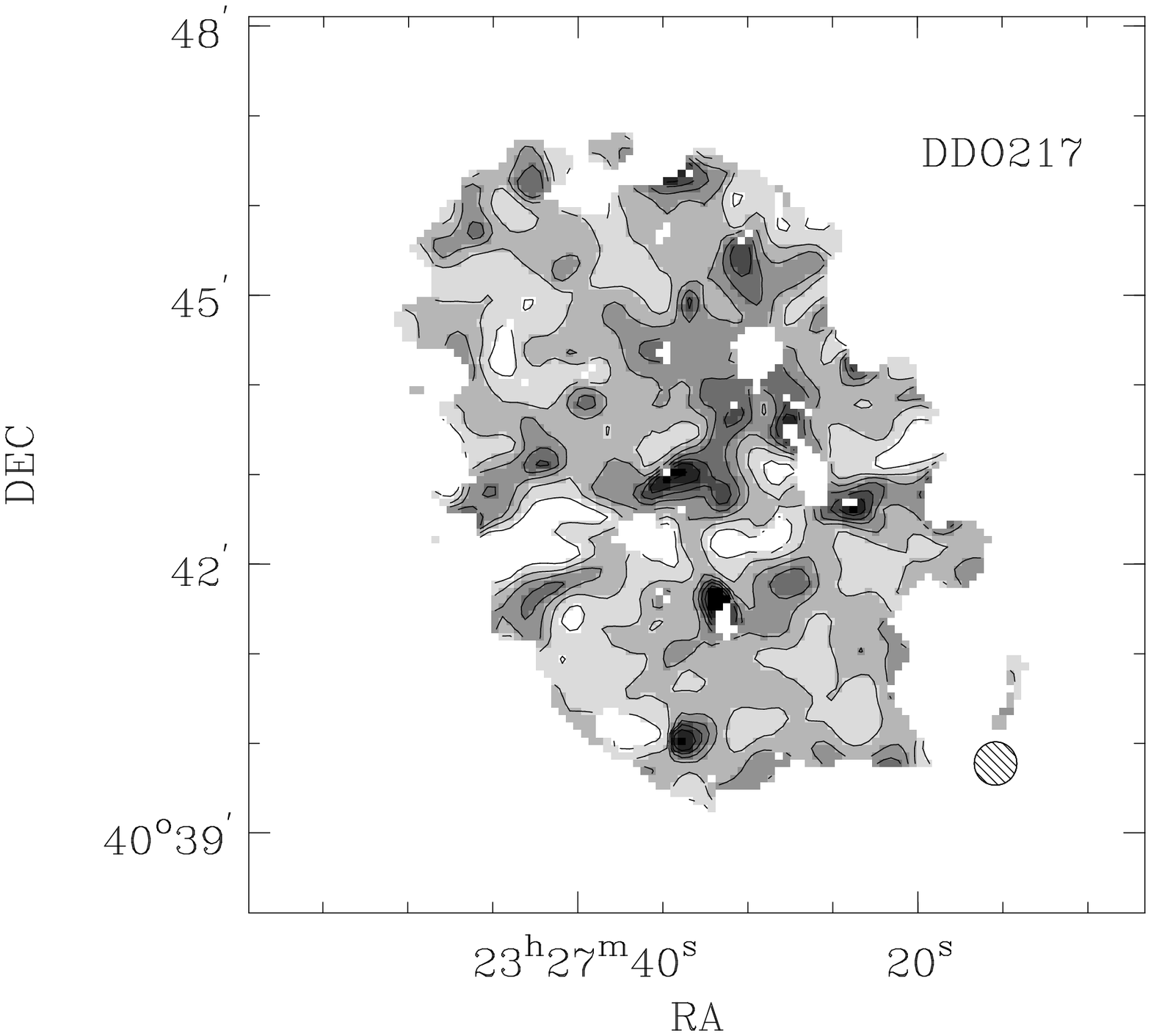}}
\end{minipage}
\caption{
\small Continued
}

\end{figure*}

{\small NGC\,}2537: The HI distribution consists of a U-shaped high 
column-density ridge. The velocity field, irregular in the north, exhibits 
the characteristic spider shape of a flattening rotation curve, also 
evident in the XV map. 
{\small NGC\,}2537 is sometimes classified as a Blue Compact
Dwarf (BCD) galaxy and is also known as the {\it Bear Paw Galaxy} 
(Schorn 1988). {\small UGC\,}4278 is a nearby companion. 
{\small NGC\,}2537A occurring just east of {\small NGC\,}2537 on PSS
plates is not visible in the HI data. The large linewidths in the center
are an artifact of the unresolved velocity gradient.

{\small UGC\,}4278: The rotation curve may show a turnover. 
{\small NGC\,}2537 is $16'.6$ away at PA = $328^\circ$.

{\small DDO\,}52: Most of the HI is in a low-column-density disk, with
a peak $N_{\rm HI}= 1.5\cdot 10^{21} \cm2$. The velocity field is regular 
with a hint of a flattening rotation curve. The XV diagram shows a 
rather high ratio of rotational to random velocity, despite its small 
amplitude.

{\small DDO\,}63: The HI is concentrated in a ring with a high column
density, with a five times lower central minimum ($N_{\rm HI}= 2.4\cdot 
10^{20} \cm2$) at $\alpha=\rm 9^h36^m 3^s.1$ $\delta=\rm 71^\circ24'44''$. 
The kinematic and HI major axes are misaligned by about $30^\circ$. 
The rotation velocity is comparable to the HI velocity dispersion. The 
ring dominates the XV map. {\small DDO\,}63 has also been studied by 
Puche \& Westpfahl (1994) and Tully et al. (1978). The latter find the 
same kinematic/HI axis misalignment, and a comparable well-ordered, 
small-amplitude rotation velocity.

{\small NGC\,}2976: The outer HI isophotes are well represented by
ellipses. The two continuum sources (Paper I) are near to but not
precisely coincident with the high HI column density regions 
($N_{\rm HI} \approx 3.5 \cdot 10^{21} \cm2$) on either side of the 
galaxy at $\alpha=\rm 9^h43^m1^s.2$, $\delta=\rm 68^\circ9'44''$ 
(NW) and $\alpha=\rm 9^h43^m19^s.1$, $\delta=\rm 68^\circ7'49''$ (SE). 
The emission at upper left in the velocity field map is unrelated G
alactic foreground HI. The rotation curve flattens near the edge of 
the HI disk. A detailed optical 
study of {\small NGC\,}2976 was presented  by Bronkala et al. (1992).

{DDO\,}64: A high-column-density ridge dominates the HI structure of
this probably nearly edge-on galaxy. The velocity 
field is poorly resolved along the minor axis. The XV map shows nearly 
solid-body rotation with only a hint of flattening. The feature south
of {\small DDO\,}64 is {\small UGC\,}5272B, the feature north of it
is probably noise.

%
\begin{figure}
\centerline{\resizebox{9cm}{!}{\includegraphics[angle=-90]{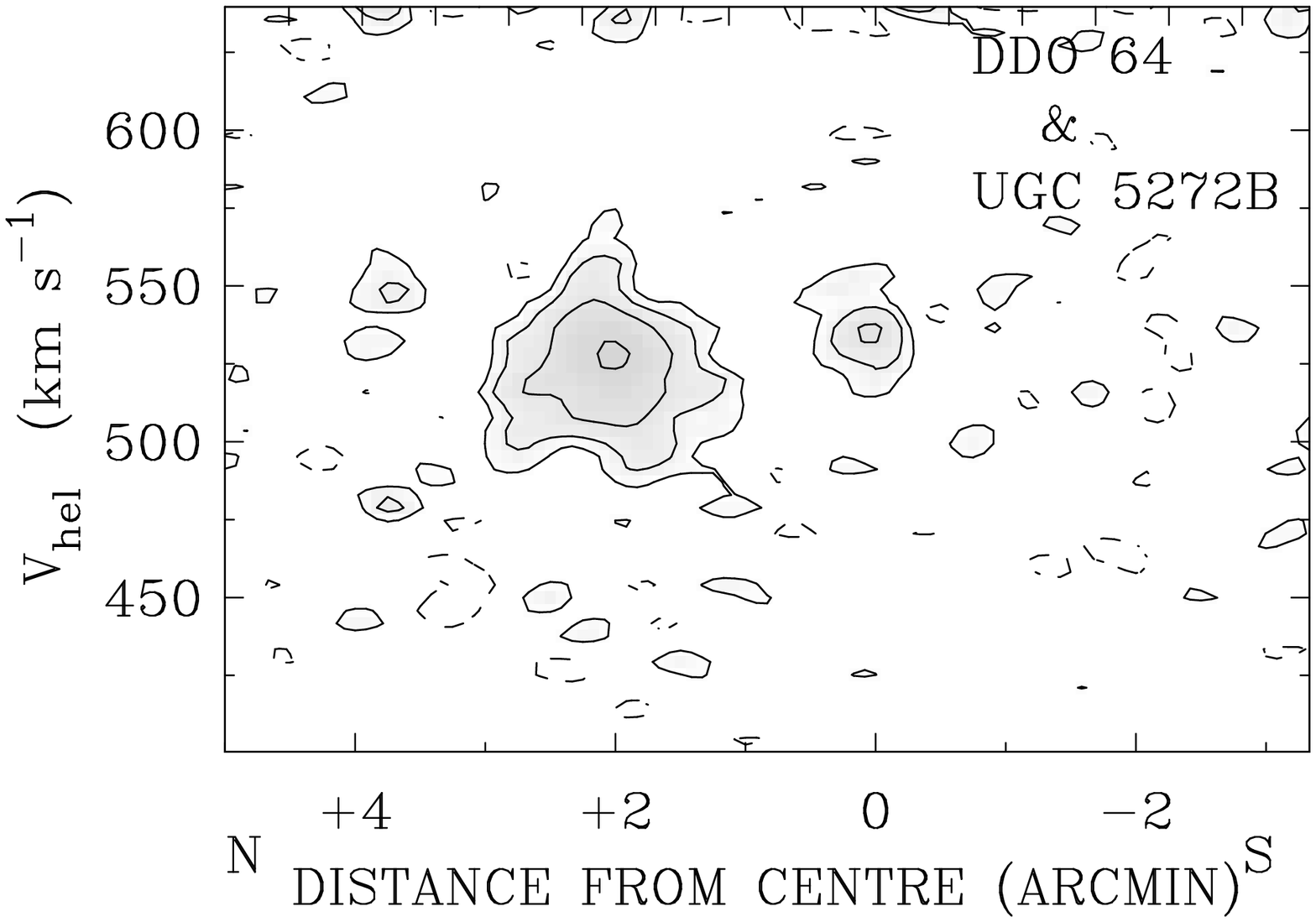}}}
\parbox[t]{\hsize}{
\caption{\small Position-velocity map through {\small UGC\,}5272B in 
position angle 21 degrees, with contourlevels at $-5.07$, $5.07$
($2\sigma$), $10.14$, $20.28$ and $40.56$ mJy per beam. {\small
UGC\,}5272B is visible at the center of the slice ($0'$); {\small
DDO\,}~64 is the larger patch to the left (north) of {\small
UGC\,}5272B. The slice also intersects the feature which can be seen
north of {\small DDO\,}64 in the HI column density map. Not much is
evident in the X-V map.
}} 
\label{U5272B-XV}
\end{figure}

{\small DDO\,}68: At low column density levels, {\small DDO\,}68 is
reasonably symmetrical.  High-column-density regions are found in the
north and east of the galaxy. A deep hole devoid of HI emission is 
located at $\alpha=\rm 9^h53^m52^s.2$, $\delta=\rm 29^\circ 4'15''$. 
Low signal-to-noise regions contribute to the irregular appearance of 
the velocity field. The ratio of rotational to random velocity is low.

{\small DDO\,}73: The HI isophotes suggest a nearly face-on orientation. 
However, both velocity field and XV map indicate projected rotational 
velocities considerably in excess of the HI velocity dispersion.

{\small DDO\,}83: The HI column density is high throughout the
galaxy. The velocity field is regular and spider-like. The rotation
curve rises strongly near the center and flattens at the edge of
the HI disk.

{\small DDO\,}87: The HI is distributed over a low-column-density disk 
with a number of small high column density regions, unresolved at $13''$ \
resolution. The fragmented appearance of the velocity field is the result 
of low signal-to-noise ratios. The XV map shows a nearly flat rotation 
curve in the outer regions. The low luminosity of {\small DDO\,}87 is 
inferred from its association with the M81 group (Huchtmeier $\&$ 
Skillman 1998).

Mk\,178: This galaxy is poorly resolved spatially. Its HI
structure, kinematics and luminosity are similar to {\small DDO\,}63,
{\small DDO\,}125 and {\small DDO\,}165. 

{\small NGC\,}3738: The HI column density is high everywhere, with
a central peak $N_{\rm HI}=4.5 \cdot10^{21} \cm2$. The velocity profile 
and the XV map indicate a steep velocity gradient, which is difficult to 
fit with the tilted ring method.  The indicated rotation velocity was 
fitted by eye to the XV map. The high velocity dispersion is an artifact
of the large velocity gradient in this marginally resolved galaxy.

{\small DDO\,}101: HI extent is too limited to show structure.

{\small DDO\,}123: The HI is distributed evenly throughout the
disk of this face-on galaxy. Irregularities in the velocity field 
coincide with low signal-to-noise regions. The rotation curve rises out 
to the edge of the HI disk.

Mk\,209: The peak of the high-column-density region is $N_{\rm
HI}=2.9\cdot 10^{21} \cm2$. The velocity field appears regular
but could not be fitted properly because of insufficient resolution.
The velocity gradient was fitted manually to the XV map; it is 
consistent with a solid body rotation curve. HI in Mk\,209 (IZw\,36) 
has also been observed with the VLA by Viallefond et al. (1987).

{\small DDO\,}125: The HI is mainly concentrated in two
high-column-density regions, separated by a low-column density
center, suggesting a fragmented ring. The velocity field shows a 
velocity gradient along the major axis. Rotation is clearly demonstrated 
by the XV map. HI in {\small DDO\,}125 was studied in detail by Tully et
al. (1978).

{\small DDO\,}133: The HI isophotes are well-represented by ellipses at 
the level of $N_{\rm HI}=3 \cdot10^{20} \cm2$. The rotation curve
flattens slightly outwards.

{\small DDO\,}165: Most of the HI is located in a ring. The velocity
field is highly irregular. The maximum velocity gradient is along a
line from SE to NW in position angle $120^\circ - 140^\circ$. The 
rotation velocity was fitted to the XV map assuming solid-body rotation. 
However, the emission in the XV map is mainly due to the ring.

{\small DDO\,}166: A ridge of high-column-density HI extends
over the eastern side of this face-on galaxy. The velocity field has a
strong gradient, in spite of the small inclination suggested by the HI 
isophotes. The isovelocity contours are twisted into an S-shape at 
the eastern side of the galaxy. Thean et al. (1997) have published VLA
HI maps of {\small NGC\,}5033, {\small DDO\,}166 and {\small UGC\,}8314.

{\small DDO\,}168: Two very high HI column density regions 
($N_{\rm HI} \geq 6 \cdot 10^{21} \cm2$) occur near the center of the 
galaxy at $\alpha=\rm 13^h12^m16^s.8$, $\delta=\rm 46^\circ 11'0''$
and $\alpha=\rm 13^h12^m15^s.8$, $\delta=\rm 46^\circ11'30''$. The 
position angle of the velocity gradient changes over the disk by 
approximately $20^\circ$. The peculiar structure of {\small DDO\,}168 
is not unique. Similar very-high column density regions combined with 
twisted velocity fields have been observed in other `amorphous galaxies' 
such as {\small NGC\,}1140 (Hunter et al. 1994) and IZw~18 (Viallefond 
et al. 1987, Van Zee et al. 1998).  A detailed study of the mass 
distribution in {\small DDO\,}168 was performed by Broeils (1992).

{\small DDO\,}185: The HI column density map suggests a disk seen at a high
inclination. The isovelocity contours are regularly spaced, consistent
with the rising rotation curve shown by the position-velocity
map. {\small DDO\,}185 (=Holmberg IV) was used as a calibrator galaxy for the
Tully-Fisher relation by Kraan-Korteweg et al. (1988).

{\small DDO\,}190: The highest column densities are found on the west. 
The velocity field is somewhat irregular, but rotation is clearly visible. 
The XV map shows a hint of flattening of the rotation curve on the NW side.

{\small DDO\,}216: The HI is located in the southern half of the (optical)
galaxy (Sandage 1986, Lo et al 1993). There is little sign of rotation 
in the velocity field and XV maps. In fact, the velocity gradient 
suggested by the XV map may represent a single HI cloud at a discrepant 
velocity. For a VLA study of {\small DDO\,}216, see Lo et al. (1993).

{\small DDO\,}217: The HI is distributed relatively evenly over the
disk. The velocity field shows differential rotation over most of
the galaxy. The difference in position angle between the inner region
and the outer disk is clearly visible in the velocity field map. The 
XV map shows a rapid rise of the rotation velocity near the center and 
a gradual increase in the outer disk.

\vskip 0.8cm
\noindent
{\bf Appendix A: Correction for the velocity gradient}
\vskip 0.8cm
\label{dispcor-sec}

\begin{figure}
\mbox{}\\
\resizebox{8cm}{!}{\includegraphics[angle=-90]{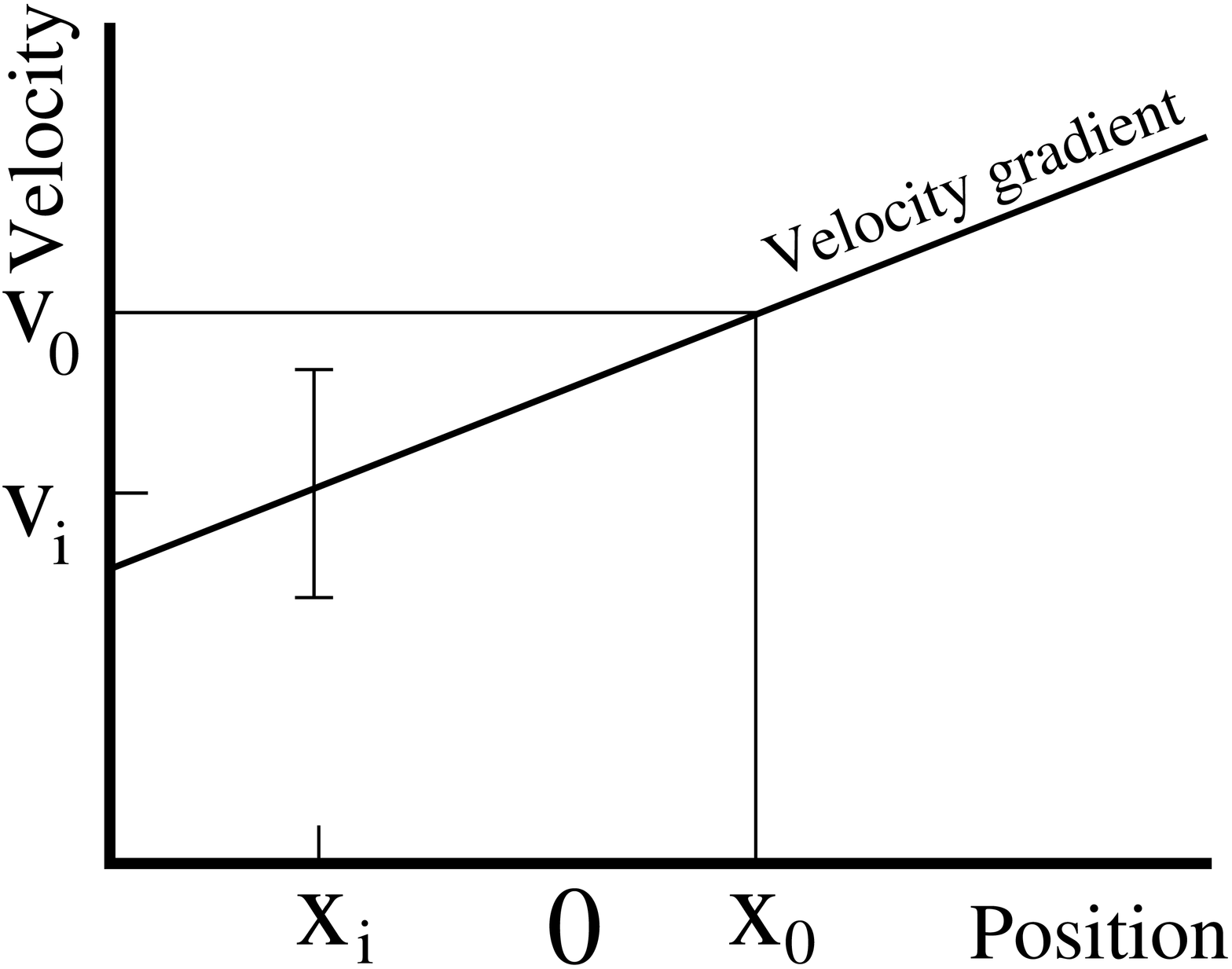}}
\caption{\small Definition of symbols used in the calculation of the
broadening of the line profile by a velocity gradient over the
synthesized beam. The bar at the position $x_i,v_i$ indicates the
velocity dispersion $\sigma_i$ at that position. The center of the
beam is $x=0$ by definition.
\label{velgrad}
}
\end{figure}

We assume a well-resolved velocity field so that second and higher order
derivatives of the velocity field can be neglected. We also assume 
an HI disk of negligible thickness so that all lines of sight cross 
the disk at a single radius, implying the one-dimensional linear situation 
sketched in Fig.~\ref{velgrad}. We choose the x-axis along the 
velocity gradient, with the zeropoint at the center of the Gaussian 
beam. The intensity-weighted mean velocity over the beam corresponds 
to that of position $x_0$, which is not necessarily the center of the 
beam. 

We now divide the beam into many ($N$) lines of sight, 
each with a large number of identical elements ($M$) with velocities 
$v_{ik}$, $k=1 \ldots M$ at position $x_i$, $i=1 \ldots N$. This definition 
includes an implicit integration over the coordinate perpendicular to the 
velocity gradient. The elements are identified with individual HI clouds of
very small intrinsic velocity dispersion.  At every position $x_i$ we 
define the mean velocity $v_i$ as $v_i={1 \over M} \sum_k v_{ik}$, which is
related to the intensity-weighted mean velocity $v_0$ and the velocity 
gradient $\nabla v$ through
$$
v_i - v_0 = (x_i-x_0) \nabla v
$$ 
The velocity dispersion of the elements at position $x_i$ is 
$\sigma_i={1 \over {M-1}} \sum_k (v_{ik}-v_i)^2$. 
Substitution of $v_i=v_0+(x_i-x_0) \nabla v$ and evaluation of the 
cross-product yields
$$
\sigma^2_i={1 \over {M-1}} \sum_{k=1}^M \{(v_{ik}-v_0)^2 
$$
$$
-2(v_{ik}-v_0)(x_i-x_0)\nabla v +(x_i-x_0)^2(\nabla v)^2 \} 
$$
The third term is independent of $k$. For the second term we may write
$$
-2 (x_i-x_0)\nabla v \sum_{k=1}^M(v_{ik}-v_0)=
$$
$$
-{{2M}\over{M-1}}(v_i-v_0)(x_i-x_0)\nabla v =
$$
$$
-{{2M}\over{M-1}}(x_i-x_0)^2(\nabla v)^2
$$
Therefore, we have 
$$
\sigma^2_i=
$$
$$
{1 \over {M-1}} \sum_{k=1}^M (v_{ik}-v_0)^2-{{M}\over{M-1}}(x_i-x_0)^2(\nabla v)^2
$$
The intensity-weighted mean velocity dispersion over the beam is
$$
\langle \sigma^2 \rangle ={ {\sum_{i=1}^N w_i \sigma^2_i} \over {\sum_{i=1}^N w_i}}
$$ 
with weight $w_i=e^{-x^2/b^2} I_i$. Therefore, with $M \gg 1$
$$
\langle \sigma^2 \rangle = 
$$
$$
{{\sum_{i=1}^N w_i \sum_{k=1}^M (v_{ik}-v_0)^2} \over{(M-1) \sum_{i=1}^N w_i}}-(\nabla v)^2{ {\sum_{i=1}^N w_i (x_i-x_0)^2} \over {\sum_{i=1}^N w_i}}
$$
This equation is of the general form
$$
\langle \sigma^2 \rangle = \sigma^2_{obs} - \Omega_I b^2 (\nabla v)^2
$$
where $\sigma_{obs}$ is the observed dispersion of a local line
profile, corrected for the instrumental spectral resolution.
The second term is the line broadening due to the velocity gradient
over the beam. Note that the velocity gradient $\nabla v$ is a function 
of position if the galaxy is not in solid-body rotation, necessitating 
use of a model velocity field constructed from the rotation curve in 
order to calculate $\nabla v$ at every position. The coefficient 
$\Omega_I > 0$ is a weighted mean of the intensity distribution over the 
beam. If we assume a constant intensity (i.e. $I_i=I_0$) over the Gaussian 
beam $e^{-(x^2/b^2)}$, we have $\Omega_I = {1 \over 2}$ :
$$
\langle \sigma^2 \rangle = \sigma^2_{obs} - {1 \over 2} b^2 (\nabla v)^2
$$ 
The error introduced by the assumption of a constant intensity can be
estimated by calculating the correction for simple analytical
intensity distributions. For any assumed intensity distribution
$$
x_0={{\sum_{i=1}^N w_i x_i} \over {\sum_{i=1}^N w_i}}
$$
and $\langle \sigma^2 \rangle$ can be calculated. The coefficients
$\Omega_I$ are given for three types of intensity distribution
in Table~\ref{velgrad-tab}).

\begin{figure}
\begin{minipage}[t]{8cm}
\mbox{}\\
\resizebox{8cm}{!}{\includegraphics[angle=0]{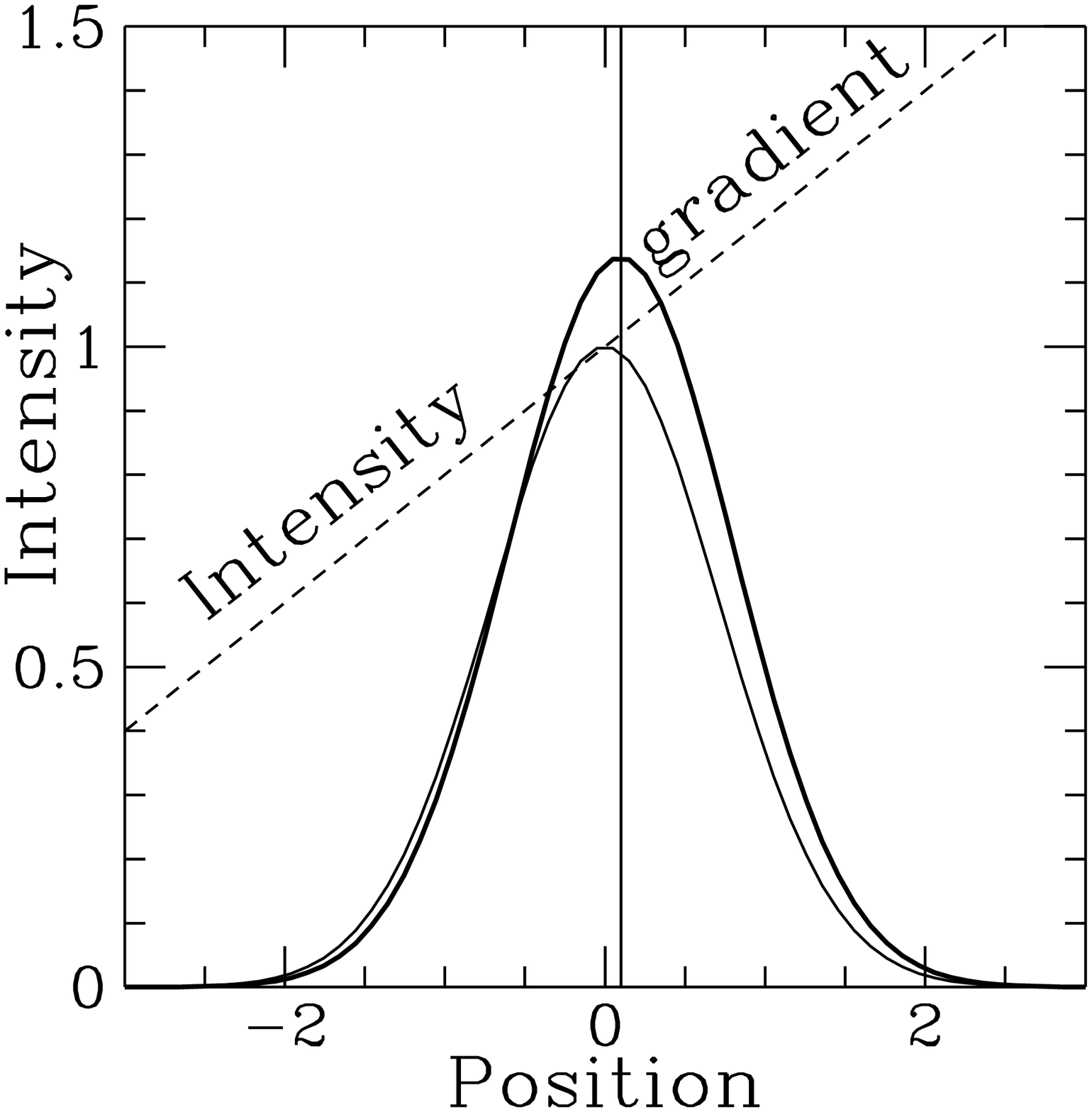}}
\end{minipage}
\caption{\small The effect of an intensity gradient $I(x)=1+0.2\cdot x$ 
(dashed line; $a = 0.2$ in Table~\ref{velgrad-tab}) over the beam (thin
curve). The thick curve is the product of the intensity and the
beamshape function with an arbitrary scaling. The vertical solid line
marks the position $x_0 = 0.1$. The small difference between the thin
and the thick curve (scaling in intensity is free) is the reason that
the value of $\Omega_I$ is not sensitive to an intensity gradient over
the synthesized beam. In this case, $\Omega_I=0.490$ although \newline
$I(x=1)/I(x=-1)=1.5$.
\label{intgrad}
}
\end{figure}

Shallow intensity gradients do not make much of a difference. To 
first order, the effect of an intensity gradient is to
shift the distribution of $w_i$ in the direction of the intensity
gradient. Since the beam function falls off rapidly for large x, only
large gradients produce a significant difference with constant
intensity. The greatest effect on the correction for the velocity
gradient is brought about by the symmetric distribution. If the
emission is highly concentrated towards the center of the beam, the
velocity gradient has no effect. On the other hand, if the emission is
concentrated in the wings of the beam, the effect of the velocity
gradient is maximal. If the intensity does not change more than 50\%
over the beam, the error in $\Omega_I$ introduced by the assumption of
constant intensity is of the order of 10\%.

\begin{table}
\caption{Values of $\Omega_I$ for a gradient, a minimum and
an jump in the intensity distribution at the center of the 
synthesized beam}
\begin{center}
\begin{tabular}{|  c |  c |  c  | c |} 
\hline 
    & Int. gradient    & Int. minimum  & Int. jump \\
\cline{2-4} 
 a  &  $I(x)=1+ax$  &  $I(x)=1+ax^2$  &  $I(x)=1-aH(x)$ \\ 
\hline 
0.0  &  0.500  &  0.500  &  0.500 \\
0.1  &  0.498  &  0.548  &  0.497 \\ 
0.2  &  0.490  &  0.591  &  0.487 \\ 
0.5  &  0.439  &  0.700  &  0.421 \\ 
1.0  &  0.349  &  0.833  &  0.182 \\ 
\hline 
\end{tabular} 
\end{center}
Note: Scale factor $a$ defines the magnitude of the intensity 
change over the beam. The position $x$ is in units of beamsize 
$b$. At positions where the indicated functional forms are negative, 
the intensity was set to zero. The symbol $H(x)$ is used for the 
heaviside function $H(x)={x \over {\mid x \mid}}$. 
\label{velgrad-tab}
\end{table}

\end{document}